\documentclass[11pt]{article}
\usepackage{a4wide}
\usepackage{amsmath,amssymb}
\usepackage{graphicx}
\usepackage{url}
\usepackage{mathrsfs}
\usepackage[bookmarks=true,bookmarksnumbered=true,setpagesize=false]{hyperref}
\usepackage{bm}
\usepackage{braket}
\usepackage{comment}
\usepackage{here}

\usepackage{mathrsfs}
\newcommand{\ms}{\mathscr}

\usepackage{color}

\newcommand{\tx}{\text}
\newcommand{\ol}{\overline}
\newcommand{\fn}[1]{\!\left(#1\right)}
\newcommand{\al}[1]{\begin{align}#1\end{align}}
\newcommand{\als}[1]{\begin{align*}#1\end{align*}}
\newcommand{\nn}{\nonumber\\ }
\newcommand{\ov}{\over}
\newcommand{\pn}[1]{\left(#1\right)}
\newcommand{\Pn}[1]{\bigl(#1\bigr)}

\newcommand{\bs}{\boldsymbol}
\newcommand{\mc}{\mathcal}
\newcommand{\mf}{\mathfrak}
\newcommand{\paren}[1]{\left(#1\right)}
\newcommand{\sqbr}[1]{\left[#1\right]}
\newcommand{\ab}[1]{\left|#1\right|}
\newcommand{\br}[1]{\left\{#1\right\}}

\newcommand{\Ab}[1]{\bigl|#1\bigr|}

\newcommand{\Fn}[1]{\!\bigl(#1\bigr)}

\newcommand{\cblue}[1]{{\color{blue}#1}}
\newcommand{\pal}{\partial}
\newcommand{\Tin}{T_\text{in}}
\newcommand{\Tout}{T_\text{out}}
\newcommand{\half}{\frac{1}{2}}

\newcommand{\df}{\text{d}}
\newcommand{\wt}{\widetilde}
\newcommand{\wh}{\widehat}

\newcommand{\bmf}[1]{\boldsymbol{\mathfrak#1}}

\DeclareMathOperator{\Ei}{Ei}

\begin{document}
\title{
Wave-Packet Effects: A Solution for Isospin Anomalies\\
in Vector-Meson Decay
}
\author{Kenzo Ishikawa,${}^{1,2,}$\thanks{E-mail: \tt ishikawa@particle.sci.hokudai.ac.jp} \mbox{} 
Osamu Jinnouchi,${}^{3,}$\thanks{E-mail: \tt jinnouchi@phys.titech.ac.jp} \mbox{} 
Kenji Nishiwaki,${}^{4,}$\thanks{E-mail: \tt kenji.nishiwaki@snu.edu.in} \mbox{}  and
Kin-ya Oda${}^{5,}$\thanks{E-mail: \tt odakin@lab.twcu.ac.jp} \bigskip\\
\normalsize\it
${}^1$Department of Physics, Faculty of Science, Hokkaido
University,\\
\normalsize\it
Sapporo 060-0810, Japan\smallskip\\
\normalsize\it
${}^2$Natural Science Center, Keio University, Yokohama 223-8521, Japan\smallskip\\
\normalsize\it
${}^3$Department of Physics, Faculty of Science, Tokyo Institute of Technology,\\
\normalsize\it
Tokyo 152-8550, Japan\smallskip\\
\normalsize\it
${}^4$Department of Physics, Shiv Nadar Institution of Eminence,\\
\normalsize\it
Gautam Buddha Nagar, 201314, India\smallskip\\
\normalsize\it
${}^5$Department of Mathematics, Tokyo Woman's Christian University,\\
\normalsize\it
Tokyo 167-8585, Japan
}
\maketitle

\begin{abstract}\noindent
There is a long-standing anomaly in the ratio of the decay width for $\psi(3770)\to D^0\ol{D^0}$ to that for $\psi(3770)\to D^+D^-$ at the level of $9.5\,\sigma$. A similar anomaly exists for the ratio of $\phi(1020)\to K_\tx{L}^0K_\tx{S}^0$ to $\phi(1020)\to K^+K^-$ at $2.1\,\sigma$.
In this study, we reassess the anomaly through the lens of Gaussian wave-packet formalism. Our comprehensive calculations include the localization of the overlap of the wave packets near the mass thresholds as well as the composite nature of the initial-state vector mesons. The results align within $\sim 1 \sigma$ confidence level with the Particle Data Group's central values for a physically reasonable value of the form-factor parameter, indicating a resolution to these anomalies. We also check the deviation of a wave-packet resonance from the Briet-Wigner shape and find that wide ranges of the wave-packet size are consistent with the experimental data.
\end{abstract}

\newpage
\tableofcontents

\newpage
\section{Introduction} 
There is a long-standing anomaly (discrepancy between experimental and theoretical results) in the ratio of the decay width for $\psi(3770)\to D^0\overline{D^0}$ to that for $\psi(3770)\to D^+D^-$.
A similar but weaker anomaly exists for the ratio of $\phi(1020)\to K_\tx{L}^0K_\tx{S}^0$ to $\phi(1020)\to K^+K^-$.
On the other hand, the ratio of $\Upsilon(4S)\to B^+B^-$ to $\Upsilon(4S)\to B^0\ol{B^0}$ is consistent with the standard theoretical predictions.

At the quark level, these processes are\footnote{
Here and hereafter, we omit $(3770)$, $(1020)$, and $(4S)$.
$\Upsilon(4S)$ is sometimes written as $\Upsilon(10580)$.
We do not distinguish the weak-interaction eigenstates $K^0\ol{K^0}$ and the mass eigenstates $K^0_\tx{L}K^0_\tx{S}$, neglecting the small $CP$ violating effects. Other processes have even smaller $CP$ violating effects and we neglect them too.
}
\als{
\phi\fn{s\ol s}
	&\to K^+\fn{u\bar s}K^-\fn{s\bar u},&
\psi\fn{c\ol c}
	&\to D^+\fn{c\ol d}D^-\fn{d\ol c},&
\Upsilon\fn{b\ol b}
	&\to	B^+\fn{u\ol b}B^-\fn{b\bar u},\\
\phi\fn{s\ol s}
	&\to K^0\fn{d\bar s}\ol{K^0}\fn{s\bar d}
	\to K^0_\tx{L}K^0_\tx{S},&
\psi\fn{c\ol c}
	&\to D^0\fn{c\ol u}\overline{D^0}\fn{u\ol c},&
\Upsilon\fn{b\ol b}
	&\to	B^0\fn{d\ol b}\ol{B^0}\fn{b\ol d},
}
which can be summarized as
$V\fn{Q\ol Q}\to P\fn{Q\ol q}+\ol P\fn{q\ol Q}$,
where $V$ and $P$ are vector and pseudo-scalar mesons, respectively, and $Q$ and $q$ are heavy ($s$, $c$, $b$) and light ($u$, $d$) quarks, respectively. This ratio of decay widths is theoretically clean because most of the quantum-chromodynamics (QCD) corrections cancel out between the numerator and the denominator.
These decay processes are via strong interaction, and hence in the limit of exact isospin symmetry $u\leftrightarrow d$, the ratio becomes unity.
The isospin violation makes a deviation from unity.

We name the ratio of the widths as\footnote{
In the original Ref.~\cite{ParticleDataGroup:2022pth}, the first two of Eq.~\eqref{PDG ratio result} are given in its inverse
\als{
\pn{R_\phi^{-1}}{}^\tx{PDG}
	&=	0.690\pm0.015,&
\pn{R_\psi^{-1}}{}^\tx{PDG}
	&=	1.253\pm0.016,
}
and we have inverted them in Eq.~\eqref{PDG ratio result}.
In the theoretical literature, the ratio~\eqref{R defined} of charged to neutral modes is mainly used, and we follow it for ease of comparison.
}
\al{
R_\phi
	&:=	{\Gamma\fn{\phi\to K^+K^-}\ov\Gamma\fn{\phi\to K_\tx{L}^0K_\tx{S}^0}},&
R_\psi
	&:=	{\Gamma\fn{\psi\to D^+D^-}\ov\Gamma\fn{\psi\to D^0\ol{D^0}}},&
R_\Upsilon
	&:=	{\Gamma\fn{\Upsilon\to B^+B^-}\ov\Gamma\fn{\Upsilon\to B^0\ol{B^0}}}.
	\label{R defined}
}
The experimental results are combined by the Particle Data Group~(PDG)~\cite{ParticleDataGroup:2022pth}:\footnote{
In the evaluation of $R_\Upsilon = 1.058 \pm 0.024$~\cite{ParticleDataGroup:2022pth}, the following isospin symmetry among, e.g., $B^0 \to J/\psi\, K_\tx{S}$ and $B^+ \to J/\psi\, K^+$ is assumed. The extent to which the isospin symmetry is valid in hadronic decays is debatable [Private communication with Dr.~Akimasa~Ishikawa].
}
\al{
R_\phi^\tx{PDG}
	&=	1.45\pm0.03,&
R_\psi^\tx{PDG}
	&=	0.798\pm0.010,&
R_\Upsilon^\tx{PDG}
	&=	1.058\pm0.024.
		\label{PDG ratio result}
}

The theoretical prediction of the decay rates for $V\to P^+P^-$ and $V\to P^0\ol{P^0}$ is based on the plane-wave formalism so far. The tree-level result of the chiral perturbation theory reads
\al{
R_V^\tx{plane}
	=	{g_{V+}^2\ov g_{V0}^2}\pn{m_V^2-4m_{P^+}^2\ov m_V^2-4m_{P^0}^2}^{3/2},
}
where $g_{V+}$ ($g_{V0}$) is the coupling between $V$ and $P^+P^-$ ($P^0\ol{P^0}$) and $m_V$, $m_{P^+}$, and $m_{P^0}$ are the masses of $V$, $P^+$, and $P^0$, respectively. Even if we assume an isospin-symmetric coupling $g_{V+}=g_{V0}$, the difference in the pseudo-scalar-meson masses $m_{P^+}\neq m_{P^0}$ results in a deviation in $R_V$ from unity:
Putting the mass values in Ref.~\cite{ParticleDataGroup:2022pth},\footnote{
Concretely,
\als{
m_\phi
	&=	\pn{1019.461\pm0.016}\tx{MeV},	&	
m_\psi
	&=	\pn{3773.7\pm0.4}\tx{MeV},&
m_\Upsilon
	&=	\pn{10579.4\pm1.2}\tx{MeV},\\
2m_{K^+}
	&=	\pn{987.354\pm0.032}\tx{MeV}, &
2m_{D^+}
	&=	\pn{3739.32\pm0.10}\tx{MeV},&
2m_{B^+}
	&=	\pn{10558.7\pm0.24}\tx{MeV},\\
2m_{K^0}
	&=	\pn{995.222\pm0.026}\tx{MeV},&
2m_{D^0}
	&=	\pn{3729.68\pm0.10}\tx{MeV},&
2m_{B^0}
	&=	\pn{10559.3\pm0.24}\tx{MeV},
}
assuming the standard error propagation for the twice pseudo-scalar mass.
The total decay widths are
\als{
\Gamma_\phi
	&=	\pn{4.249\pm0.013}\tx{MeV},&
\Gamma_\psi
	&=	\pn{27.2\pm1.0}\tx{MeV},&
\Gamma_\Upsilon
	&=	\pn{20.5\pm2.5}\tx{MeV}.
}
}
we obtain
\al{
R_\phi^\tx{plane}
	&=	{g_{\phi+}^2\ov g_{\phi0}^2}\pn{1.5156\pm0.0033},&
R_\psi^\tx{plane}
	&=	{g_{\psi+}^2\ov g_{\psi0}^2}\pn{0.6915\pm0.0046},&
R_\Upsilon^\tx{plane}
	&=	{g_{\Upsilon+}^2\ov g_{\Upsilon0}^2}\pn{1.047\pm0.026}.
		\label{tree plane}
}
Comparing Eqs.~\eqref{PDG ratio result} and \eqref{tree plane}, we see that the isospin-symmetric limit for the coupling $g_{V+}=g_{V0}$ results in the anomaly at the level of $2.1\,\sigma$, $9.5\,\sigma$, and $0.32\,\sigma$ for $\phi$, $\psi$, and $\Upsilon$, respectively.

We briefly review the theoretical accounts for the anomaly within plane-wave formalism.
For $R_\phi$, it turned out that radiative corrections make the anomaly more significant~\cite{Bramon}:
The standard quantum-electrodynamics (QED) corrections make the theoretical prediction of the ratio 4\,\% larger, and isospin-breaking corrections to the ratio $g^2_{\phi+}/g^2_{\phi0}$ further make it ``some 2\,\%''~\cite{Bramon} larger, leading to a larger anomaly of roughly $5.2\,\sigma$ assuming that the error is dominated by that in Eq.~\eqref{PDG ratio result}.
In Ref.~\cite{Fischbach} the authors introduce a smeared decay rate that is a function of the energy difference between the initial and final plane-wave states;
this smearing is by the Lorentzian distribution due to the inclusion of the width as well as by a phenomenological form factor put by hand to regularize an ultraviolet (UV) divergence;
the anomaly for $\phi$ can be explained with a mass parameter $M\simeq1.5\,\tx{GeV}$ in the phenomenological form factor.
In Ref.~\cite{FloresBaez:2008jp}, the authors have estimated the effects of the electromagnetic structure of kaons and other model-dependent contributions to the radiative corrections, and the resultant corrections have turned out to be tiny.
In Ref.~\cite{Kuksa:2009zz}, two (a Breit-Wigner and a non-relativistic Lorentzian) types of averaged decay widths over the initial-state energy are introduced with two phenomenologically chosen energy intervals 1.010--1.060\,GeV and 1.000--1.100\,GeV, to relax the anomaly.

For $R_\psi$, another type of averaged decay width is introduced in Ref.~\cite{Coito:2017ppc}, and the resultant anomaly has become even more significant. There is no explanation for this $9.5\,\sigma$ anomaly so far.

The above smearing/averaging over the energy provides significant effects because the decay $V\to P\ol P$ is near the threshold $m_V\simeq 2m_P$.
In situations near the threshold, it is desirable to treat the decay more rigorously by using wave packets for the initial and final states. Recall that the $S$-matrix in the plane-wave formalism contains the energy-momentum-conserving delta function and is theoretically ill-defined when computing the probability rather than the rate. A well-defined decay probability can be calculated only as a transition from a wave packet to a pair of wave packets. This is theoretically more reliable.

In the previous analyses~\cite{Fischbach,FloresBaez:2008jp,Kuksa:2009zz}, it has been assumed that the transition processes are described by the (plane-wave) rates alone.\footnote{
See also Refs.~\cite{Joichi:1997xn,Yoshimi:2021qcs} for non-standard approaches within the plane-wave formalisms. 
}
In this paper, we present an analysis based on the transition probability of the normalized states, wave packets, without the divergence of the delta-function squared.
Concretely, we compute the decay $V\to P\ol P$ in the Gaussian wave-packet formalism~\cite{Ishikawa-Shimomura,Ishikawa-Tobita,Ishikawa-Oda,Ishikawa-Nishiwaki-Oda2,ishikawa-nishiwaki-oda1}; see also Refs.~\cite{Ishikawa-Tobita-AN,Ishikawa-Tajima-Tobita,Maeda}.\footnote{
There is an ongoing experimental project directly to confirm this wave-packet effect~\cite{Ishikawa-Jinnouchi,Ushioda-}; see also Ref.~\cite{RevModPhys.29.94}.
}
In particular, we include a wave packet effect, called the \emph{in-time-boundary effect} for the decay, by simply limiting the time-integral of the decay-interaction point to $t>T_\tx{in}$~\cite{Ishikawa-Tobita}. Here, $T_\tx{in}$ is the time from which the interaction is switched on.
This procedure is proven to provide approximate modeling of the full production process of $V$ in the corresponding two-to-two wave-packet scattering, say, $e^+e^-\to V\to P\ol P$~\cite{ishikawa-nishiwaki-oda1}; see also Refs.~\cite{Nauenberg:1998vy,Achasov:2003cn,Ishikawa-Oda,Ishikawa-Nishiwaki-Oda2,Oda:2021tiv,Oda:2023qek,Mitani:2023hpd} for related discussions.

The organization of this paper is as follows:
In Section~\ref{sec:basics}, we will introduce the minimum basics of calculating the (generalized) $S$-matrix that describes wave-packet-to-wave-packet transitions considering the initial state's decaying nature when wave packets take the Gaussian form.
In Section~\ref{magnitude section}, we will review significant properties of the Gaussian wave-packet $S$-matrix.
In Section~\ref{sec:R_V}, we will compare the theoretical predictions for the ratios of $R_\phi$, $R_\psi$, and $R_\Upsilon$ in the wave-packet and the plane-wave formalisms taking into account the form factor of the vector mesons.
In Section~\ref{sec:resonance}, we will discuss the constraint from the resonant shape in the electron-positron-collider experiments for $\phi$ and $\psi$.
In Section~\ref{sec:summary}, we will provide a summary and further discussions.
In Appendix~\ref{form factor session}, we will review the form-factor details for vector mesons used for our analysis.
In Appendix~\ref{detailed decay probability section}, we will provide the details on how to derive the total probability of $V \to P \ol{P}$ under non-relativistic approximations.
In Appendix~\ref{decay rate section}, a brief review on how to derive the plane-wave decay rate for $V \to P \ol{P}$ will be provided.
In Appendix~\ref{sec:Gamma-zero-limit}, we will comment on a specific formal limit where the wave-packet decay rate coincides with the plane-wave decay rate.
In Appendix~\ref{sec:R-rho}, we will briefly consider the isospin violation on the $\rho$ system.

\section{Basics of Gaussian wave-packet formalism\label{sec:basics}}
For the near-threshold decay, the velocities in the final state are small, and the overlap of the wave packets becomes more significant in general.
Therefore, it is important to take them into account.

Here, we spell out how to compute the probability for the $V\to P\ol P$ decay in the Gaussian wave-packet formalism. 
Throughout this paper, we work in the natural units $\hbar=c=1$.
Readers who are more interested in analyses of experimental results rather than detailed theoretical formulation may skim through this section.

\subsection{Wave-packet $S$-matrix
\label{sec:$S$-matrix}}

In the Gaussian wave-packet formalism, a transition from an initial wave-packet state $\ket{\mathcal{WP}_0}$
to a two-body final wave-packet state $
\ket{ \mathcal{WP}_1,\mathcal{WP}_2 }$
is characterized by the following generalized $S$-matrix~\cite{Ishikawa-Shimomura}:
\al{
S_{\mathcal{WP}_0 \to \mathcal{WP}_1\mathcal{WP}_2}
&=
\braket{ \mathcal{WP}_1,\mathcal{WP}_2 | \, \widehat{U}\fn{\Tout, \Tin} | \mathcal{WP}_0 },\label{S-matrix given first time}
}
where $\wh U$ describes the unitary time evolution from the initial time $\Tin$ to the final time $\Tout$
\al{
\widehat{U}\fn{\Tout, \Tin}
	:=
		\tx{T}\, \tx{exp}\fn{ -i \int_{\Tin}^{\Tout} \df t \,\int\df^3\bs x\, \widehat{ {\cal H} }_\tx{int}^\tx{(I)}\fn{t,\bs x}},
		\label{Dyson series}
}
in which $\tx{T}$ denotes the time-ordering and $\widehat{ {\cal H} }_\tx{int}^\tx{(I)}$ 
is the interaction Hamiltonian density in the interaction picture.
The local interaction point $\paren{t,\bs x}$ is integrated in the four-dimensional spacetime.
It is noteworthy that the wave-packet states $\ket{\mathcal{WP}_0}$ and $\ket{ \mathcal{WP}_1,\mathcal{WP}_2 }$ are normalizable and hence the transition amplitude~\eqref{S-matrix given first time} is finite, unlike in the ordinary plane-wave formalism.\footnote{
See ``Discussion'' subsection in Ref.~\cite{ishikawa-nishiwaki-oda1} for further discussion.
}
Through the Dyson series expansion of $\widehat{U}\fn{\Tout, \Tin}$,
a perturbative $S$-matrix can be systematically constructed at any order of perturbation using Wick's theorem, as in the plane wave case~\cite{Ishikawa-Oda}.
Throughout this paper, the subscripts 0, 1, and 2 denote $V$, $P$, and $\ol P$, respectively.

A free Gaussian wave packet is characterized by a set of parameters
$\Set{ m, \sigma, X^0, \bs{X}, \bs{P} }$, where $m$ is the mass;
$\sigma$ is the width-squared; and
$X^0$ is a reference time at which the wave packet takes the Gaussian form with the central values of the peak position $\bs X$ and momentum~$\bs P$.

Within the chiral perturbation theory, the effective-interaction-Hamiltonian density is
\al{
\widehat{{\cal H}}^\tx{(I)}_\tx{int,eff}
	= i g_{V+} {\cal V}^\mu \sqbr{ {\cal P}^+ \pal_\mu {\cal P}^- - {\cal P}^- \pal_\mu {\cal P}^+ }
+  i g_{V0} {\cal V}^\mu \sqbr{ {\cal P}^0 \pal_\mu \overline{{\cal P}^0} - \overline{{\cal P}^0} \pal_\mu {\cal P}^0 },
	\label{eq:effective-Hamiltonian}
}
where $\cal V$, $\cal P^\pm$, ${\cal P}^0$, and $\overline{{\cal P}^0}$ are the fields representing the vector meson, the charged pseudo-scalar mesons, the neutral pseudo-scalar meson, and its antiparticle, respectively, 
and $g_{V+}$ and $g_{V0}$ are the vector-meson effective couplings to the charged pseudo-scalars 
and to the neutral pseudo-scalars, respectively. 
In this paper, we take the isospin-symmetric limit,
\al{
g_{V+} = g_{V0} \quad(=:g_V),
	\label{isospin-symmetric limit}
}
with which the effective coupling $g_\tx{eff}$ takes the form in the momentum space
\al{
g_\text{eff}\fn{\lambda_0,P_0,P_1,P_2}
	:=
		g_V \, \varepsilon_\mu\fn{P_0,\lambda_0} \paren{P_1^\mu - P_2^\mu},
		\label{effective coupling}
}
where 
$P_0$, $P_1$, and $P_2$ are the four-momenta of the vector meson $V$, the pseudo-scalar meson~$P$, and its antiparticle $\ol P$, respectively, and
$\varepsilon_\mu$ is the polarization vector of the vector meson with $\lambda_0$ being its helicity.\footnote{
In Eq.~\eqref{effective coupling}, $P_0=\pn{P_0^\mu}_{\mu=0,\dots,3}=\pn{P_0^0,\bs P_0}$ stands for the four-momentum of $V$, with its subscript denoting the initial particle. We use the same letter $P^0$ for the particle label of the neutral pseudo-scalar, with its zero denoting its charge. The distinction should be apparent from the context.
}


In this paper, we investigate the transition from an off-shell initial state for $V$ to an on-shell final state for $P\overline{P}$, having an off-shell energy $\wt E_0$ and on-shell ones $E_1$, $E_2$, respectively:\footnote{\label{WW}
This procedure of introducing $\wt E_0$ is equivalent to the Weisskopf-Wigner approximation~\cite{Weisskopf:1930au,Weisskopf:1930ps}.
See Ref.~\cite{Ishikawa:2014ela} for its inclusion in the Gaussian wave-packet formalism.
}
\al{
\wt{E}_0
	&:=
		\sqrt{m_V^2 + \bs{P}_0^2  -i \, \Gamma_V m_V}
	= \sqrt{E_0^2  -i \, \Gamma_V m_V}
	\simeq
		E_0 - i \frac{m_V}{2 E_0} \Gamma_V,\nn
E_1 &:= \sqrt{m_P^2 + \bs{P}_1^2 },\nn
E_2 &:= \sqrt{m_P^2 + \bs{P}_2^2 },
				\label{E0 tilde}
}
where
$m_V$ and $m_P$ are the masses of $V$ and $P$, respectively;
$E_0:= \sqrt{m_V^2 + \bs{P}_0^2}$ is the on-shell energy of $V$;
and $\Gamma_V$ is the ``decay width'' of $V$, or more precisely, the imaginary part of its plane-wave propagator divided by $m_V$; see Ref.~\cite{ishikawa-nishiwaki-oda1} for detailed discussion; see also footnote~\ref{Gamma footnote}.
Throughout this paper, we take the narrow-width approximation for $\Gamma_V$ as in Eq.~\eqref{E0 tilde}.\footnote{
Theoretically, $\Gamma_V$ is obtained as the imaginary part of the plane-wave $V$ propagator at the loop level. See footnote~\ref{Gamma footnote}.
}
Here, the off-shell $V$ should eventually be regarded as an intermediate state for a scattering process that includes the production of $V$, which necessarily introduces the \emph{in-time-boundary effect} appearing below.

Their wave functions take the form
\al{
f_{V}\fn{x}
	&:=
		N_V 
		\paren{\frac{\sigma_0}{\pi}}^{3/4} \paren{\frac{\pi}{\sigma_0}}^{3/2}
		\frac{1}{\sqrt{2P^0_0} \paren{2\pi}^{3/2}}
		e^{ i P_0 \cdot (x - X_0) - \frac{ \paren{\bs{x} - \bs{\Xi}_0 \fn{t}}^2 }{2\sigma_0} } \Bigg|_{P_0^0 = \wt{E}_0}, 
			\label{f_V given}\nn
f_P\fn{x}
	&:=
		\phantom{N_V} 
		\paren{\frac{\sigma_1}{\pi}}^{3/4} \paren{\frac{\pi}{\sigma_1}}^{3/2}
		\frac{1}{\sqrt{2P^0_1} \paren{2\pi}^{3/2}}
		e^{ i P_1 \cdot (x - X_1) - \frac{ \paren{\bs{x} - \bs{\Xi}_1 \fn{t}}^2 }{2\sigma_1} } \Bigg|_{P^0_1 = E_1}, \nn
f_{\overline{P}}\fn{x}
	&:=
		\phantom{N_V} 
		\paren{\frac{\sigma_2}{\pi}}^{3/4} \paren{\frac{\pi}{\sigma_2}}^{3/2}
		\frac{1}{\sqrt{2P^0_2} \paren{2\pi}^{3/2}}
		e^{ i P_2 \cdot (x - X_2) - \frac{ \paren{\bs{x} - \bs{\Xi}_2 \fn{t}}^2 }{2\sigma_2} } \Bigg|_{P^0_2 = E_2}, 
}
where $N_V$ is a wave-function (field) renormalization factor for $V$ due to its offshellness\footnote{
$N_V$ shows a factor that accounts for the possible extra decrease of the norm of the initial state due to the off-shellness $\Gamma_V>0$. Anyway, $N_V$ will drop out of the final ratio of the decay probabilities.
}
and 
\al{
\bs{\Xi}_A \fn{t}
	:=
		\bs{X}_A + \bs{V}_A  \paren{t - X^0_A}\qquad(A=0,1,2),
}
describes the location of the center of the wave packet at a time $t$ with the central velocity
\al{
\bs{V}_A
	:=
		\frac{ \bs{P}_A }{ E_A }.
}

Now, it is straightforward to compute the $S$-matrix from Eq.~\eqref{S-matrix given first time} at the leading order in the Dyson series~\eqref{Dyson series} with the effective Hamiltonian~\eqref{eq:effective-Hamiltonian} from the wave functions~\eqref{f_V given}~\cite{Ishikawa-Oda}:
\al{
S_{V \to P \ol P}
	&= 	i g_\tx{eff}\fn{\lambda_0,P_0,P_1,P_2}
		\paren{ \prod_{A=0}^2 \frac{1}{\sqrt{2E_A}} \paren{\frac{1}{\pi\sigma_A}}^{3/4} }
		e^{ - \frac{\sigma_t}{2}(\delta \omega)^2 - \frac{\sigma_s}{2} (\delta \bs P)^2 - \frac{\cal R}{2} } \notag \\
	&\quad
		\times \int_{\Tin}^{\Tout} \df t \,
		e^{ -\frac{1}{2\sigma_t} \sqbr{ t - \paren{ \mf T + i \sigma_t \delta\omega } }^2      
		}
		\int \df^3 \bs x \,
		e^{ -\frac{1}{2\sigma_s} \sqbr{ \bs x - \paren{ \overline{\bmf X} + \overline{\bs V} t - i \sigma_s \delta \bs P } }^2
		} \notag \\
	&\quad
		\times N_V \, e^{ - {\Gamma_V\ov2}\pn{t - {T_0}}}  \widetilde{F} \fn{\ab{\bs V_1 - \bs V_2}},
	\label{eq:first-S-form}
}
where the notation follows Eq.~$(27)$ of Ref.~\cite{Ishikawa-Oda} (see also below for a short summary).\footnote{
In Eq.~\eqref{eq:first-S-form}, we have dropped the overall phase factor which is irrelevant to the calculation of the probability, while properly taking into account the real damping factor $e^{-{\Gamma_V\ov2}\pn{t-T_0}}$ coming from the imaginary part of $\wt E_0$; see Eq.~\eqref{f_V given}.
}
Differences from the previous calculation~\cite{Ishikawa-Oda} are the following four points:
First, the coupling is changed to $\kappa/\sqrt2\to g_\tx{eff}$.
Second, the ``decay width'' of $V$ is included as the phenomenological factor $e^{ - {\Gamma_V\ov2}\pn{t - {T_0}}}$,\footnote{\label{Gamma footnote}
When one includes the production process in the amplitude, e.g., as $e^+e^-\to V\to P\ol P$, the result does not change whether we expand the complete set of intermediate states of $V$ by the Gaussian-wave-packet or plane-wave bases; see Sec.~2.3 in Ref.~\cite{Ishikawa-Nishiwaki-Oda2}.
The imaginary part $m_V\Gamma_V$ of the plane-wave propagator of $V$ appears through loop corrections, and when translated to the decay process $V\to P\ol P$, its effect can be expressed as the phenomenological factor $e^{-\Gamma_V\pn{t-T_0}/2}$ in the plane-wave formalism.
Here, we also phenomenologically take into account the exponentially decaying nature of the initial wave packet of $V$ through the channel that is common to the plane-wave decay, namely, through the bulk effect that appears below. See also footnote~\ref{WW}.
}
where $T_0 := X^0_0$ is the initial time from which $V$ starts to exist.\footnote{\label{footnote on T0}
$T_0$ is indeed irrelevant in the sense that the time-translational invariance results in the dependence of the final result only on the difference $T_\tx{in}-T_0$.
Furthermore, this dependence on $T_\tx{in}-T_0$ cancels out between the numerator and denominator of the final ratio of the decay probabilities, as we will see. (Physically, we would expect $T_\tx{in}\simeq T_0$.)
}
Third, $N_V$ in Eq.~\eqref{f_V given} is introduced.
Fourth, we have included a phenomenological form factor $\widetilde{F}$ due to the composite nature of $V$: 
\al{
\widetilde{F} \Fn{\ab{\bs V_1-\bs V_2}} 
	:=
		\frac{1}{1+\paren{\frac{R_0 m_P\ab{\bs V_1-\bs V_2}}{2}}^2 },
		\label{eq:form-factor-for-calc}
}
where $R_0$ describes a typical length scale of the compositeness of $V$; see Appendix~\ref{form factor section}.
The normalization is such that $\widetilde{F}$ becomes unity for $\bs V_1 = \bs V_2$.


\begin{figure}[t]\centering
\hfill
 \includegraphics[width=.58\textwidth]{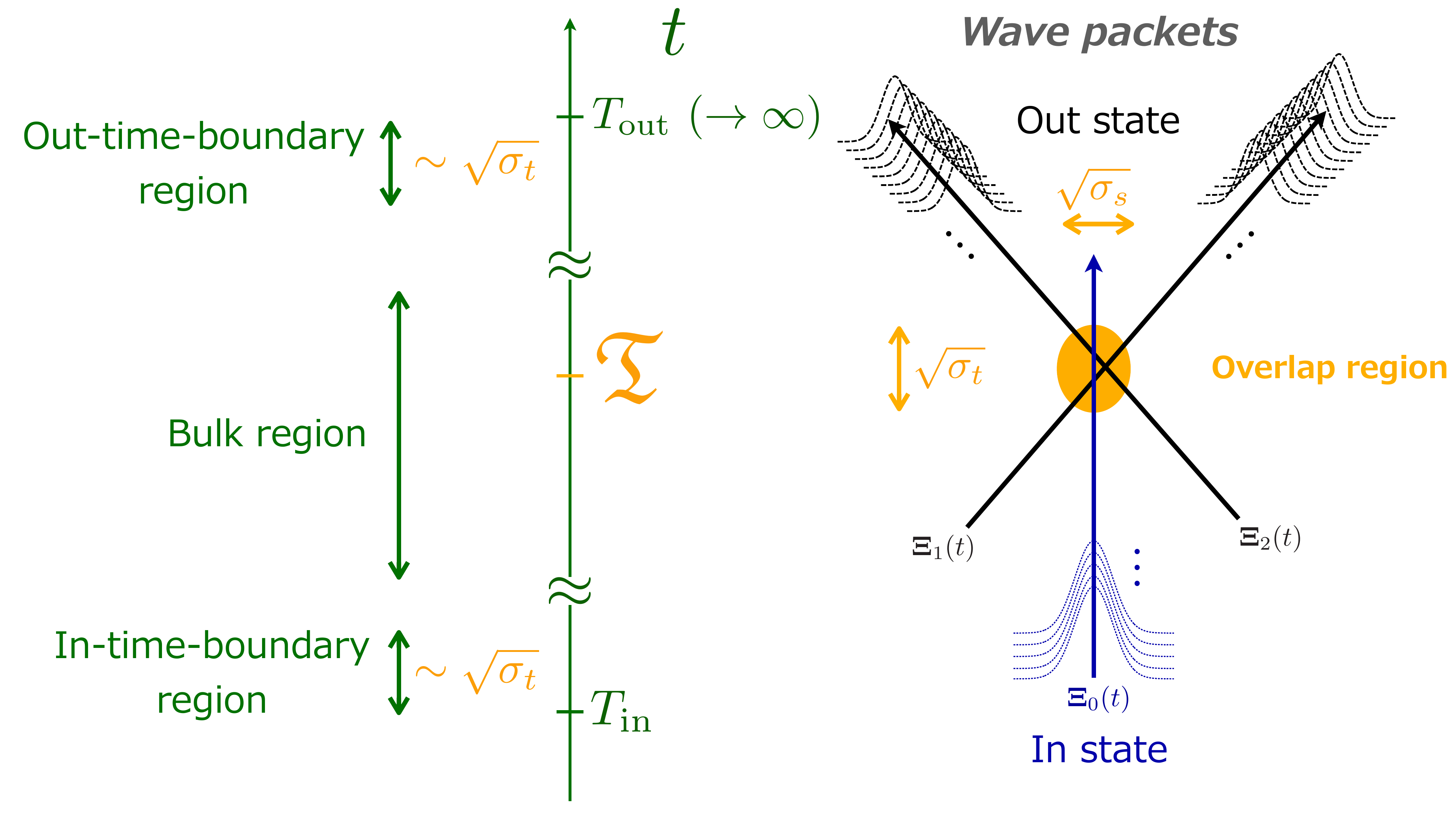}\hfill
 \includegraphics[width=.35\textwidth]{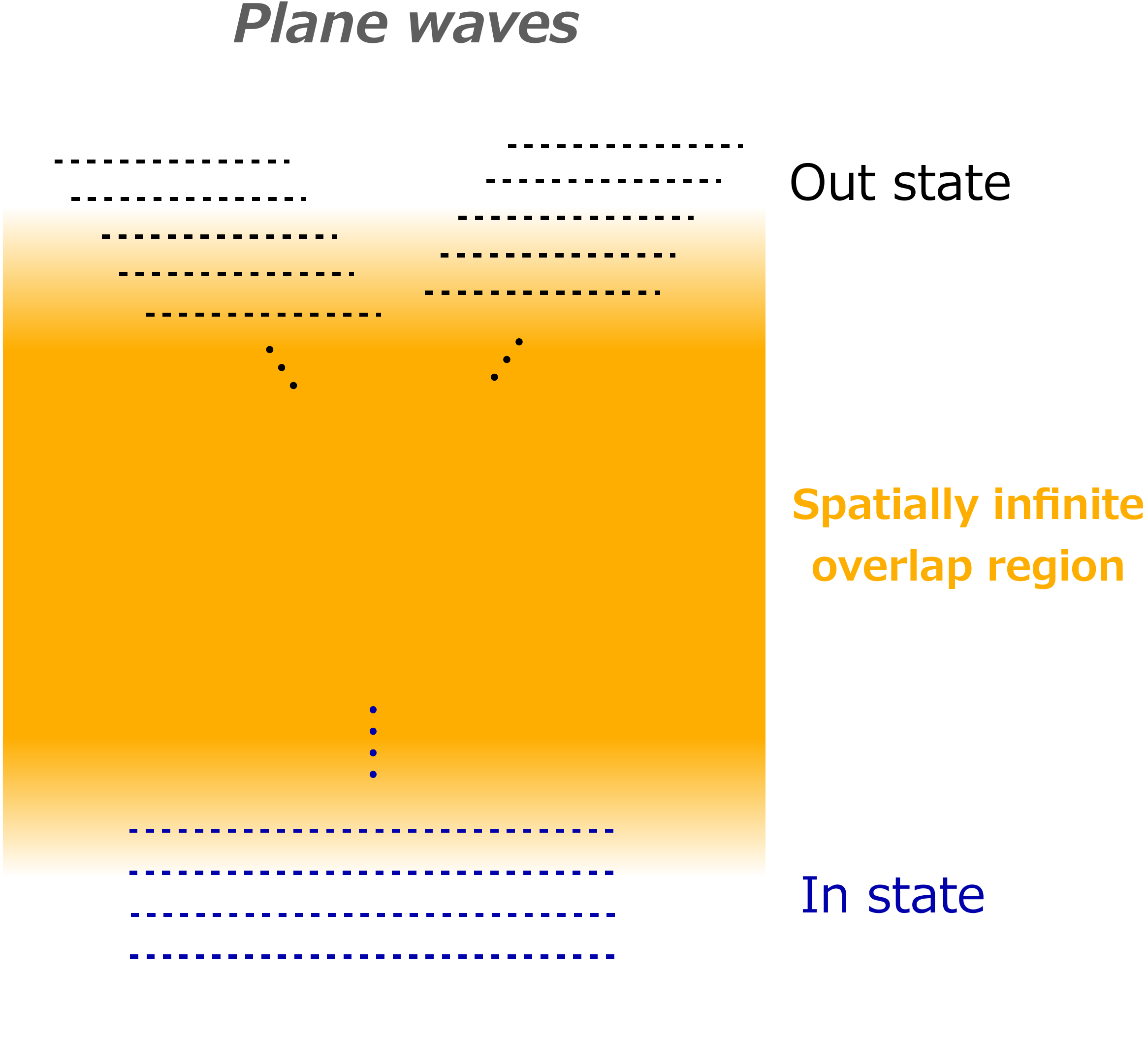}\hfill
\caption{Schematic figure for the finite wave-packet process (left) and the infinite plane-wave process (right), without taking into account the decay width $\Gamma_V$.
In the left, we have shown the time of intersection $\mf T$; the spatial and temporal sizes of the overlap $\sqrt{\sigma_s}$ and $\sqrt{\sigma_t}$, respectively; the center of wave packets $\bs{\Xi}_A$ ($A=0,1,2$); and the initial and final times of the scattering $\Tin$ and $\Tout$, respectively.
Also, the bulk $T_\tx{in}\ll t\ll T_\tx{out}$, in-time-boundary ($\ab{t-T_\tx{in}}\lesssim\sqrt{\sigma_t}$), and out-time-boundary ($\ab{T_\tx{out}-t}\lesssim\sqrt{\sigma_t}$) regions are shown.
(This panel corresponds to the bulk-like case $\ab{\mf T-\Tin}\gg\sqrt{\sigma_t}$; see Fig.~\ref{two limits figure}.)
In the right, the spatial overlap of the plane waves never decreases in time, and hence the interaction would be never switched off, and the scattering would be never completed; therefore the extra damping factor $e^{\mp \epsilon t\int\df^3\bs x\,\widehat{{\cal H}}_\tx{int}^\tx{(I)}\fn{t,\bs x}}$ with an infinitesimal $\epsilon>0$ is conventionally put by hand for the future and past infinite times $t\to\pm\infty$, which is depicted by the damping of the opacity of the orange region. This factor eventually results in the propagator $\propto \pn{p^2+m^2-i\epsilon}^{-1}$ in the conventional Feynman diagram calculation.
}
\label{schematic figure}
\end{figure}

Now we provide a brief introduction to other variables in the first two lines of~\eqref{eq:first-S-form} (see Section~3.1 of~\cite{Ishikawa-Oda} for more details):
\begin{itemize}
\item $\sqrt{\sigma_s}$ is a typical spacial size of the region of interaction:
\al{
\sigma_s^{-1} := \sum_{A=0}^{2} \frac{1}{\sigma_A}.
}
\item $\sqrt{\sigma_t}$ is a typical temporal size of the interaction region:\footnote{
We adopt the following notation for arbitrary scalar and vector variables $C$ and $\bs C$, respectively:
\als{
\overline{C} &:= \sigma_s \sum_{A=0}^{2} \frac{C_A}{\sigma_A}, &
\overline{\bs C} &:= \sigma_s \sum_{A=0}^{2} \frac{\bs C_A}{\sigma_A}, &
\Delta \bs C^2 &:= \overline{\bs C^2} - {\overline{\bs C}}^2.
}
}
\al{
\sigma_t := \frac{\sigma_s}{\Delta \bs V^2}.
}
\item $\mf T$ is the time of intersection of the three wave packets:
\al{
\mf T
	:=
		\sigma_t 
		\frac
		{ \overline{\bs V} \cdot \overline{\bmf X} - \overline{ \bs V \cdot \bmf X } }
		{\sigma_s},
}
where
\al{
\bmf X_A 
	:= {\bs \Xi}_A\fn{0}
	\qquad\left(\,= \bs X_A - \bs V_A X_A^0\,\right)
}
is the location of the center of each wave packet at our reference time $t=0$.
As mentioned above, each wave packet takes the Gaussian form centered at $\bs X_A$ at its reference time $X_A^0$.
\item ${\cal R}$ is called the overlap exponent, which provides the exponential suppression when wave packets are separated from each other:
\al{
{\cal R} := \frac{\Delta \bmf X^2}{\sigma_s} - \frac{\mf T^2}{\sigma_t}.
}
\item We write the deviation of energy-momentum from the conserved values (for their central values of wave packets)
\al{
\delta \bs P &:= \bs P_1 + \bs P_2 - \bs P_0, \notag \\
\delta E &:= E_1 + E_2 - E_0, &
\delta \omega &:= \delta E - \overline{\bs V} \cdot \delta \bs P,
}
where $\omega_A := E_A - \overline{\bs V} \cdot\bs P_A$ is the ``shifted energy'' of each packet. 
\end{itemize}
A schematic figure is shown in the left panel of Fig.~\ref{schematic figure}, compared with the plane-wave counterpart in the right.

After the square completion of $t$ and the analytic Gaussian integration over $\bs{x}$ in~\eqref{eq:first-S-form},
as made in~\cite{Ishikawa-Oda}, we represent the $S$-matrix as follows:
\al{
S_{V \to P \ol P}
	&= 	ig_\tx{eff} N_V
		\paren{ \prod_{A=0}^2 \frac{1}{\sqrt{2E_A}} \paren{\frac{1}{\pi\sigma_A}}^{3/4} }
		e^{ - \frac{\sigma_t}{2}(\delta \omega)^2 - \frac{\sigma_s}{2} (\delta \bs P)^2 - \frac{\cal R}{2} }
		\paren{2\pi\sigma_s}^{3/2} \sqrt{2\pi\sigma_t} \, G(\mf T) \notag \\
	&\quad
		\times e^{ - {\Gamma_V\ov2}\pn{ \mf T - {T_0} +i \sigma_t \delta \omega } + \frac{\Gamma_V^2\sigma_t}{8}  }\,
		 \widetilde{F} \fn{\ab{\bs V_1 - \bs V_2}},
		 	\label{S final}
}
where the window function $G(\mf T)$ is defined as\footnote{
In Eq.~\eqref{G given}, $\mf T$ on the right-hand side is replaced from the original definition of $G(\mf T)$ in~\cite{Ishikawa-Oda} as $\mf T \to \mf T - {\Gamma_V\sigma_t\ov2}$.
}
\al{
G(\mf T)
	&:=	\int_{\Tin}^{\Tout} \frac{\df t}{\sqrt{2\pi\sigma_t}}  
		e^{  -\frac{1}{2\sigma_t} \sqbr{ t - \paren{ \mf T - {\Gamma_V\sigma_t\ov2} + i \sigma_t \delta\omega } }^2
		} \notag \\
	&=	\frac{1}{2}
		\sqbr{
		\text{erf}\paren{ \frac{\mf T -\Tin - {\Gamma_V\sigma_t\ov2} + i \sigma_t \delta\omega}{\sqrt{2\sigma_t}} }   -
		\text{erf}\paren{ \frac{\mf T -\Tout - {\Gamma_V\sigma_t\ov2} + i \sigma_t \delta\omega}{\sqrt{2\sigma_t}} }
		},
		\label{G given}
}
with
\al{
\text{erf}(z)	:= \frac{2}{\sqrt{\pi}} \int_0^z e^{-x^2} \df x
}
being the Gauss error function. 
The window function $G\fn{\mf T}$ becomes unity for $\Tin\ll\mf T\ll\Tout$ and zero for $\mf T\ll\Tin$ and for $\Tout\ll\mf T$.
For a given configuration of in and out states, which fixes the value of $\sigma_t$, the time regions $\Tin\ll t\ll\Tout$,\footnote{
More precisely, the bulk region is the one satisfying $\ab{t-\Tin}\gg\sqrt{\sigma_t}$ and $\ab{\Tout-t}\gg\sqrt{\sigma_t}$.
}
$\ab{t-\Tin}\lesssim\sqrt{\sigma_t}$, and $\ab{t-\Tout}\lesssim\sqrt{\sigma_t}$ are called the \emph{bulk}, \emph{in-time-boundary}, and \emph{out-time-boundary} regions, respectively. In the phenomenological analysis below, we will neglect the out-time-boundary contributions as we will discuss.

\subsection{Differential decay probability}
From the $S$-matrix~\eqref{S final}, the differential decay probability can be derived as
\al{
\df P_{V \to P \ol{P}}
	&=
		\frac{ \df^3 \bs X_1 \df^3 \bs P_1}{(2\pi)^3} \frac{ \df^3 \bs X_2 \df^3 \bs P_2}{(2\pi)^3} \ab{S_{V \to P \ol P}}^2 \notag \\
	&=
		\overline{|g_\tx{eff}|^2} N_V^2 \frac{1}{2 E_0}
		\frac{\df^3 \bs P_1}{(2\pi)^3 2E_1} \frac{\df^3 \bs P_2}{(2\pi)^3 2E_2}
		(2 \pi)^4
		\paren{ \sqrt{\frac{\sigma_t}{\pi}} e^{-\sigma_t\paren{\delta \omega}^2} }
		\paren{ \paren{\frac{\sigma_s}{\pi}}^{3/2} e^{-\sigma_s\paren{\delta \bs P}^2} } \notag \\
	&\quad
		\times \sqrt{ \frac{\sigma_t}{\pi^5} \paren{ \frac{\sigma_s}{\sigma_0\sigma_1\sigma_2} }^3 }
		\df^3 \bs X_1 \df^3 \bs X_2 e^{- \mc R} \, |G(\mf T)|^2  \,
		e^{ - \Gamma_V\pn{ \mf T - {T_0}  } + \frac{\Gamma_V^2\sigma_t}{4}  }\,
		 \ab{ \widetilde{F}\fn{\ab{\bs V_1 - \bs V_2}} }^2,
		\label{eq:dP-first-form}
}
where we have taken the average over the helicity $\lambda_0$, which results in the helicity-averaged effective coupling:
\al{
\overline{|g_\tx{eff}|^2}
	&:=
		\frac{g_V^2}{3} \sum_{\lambda_0} 
		\Ab{ \varepsilon_\mu(P_0,\lambda_0) \paren{P_1^\mu - P_2^\mu} }^2 
	=
		{\frac{g_V^2}{3}} (\bs P_1 - \bs P_2)^2.
}
Here, the last equality further assumes the vanishing initial momentum $\bs P_0=0$.
We will compute the integrated decay probability under this assumption in Sec.~\ref{integrated decay probability section} and in Appendix~\ref{detailed decay probability section}.

\begin{figure}[t]\centering
\includegraphics[width=.6\textwidth]{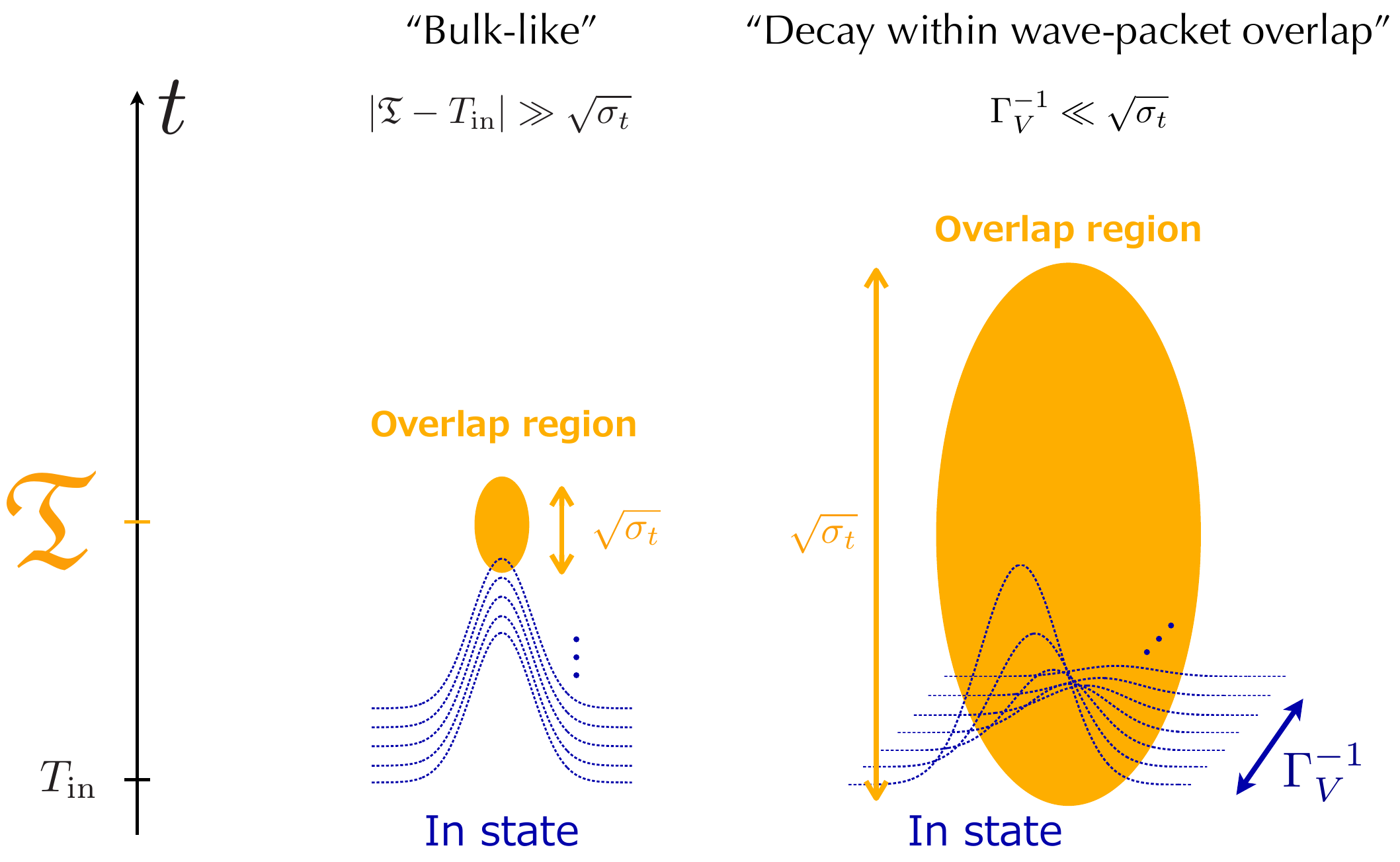}
\caption{Schematic figures for two limiting cases $\ab{\mf T-\Tin}\gg\sqrt{\sigma_t}$ (left) and $\sqrt{\sigma_t}\gg \Gamma_V^{-1}$ (right). The case $\delta\omega\gg\sigma_t^{-1/2}$ is hard to draw in the position space and is not shown here. The overlap region is determined by both the initial and final states as in Fig.~\ref{schematic figure}.
}
\label{two limits figure}
\end{figure}

Hereafter, we assume both the following conditions
\al{
\frac{\ab{\mf T -\Tin - {\Gamma_V\sigma_t\ov2} + i \sigma_t \delta\omega}}{\sqrt{2\sigma_t}} &\gg 1, &
\frac{\ab{\mf T -\Tout - {\Gamma_V\sigma_t\ov2} + i \sigma_t \delta\omega}}{\sqrt{2\sigma_t}} &\gg 1.
	\label{eq:temporal-condition-1}
}
Physically, each of these conditions is satisfied when at least one of the following three conditions is met:
\begin{itemize}
\item  $\ab{\mf T-\Tin}\gg\sqrt{\sigma_t}$ (or $\ab{\mf T-\Tout}\gg\sqrt{\sigma_t}$) when the interaction time $\mf T$ is apart enough from $\Tin$ (or $\Tout$) compared to the temporal width of the overlap $\sqrt{\sigma_t}$, which typically corresponds to the ``bulk-like'' case (Fig.~\ref{two limits figure}, left);
\item $\sqrt{\sigma_t}\gg\Gamma_V^{-1}$ when the ``mean lifetime through bulk effect'' $\Gamma_V^{-1}$ is much shorter than the temporal width of the overlap region $\sqrt{\sigma_t}$, which typically corresponds to the so-to-say ``decay within wave-packet overlap'' case (Fig.~\ref{two limits figure}, right);
\item $\delta\omega\gg\sigma_t^{-1/2}$ when the deviation from the conservation of the shifted-energy, $\delta\omega$, is much larger than the inverse of the temporal width of the overlap $1/\sqrt{\sigma_t}$, namely the ``violation of shifted-energy'' case.
\end{itemize}
This assumption~\eqref{eq:temporal-condition-1} is made for simplicity, and there is no obstacle to using the full form~\eqref{G given} in the numerical computation in principle, but the result would remain the same approximately because this is anyway satisfied in the ordinary bulk-like case as well as when anything interesting happens around the (in-)time-boundary.

Under the assumption~\eqref{eq:temporal-condition-1}, the following asymptotic form is obtained~\cite{Ishikawa-Oda}:
\al{
G(\mf T)
	&\simeq 	
		W(\mf T)
		- {\half} e^{ -\frac{ \paren{\mf T - \Tin - {\Gamma_V\sigma_t\ov2}}^2 }{2\sigma_t}  +  \frac{\sigma_t}{2}(\delta \omega)^2  
		        -i \delta \omega \paren{ \mf T -\Tin - {\Gamma_V\sigma_t\ov2} }
		}
		\sqrt{ \frac{2\sigma_t}{\pi} }
		\frac{1}{ \mf T -\Tin - {\Gamma_V\sigma_t\ov2} + i \sigma_t \delta\omega } \notag \\
	&\quad
		+ {\half} e^{ -\frac{ \paren{\mf T - {\Gamma_V\sigma_t\ov2} - \Tout}^2 }{2\sigma_t}  +  \frac{\sigma_t}{2}(\delta \omega)^2  
		        -i \delta \omega \paren{ \mf T - {\Gamma_V\sigma_t\ov2} -\Tout }
		}
		\sqrt{ \frac{2\sigma_t}{\pi} }
		\frac{1}{ \mf T -\Tout - {\Gamma_V\sigma_t\ov2} + i \sigma_t \delta\omega },
		\label{eq:GT-asymptotic}
}
where we have defined the ``bulk window function''
\al{
W(\mf T) :=
\frac{1}{2}
		\sqbr{
		\text{sgn}\paren{ \frac{\mf T -\Tin - {\Gamma_V\sigma_t\ov2} + i \sigma_t \delta\omega}{\sqrt{2\sigma_t}} }   -
		\text{sgn}\paren{ \frac{\mf T -\Tout - {\Gamma_V\sigma_t\ov2} + i \sigma_t \delta\omega}{\sqrt{2\sigma_t}} }
		},
		\label{W given}
}
in which the sign function for a complex variable is
\al{
\text{sgn}(z)
	&:=
		\begin{cases}
		+1 & \text{for }  \Re z >0 \text{ or } ( \Re z = 0 \text{ and } \Im z >0 ), \\
		-1 & \text{for }  \Re z <0 \text{ or } ( \Re z = 0 \text{ and } \Im z <0 ), \\
		 0 & \text{for } z=0.
		\end{cases}
}
Here and hereafter, $\Re$ and $\Im$ denote the real and imaginary parts, respectively.
Equation~\eqref{W given} describes the ordinary ``bulk contribution'' for the quantum transition from the in to out states in the time period $[\Tin, \Tout]$.
The second and third terms of Eq.~\eqref{eq:GT-asymptotic} show the contributions near the in and out time boundaries $T_\tx{in}$ and $T_\tx{out}$, respectively.
More explicitly,\footnote{
Precisely speaking, Eq.~\eqref{eq:bulk-window-function} is given except right at the boundary $\mf T =\Tin + {\Gamma_V\sigma_t\ov2}$ or $\mf T=\Tout + {\Gamma_V\sigma_t\ov2}$, which is rather a peculiarity of how to define a boundary value and is out of our current interest.
}
\al{
W(\mf T)
	&=
		\begin{cases}
		1 & (
		       \Tin + {\Gamma_V\sigma_t\ov2} < \mf T  < \Tout + {\Gamma_V\sigma_t\ov2} ), \\
		0 & \text{otherwise}.
		\end{cases}
\label{eq:bulk-window-function}
}

Because the contribution from the out-time-boundary $\mf T\simeq T_\tx{out}$ is suppressed by the extra dumping factor $e^{ -\Gamma_V\pn{\mf T - T_0}}$ in the differential probability~\eqref{eq:dP-first-form},\footnote{
Here, we physically assume $\Tout-\Tin\gg\Gamma_V^{-1}$ with $T_0\simeq T_\tx{in}$; see also footnote~\ref{footnote on T0}.
}
it is safe to neglect the out-time-boundary contribution, and we can take
\al{
\Tout \to +\infty,
	\label{eq:temporal-condition-2}
}
with which the second line in Eq.~\eqref{eq:GT-asymptotic} goes down to zero.
With this limit, $\ab{G(\mf T)}^2$ reads
\al{
\ab{G(\mf T)}^2
	&\to
		\sqbr{\cal GG}_\tx{bulk}(\mf T) +
	       \sqbr{\cal GG}_\tx{bdry}(\mf T) +
	        \sqbr{\cal GG}_\tx{intf}(\mf T),
}
where
\al{
\sqbr{\cal GG}_\tx{bulk}(\mf T)
	&:=
		\Ab{{W(\mf T)}}^2, \\
\sqbr{\cal GG}_\tx{bdry}(\mf T)
	&:=
		\frac{1}{4} 
		e^{ -\frac{ \paren{\mf T - \Tin - {\Gamma_V\sigma_t\ov2}}^2 }{\sigma_t}  +  {\sigma_t}(\delta \omega)^2  
		}
		\frac{2\sigma_t}{\pi} 
		\frac{1}{ \paren{ \mf T -\Tin - {\Gamma_V\sigma_t\ov2} }^2 + \paren{ \sigma_t \delta\omega }^2 }, \\
\sqbr{\cal GG}_\tx{intf}(\mf T)
	&:=
		-\frac{W\fn{\mf T}}{2}
		e^{ -\frac{ \paren{\mf T - \Tin - {\Gamma_V\sigma_t\ov2}}^2 }{2\sigma_t}  +  \frac{\sigma_t}{2}(\delta \omega)^2}
		\sqrt{ \frac{2\sigma_t}{\pi} } \notag \\
	&\ \qquad
		\times
		\br{
			\frac{ e^{ -i \delta \omega \paren{ \mf T -\Tin - {\Gamma_V\sigma_t\ov2} } } }
			       { \mf T -\Tin - {\Gamma_V\sigma_t\ov2} + i \sigma_t \delta\omega }
			+
			\frac{ e^{ +i \delta \omega \paren{ \mf T -\Tin - {\Gamma_V\sigma_t\ov2} } } }
			       { \mf T -\Tin - {\Gamma_V\sigma_t\ov2} - i \sigma_t \delta\omega }
		}.
}
The three functions $\sqbr{\cal GG}_\tx{bulk}$, $\sqbr{\cal GG}_\tx{bdry}$ and $\sqbr{\cal GG}_\tx{intf}$
describe the square of the bulk term, the square of the in-time-boundary term,
and the interference between the bulk and in-boundary terms, respectively.

With the approximation~\eqref{eq:temporal-condition-1} and the limit~\eqref{eq:temporal-condition-2},
the differential probability~\eqref{eq:dP-first-form} takes the simpler form and can be classified into the following three parts:
\al{
\df P_{V \to P \ol{P}} = \df P_{V \to P \ol{P}}^{\text{bulk}} + \df P_{V \to P \ol{P}}^{\text{bdry}} + \df P_{V \to P \ol{P}}^\tx{intf},
	\label{dP given}
}
with
\al{
\df P_{V \to P \ol{P}}^\tx{``type"}
	&:=
		|g_\tx{eff}|^2 N_V^2 \frac{1}{2 E_0}
		\frac{\df^3 \bs P_1}{(2\pi)^3 2E_1} \frac{\df^3 \bs P_2}{(2\pi)^3 2E_2}
		(2 \pi)^4
		\paren{ \sqrt{\frac{\sigma_t}{\pi}} e^{-\sigma_t\paren{\delta \omega}^2} }
		\paren{ \paren{\frac{\sigma_s}{\pi}}^{3/2} e^{-\sigma_s\paren{\delta \bs P}^2} } \notag \\
	&\quad
		\times \sqrt{ \frac{\sigma_t}{\pi^5} \paren{ \frac{\sigma_s}{\sigma_0\sigma_1\sigma_2} }^3 }\,
		\df^3 \bs X_1 \df^3 \bs X_2\, e^{- \mc R} \,
			\sqbr{\cal GG}_\tx{``type"}\fn{\mf T}\,
		e^{ - \Gamma_V\pn{ \mf T - {T_0}  } + \frac{\Gamma_V^2\sigma_t}{4}  }
		\ab{ \widetilde{F}\Fn{\ab{\bs V_1 - \bs V_2}} }^2,
		\label{master dP}
}
where the argument ``type" discriminates the three types of contributions.

\subsection{Integrated decay probability}\label{integrated decay probability section}

To compare the theoretical predictions in the Gaussian wave-packet formalism with the experimental results in Eq.~\eqref{PDG ratio result},
we integrate the differential decay probability~\eqref{dP given} over the whole position-momentum phase space of the final-state pseudoscalar mesons, namely, over $\bs X_1$, $\bs X_2$, $\bs P_1$, and $\bs P_2$:
\al{
P_{V\to P\ol P}=P_{V\to P\ol P}^\tx{bulk}+P_{V\to P\ol P}^\tx{bdry}+P_{V\to P\ol P}^\tx{intf}.
	\label{integrated probability}
}
We focus on the situation where these integrals can be performed analytically using the saddle-point approximation; see e.g.\ Ref.~\cite{Ishikawa-Oda}.
In the current setup, we can safely take non-relativistic approximations in the kinematics of the system because the mass difference $m_V - 2 m_{P}$ is small.

Here, we only list the final form of the three types of contributions to the integrated decay probability: the bulk, boundary, and interference contributions.
These calculations' details are provided in Appendix~\ref{detailed decay probability section}.
For later convenience, we define a common dimensionless factor ${\cal C}_{V \to P \ol{P}}$ for all of 
$P_{V \to P\ol{P}}^\tx{bulk}$, $P_{V \to P\ol{P}}^\tx{bdry}$, and $P_{V \to P\ol{P}}^\tx{intf}$:
\al{
{\cal C}_{V \to P \ol{P}}
	:=
		\frac{g^2_V m_P  N_V^2  e^{  -\Gamma_V\pn{\Tin-{T_0}} }   }{12 \pi m_V}.
	\label{eq:common-C-factor}
}

\subsubsection{Bulk contribution}
Integrating the bulk contribution in Eq.~\eqref{dP given}, we obtain
\al{
P_{V \to P\ol{P}}^\tx{bulk}
	&\simeq
		{{\cal C}_{V \to P \ol{P}}\,m_P\ov\Gamma_V}
		\pn{ \frac{\paren{m_V-2m_P}^2}{m_P^2} + {\Gamma_V^2\ov4m_P^2} }^{3/4}
		\notag \\
	&\quad
		\times
		\frac{1}{2}
		\sqbr{  1 + \text{erf}\paren{  \frac{  m_P\sqrt{\sigma_P}  {V_-^\text{B}}  }{\sqrt{2}}  }  }
		{\frac{e^{-F^0_\text{bulk}}}{A_\text{bulk}^{3/2}}
		\ab{ \widetilde{F}\fn{V_-^\text{B}} }^2  },
	\label{eq:P-integrated_bulk}
}
where
\al{
V_-^\text{B}
	&:=
		2 \sqbr{ \frac{\paren{m_V-2m_P}^2}{m_P^2} + {\Gamma_V^2\ov4m_P^2} }^{1/4}, 
		\label{eq:V_minus-form} \\
F_\text{bulk}^0
	&:=
		m_P \sigma_P
		\paren{ -\paren{m_V - 2m_P}  + \sqrt{ \paren{m_V - 2m_P}^2+{\Gamma_V^2\ov4} }  }, 
		\label{F0bulk}\\
A_\text{bulk}
	&:=
		\half + \frac{ \paren{m_V - 2m_P}}{2\sqrt{\paren{m_V-2m_P}^2+{\Gamma_V^2\ov4}}}.
}
We note that the wave packet size of the decaying particle $\sigma_V$ drops out of this expression at this order of the saddle-point approximation.

\subsubsection{Boundary contribution}
Integrating the boundary contribution in Eq.~\eqref{dP given}, we obtain
\al{
P_{V \to P\ol{P}}^\tx{bdry}
	&\simeq
		{\cal C}_{V \to P \ol{P}}
		\frac{I_\text{bdry}}{2\pi} ,
	\label{eq:P-integrated_bdry}
}
where $I_\text{bdry}$ is written as an integration of a function of $V_-$:
\al{
I_\text{bdry}
	&:=
		\int_0^\infty \df V_- \widetilde{f}_\text{bdry}\fn{V_-},
		\label{ingegral I for boundary}
}
in which
\al{
\widetilde{f}_\text{bdry}\fn{V_-}
	&=
			\frac
			{V_-^4}
			{\pn{V_-^2 - 4\frac{m_V-2m_P}{m_P}}^2 
				+ \frac{4^2}{m_P^2} 
					{ 
							{\Gamma_V^2\ov4} 
						}
			}
			\ab{ \widetilde{F}\fn{V_-} }^2 \notag \\
	&\quad
			-
			\frac
			{  V_-^4 \sqbr{\pn{V_-^2 - 4\frac{m_V-2m_P}{m_P}}^2 -3 \frac{4\Gamma_V^2}{m_P^2}} }
			{
				2{ \frac{2 \sigma_P}{V_-^2} }
				\paren{ \frac{m_P}{4} }^2
				\sqbr{ \pn{V_-^2 - 4\frac{m_V-2m_P}{m_P}}^2 + \frac{4\Gamma_V^2}{m_P^2} }^3
			}
			\ab{ \widetilde{F}\fn{V_-} }^2.
}
The integral~\eqref{ingegral I for boundary} will be evaluated numerically.

\subsubsection{Interference contribution}
Integrating the interference contribution in Eq.~\eqref{dP given}, we obtain
\al{
P^\tx{intf}_{V \to P\ol{P}}
	&\sim
		- \,
		{\cal C}_{V \to P \ol{P}}
		\frac{ m_P \sqrt{\sigma_P}}{2\sqrt{2} \sqrt{\pi}}
		\paren{ V_-^\text{I} }^2
		\frac{1}{2}
		\sqbr{  1 + \text{erf}\paren{  \frac{  m_P\sqrt{ \sigma_P}  {V_-^\text{I}}  }{2}  }  }
		{\frac{e^{-F^0_\tx{intf}}}{A_\tx{intf}^{3/2}}}\,
		 \ab{ \widetilde{F}\fn{V^\text{I}_-} }^2 \notag \\
	&\quad
		\times
		\frac
		{ \paren{\Gamma_V^2 - \paren{\wt{\delta\omega}}^2 }
			\cos\fn{2 \Gamma_V\wt{\sigma_t} \wt{\delta\omega}}
		  	- 2 \Gamma_V \wt{\delta\omega} \sin\fn{2 \Gamma_V\wt{\sigma_t} \wt{\delta\omega}} }
		{\wt{\sigma_t}\paren{  \Gamma_V^2 + \paren{\wt{\delta\omega}}^2     }^2},
	\label{eq:P-integrated_intf}
}
where the definition of new parameters is as follows:
\al{
V_-^\text{I}
	&:=
		\frac{ 2 \sqbr{\paren{m_V-2m_P}^2+{\Gamma_V^2\ov2}}^{1/4}  }{\sqrt{m_P }}, \\
{F_\tx{intf}^0}
	&:= 
		\frac
		{ m_P \sigma_P  }
		{2}
		\sqbr{ -{\paren{m_V - 2m_P}}  + \sqrt{\pn{m_V - 2m_P}^2+{\Gamma_V^2\ov2}}  }, \\
A_\tx{intf}
	&:=
		\frac{1}{4}
		\sqbr{
		3 + \frac{ m_V - 2m_P}{\sqrt{\paren{m_V-2m_P}^2+{\Gamma_V^2\ov2}}}
		}, \\
\widetilde{\sigma_t}
	&:=
		\frac{2 \sigma_P}{ \paren{V_-^\text{I}}^2 }, \\
\widetilde{\delta\omega}
	&:=
		\frac{1}{4} m_P \paren{V_-^\text{I}}^2 - \paren{m_V - 2m_P}.
}
The tilde denotes that the values are evaluated at the saddle point for the interference contribution.

\section{Magnitudes of three kinds of contributions}\label{magnitude section}

We have seen the magnitudes of the three kinds of contributions to the integrated probability
from the bulk part $P_{V \to P\ol{P}}^\tx{bulk}$ in Eq.~\eqref{eq:P-integrated_bulk},
from the boundary part $P_{V \to P\ol{P}}^\tx{bdry}$ in Eq.~\eqref{eq:P-integrated_bdry}, and
from the bulk-boundary interference $P_{V \to P\ol{P}}^\tx{intf}$ in Eq.~\eqref{eq:P-integrated_intf}.
Hereafter, we call the following three ratios the P-factors:
\al{
\ms P_\tx{bulk}
	&:=	{P_{V \to P\ol{P}}^\tx{bulk}\ov{\cal C}_{V \to P\ol{P}}},&
\ms P_\tx{bdry}
	&:=	{P_{V \to P\ol{P}}^\tx{bdry}\ov{\cal C}_{V \to P\ol{P}}},&
\ms P_\tx{intf}
	&:=	{P_{V \to P\ol{P}}^\tx{intf}\ov{\cal C}_{V \to P\ol{P}}},
	\label{P-factors}
}
where the common ${\cal C}_{V \to P\ol{P}}$ given in Eq.~\eqref{eq:common-C-factor} is factored out.

\begin{figure}[tp]
\centering
\includegraphics[width=.45\textwidth]{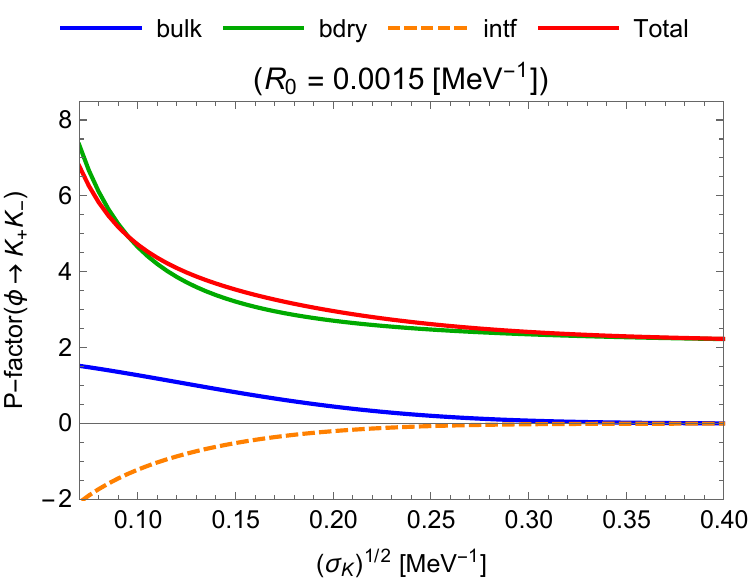} \ \
\includegraphics[width=.45\textwidth]{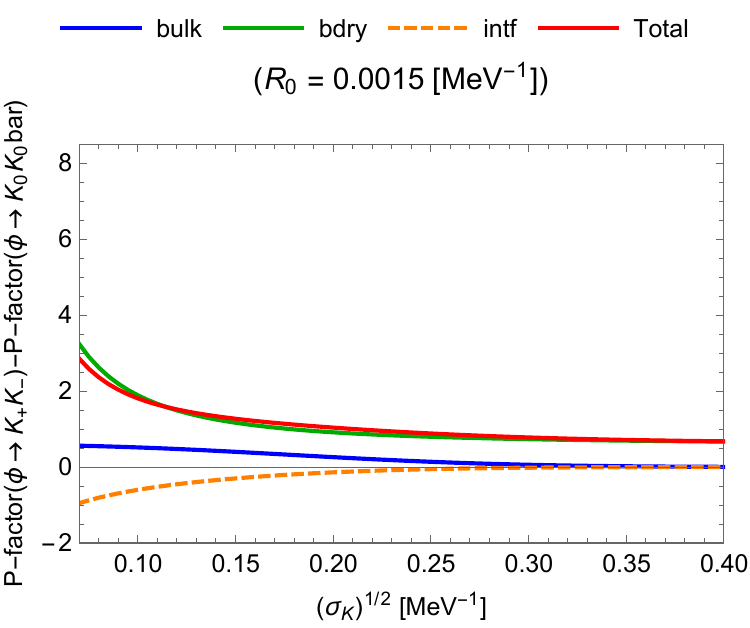} \\
\includegraphics[width=.45\textwidth]{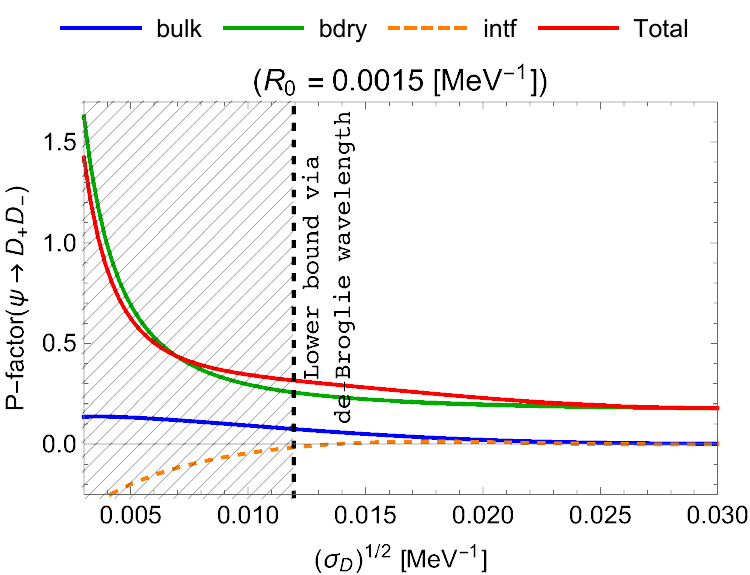} \ \
\includegraphics[width=.45\textwidth]{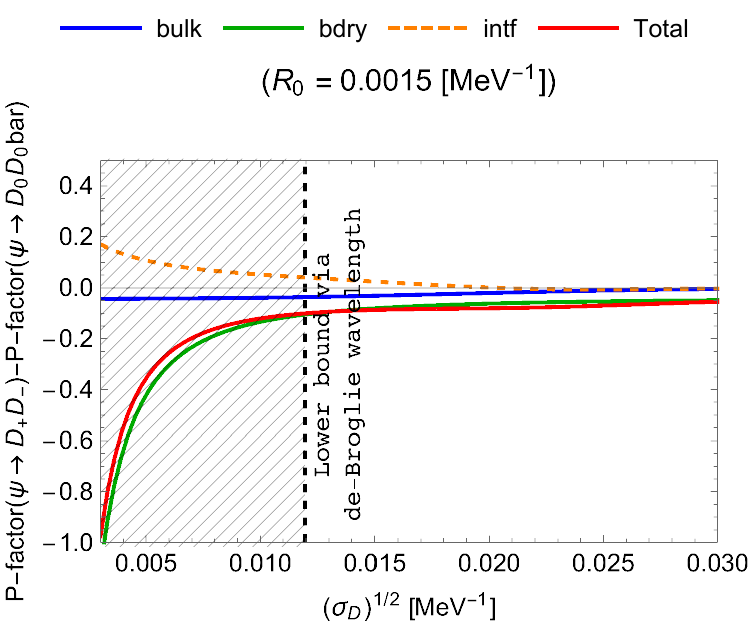} \\
\includegraphics[width=.45\textwidth]{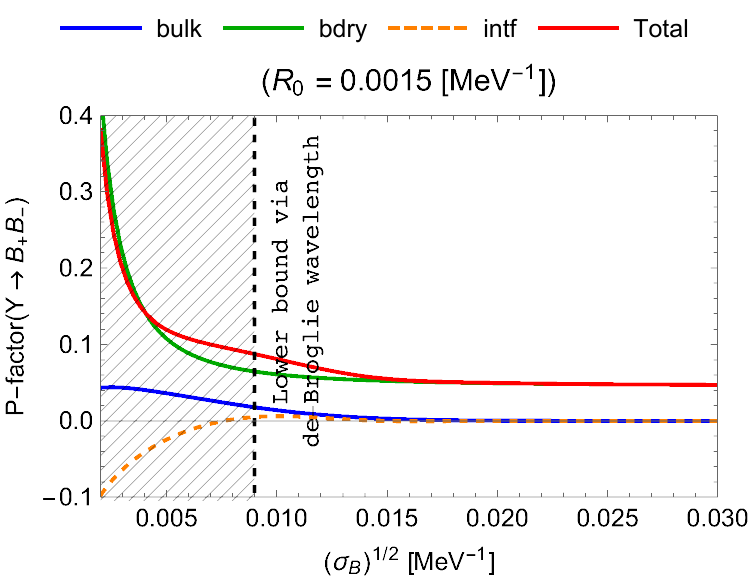} \ \
\includegraphics[width=.45\textwidth]{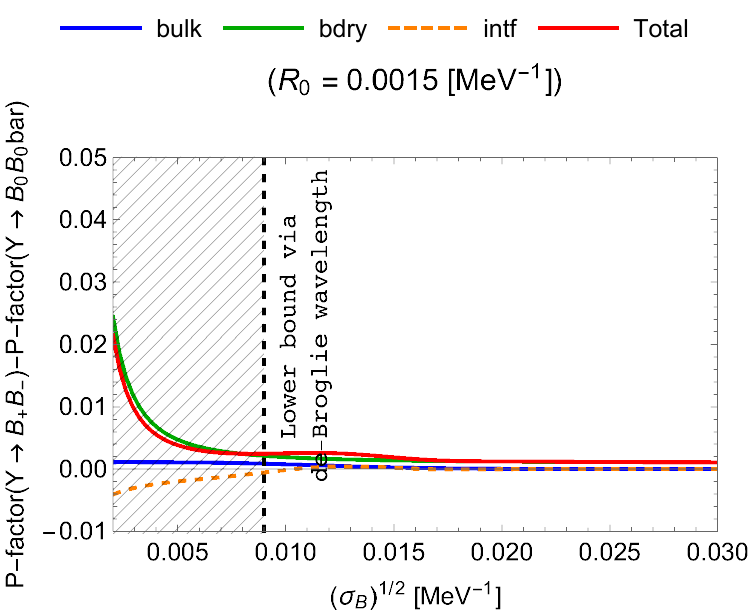} 
\caption{
The P-factors~\eqref{P-factors}
are shown
for
$\phi \to K^+ K^-$ and $\phi \to K^0 \overline{K^0}$ (in the top row),
$\psi \to D^+ D^-$ and $\psi \to D^0 \overline{D^0}$ (in the middle row),
and $\Upsilon \to B^+ B^-$ and $\Upsilon \to B^0 \overline{B^0}$ (in the bottom row),
where we take a typical value $0.0015\,\text{MeV}^{-1}=\pn{0.67\,\tx{GeV}}^{-1}$ for $R_0$; see Appendix~\ref{form factor section}.
The wave-packet treatment breaks down when the wave-packet size of the decay product $\sqrt{\sigma_P}$ is smaller than the de-Broglie wavelength of $P$, which is depicted by the hatched region.
}
\label{fig:P-factor}
\end{figure}

In Fig.~\ref{fig:P-factor}, the P-factors~\eqref{P-factors} are shown
for the following cases, with a typical value $0.0015\,\text{MeV}^{-1}=\pn{0.67\,\tx{GeV}}^{-1}$ for $R_0$ (see Appendix~\ref{form factor section}):
\begin{itemize}
\item $\phi \to K^+ K^-$ and $\phi \to K^0 \overline{K^0}$ (1st row),
\item $\psi \to D^+ D^-$ and $\psi \to D^0 \overline{D^0}$ (2nd row),
\item $\Upsilon \to B^+ B^-$ and $\Upsilon \to B^0 \overline{B^0}$ (3rd row).
\end{itemize}
We remind that all of $P_{V \to P\ol{P}}^\tx{bulk}$, $P_{V \to P\ol{P}}^\tx{bdry}$, and $P_{V \to P\ol{P}}^\tx{intf}$ do not depend on $\sigma_V$ within the non-relativistic saddle-point approximation; see Appendix~\ref{detailed decay probability section} for details.

From Fig.~\ref{fig:P-factor}, we can read the following properties:
\begin{itemize}
\item
If the magnitude of $\sqrt{\sigma_P}$ is relatively low, the three kinds of the P-factors are of the same order.
\item
When $\sqrt{\sigma_P}$ is a certain magnitude, the bulk contribution is exponentially suppressed, while
the boundary contribution takes almost the same value, where the interference part is negligible.
For such a $\sqrt{\sigma_P}$ and other higher choices of it, the boundary part dominates.
\item
Physically, the wave-packet treatment of the decay product $P$ breaks down when the wave-packet size $\sqrt{\sigma_P}$ is shorter than the de-Broglie wavelength of $P$:
\al{
\lambda_\tx{de-Broglie}
	:=
		\frac{2\pi}{ m_P V_{-}^\tx{B}},
}
where $V_{-}^\tx{B}$ is the expectation value in the bulk part in Eq.~\eqref{eq:V_minus-form}.
It is straightforward to estimate it for each decay process:
\al{
\lambda_\tx{de-Broglie}|_{\phi \to K^+K^-}
	&=
		0.025\,\text{MeV}^{-1}
	=	{1\ov40\,\tx{MeV}},&
\lambda_\tx{de-Broglie}|_{\phi \to K^0 \ol{K^0}}
	&=
		0.029\,\text{MeV}^{-1}
	=	{1\ov34\,\tx{MeV}},& \notag \\
\lambda_\tx{de-Broglie}|_{\psi \to D^+D^-}
	&=
		0.012\,\text{MeV}^{-1}
	=	{1\ov83\,\tx{MeV}},&
\lambda_\tx{de-Broglie}|_{\psi \to D^0 \ol{D^0}}
	&=
		0.011\,\text{MeV}^{-1}
	=	{1\ov91\,\tx{MeV}},& \notag \\
\lambda_\tx{de-Broglie}|_{\Upsilon \to B^+B^-}
	&=
		0.0090\,\text{MeV}^{-1}
	=	{1\ov0.11\,\tx{GeV}},&
\lambda_\tx{de-Broglie}|_{\Upsilon \to B^0 \ol{B^0}}
	&=
		0.0091\,\text{MeV}^{-1}
	=	{1\ov0.11\,\tx{GeV}}.&
}
The theoretically excluded region $\sqrt{\sigma_P}<\lambda_\tx{de-Broglie}$ is depicted by the hatched region.
\item
The differences in the P-factors between the charged-meson and neutral-meson final states are sizable for higher-$\sqrt{\sigma_P}$ regions.
\end{itemize}

\begin{figure}[tp]
\centering
\includegraphics[width=.60\textwidth]{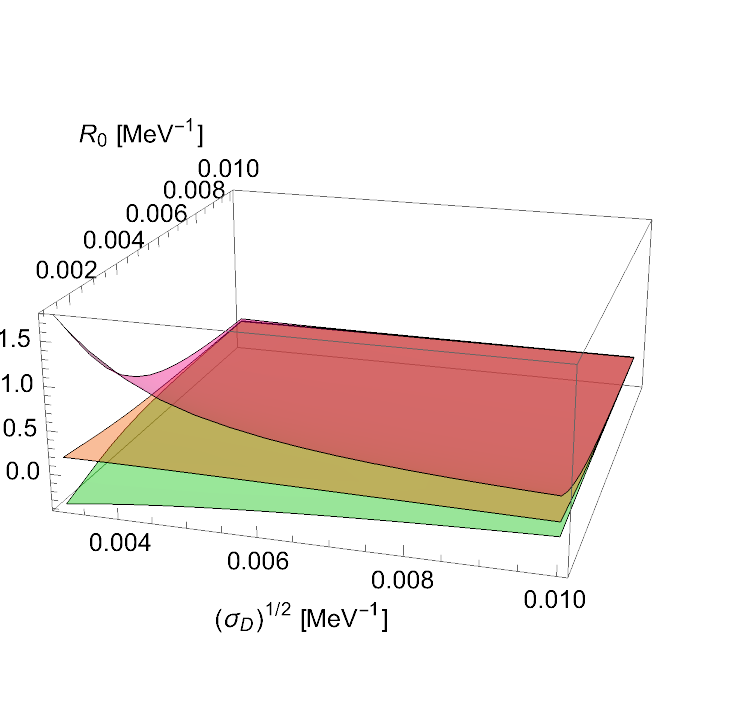}
\caption{
The distributions of the P-factors~\eqref{P-factors} representing $\psi \to D^+ D^-$ are shown as functions of $\paren{\sigma_D}^{1/2}$ and $R_0$,
where the bulk, boundary, and interference ones are shown by the orange, magenta, and green color, respectively. 
}
\label{fig:P-factor-region}
\end{figure}

\newpage

\begin{figure}[tp]
\centering
\includegraphics[width=.45\textwidth]{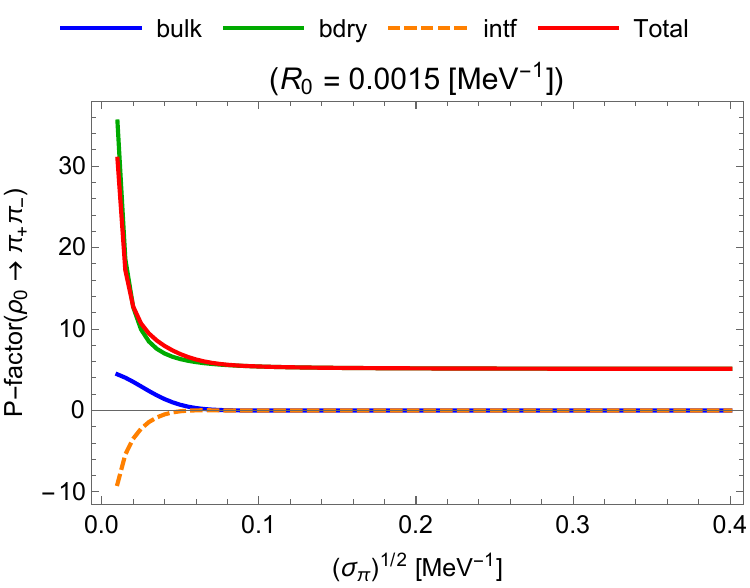} \ \
\includegraphics[width=.45\textwidth]{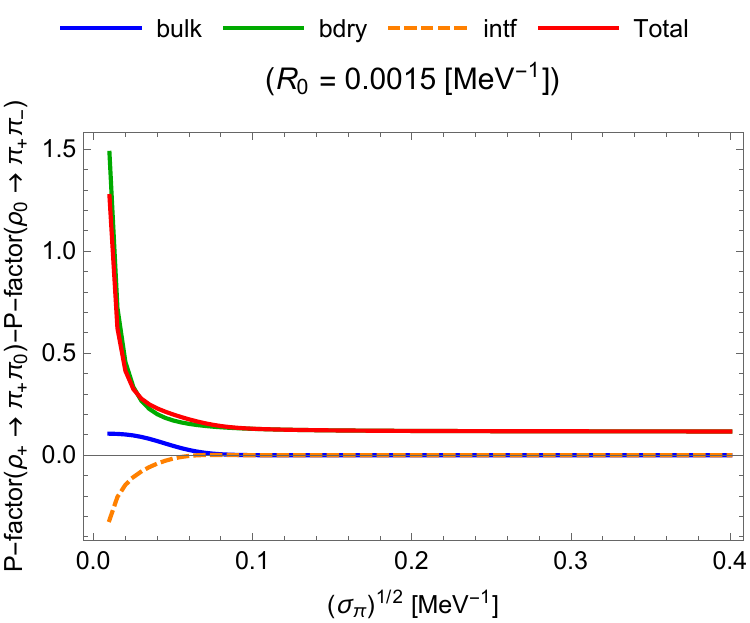}
\caption{
We show the P-factors~\eqref{P-factors} for $\rho^0 \to \pi^+ \pi^-$ (left panel) and the difference between the P-factors for $\rho^0 \to \pi^+ \pi^-$ and $\rho^+ \to \pi^+ \pi^0$ (right panel),
with a typical value $0.0015\,\text{MeV}^{-1}=\pn{0.67\,\tx{GeV}}^{-1}$ for $R_0$; see Appendix~\ref{form factor section}.
We set the lowest $\sqrt{\sigma_\pi}$ near the bound from the de-Broglie wavelength $\simeq 0.012\,\text{MeV}^{-1}$.
}
\label{fig:P-factor-rho}
\end{figure}

To clarify the dependence on $R_0$, we prepare the surface plots for the bulk, boundary, and interference parts of the P-factors for $\Psi \to D^+ D^-$ as a typical example as Fig.~\ref{fig:P-factor-region}.
Here, the following properties are observed:
(i) The magnitude of each part becomes larger for a smaller $\sqrt{\sigma_D}$ and a smaller $R_0$;
(ii) In the entire domain of $\sqrt{\sigma_D}$ and $R_0$, the boundary part exceeds the bulk part in magnitude.

In Fig.~\ref{fig:P-factor-rho}, we have also plotted the P-factors for $\rho^0 \to \pi^+ \pi^-$ and $\rho^+ \to \pi^+ \pi^0$ by adopting the same formulas~\eqref{eq:P-integrated_bulk}, \eqref{eq:P-integrated_bdry} and \eqref{eq:P-integrated_intf} for a purpose of qualitative comparison between the decays with narrow phase spaces ($\phi$, $\psi$, and $\Upsilon$) and that with broad phase spaces ($\rho$), knowing that it is speculative whether we can still use the non-relativistic approximation.\footnote{
We set $m_{\rho^+} \simeq m_{\rho^0} \simeq 770\,\tx{MeV}$ and remind $m_{\pi^0} \simeq 135\,\tx{MeV}$, and $m_{\pi^+} \simeq 140\,\tx{MeV}$. We also remind the property that the decay channel $\rho^0 \to \pi^0 \pi^0$ is prohibited by the conservation of the isospin.
}
As expected, the difference between $\rho^0 \to \pi^+ \pi^-$ and $\rho^+ \to \pi^+ \pi^0$ is small since the magnitude of the isospin breaking is much smaller even for a smaller $\sqrt{\sigma_\pi}$.

\section{Analysis of ratio of decay probabilities $R_V$
\label{sec:R_V}}

In this section, we discuss the ratio $R_V$ of decay probabilities for three vector mesons $\phi$, $\psi$, and $\Upsilon$ in the wave-packet formalism. We compare each with the PDG result and find an agreement around a reasonable value of $R_0$ of the form factor (for the compositeness of $V$). In particular, the 9.5\,$\sigma$ discrepancy for $\psi$ is dramatically ameliorated. We find that the effect of the form factor is significant in both the wave-packet and plane-wave formalisms.

\subsection{Wavepacket analysis}

We estimate the wave-packet counterparts of the ratio of the decay rates defined in Eq.~\eqref{R defined} for the three vector mesons:
\al{
R_\phi^\tx{WP}
	&:=
		{P_{\phi\to K^+K^-}\ov P_{\phi\to K^0 \ol{K^0}}},&
R_\psi^\tx{WP}
	&:=	{P_{\psi\to D^+D^-}\ov P_{\psi\to D^0\ol{D^0}}},&
R_\Upsilon^\tx{WP}
	&:=	{P_{\Upsilon\to B^+B^-}\ov P_{\Upsilon\to B^0\ol{B^0}}},
	\label{eq:R-WP-definition}
}
where we ignored the tiny CP-violation effect in $\phi \to K^0_\tx{L}K^0_\tx{S}$.
We note that the ratio does not depend on the wave-function (field) renormalization factor $N_V$ (accounting for the offshellness of the vector meson $V$), nor on the decay factor $e^{  -\Gamma_V\pn{ \Tin - {T_0} }}$.
Also, we take the isospin-symmetric limit in the couplings as introduced in Eq.~\eqref{isospin-symmetric limit}, and the dependence on the coupling is dropped off.
For further comparison, 
we also define a ``bulk'' ratio:
\al{
R_V^\tx{bulk}
	:=& \
		\frac{P^\tx{bulk}_{V \to P^+ P^-}}{P^\tx{bulk}_{V \to P^0 \ol{P^0}}},
}
which contains the wave-packet contribution only from the bulk part.
Also, we introduce the ratio without interference:
\al{
R_V^\tx{without interference}
	:=& \
		\frac{P^\tx{bulk}_{V \to P^+ P^-} + P^\tx{bdry}_{V \to P^+ P^-}}{P^\tx{bulk}_{V \to P^0 \ol{P^0}} + P^\tx{bdry}_{V \to P^0 \ol{P^0}}}.
}

\begin{figure}
\centering
\includegraphics[width=.43\textwidth]{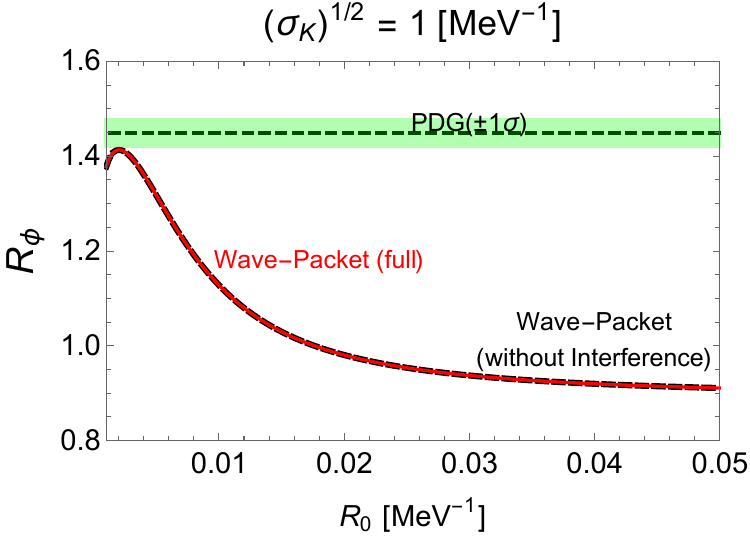} \ \
\includegraphics[width=.43\textwidth]{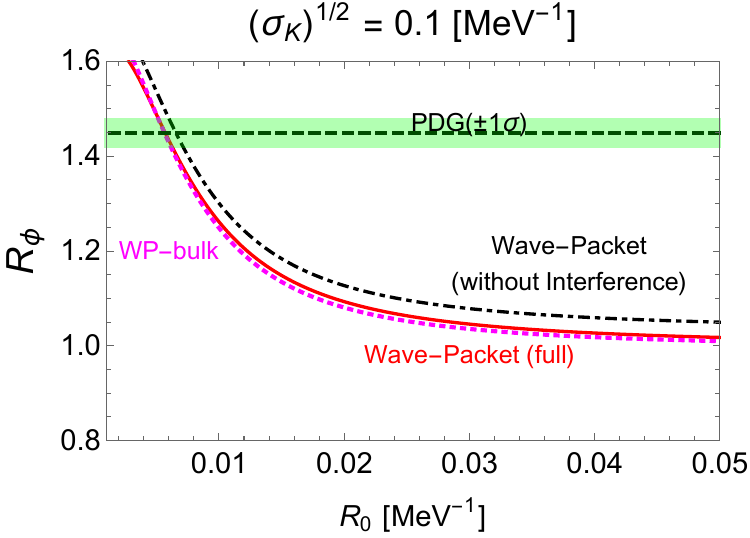} \\
\caption{
The ratio comparing the decay rates of $\phi \to K^+K^-$ to $\phi \to K^0\ol{K^0}$ is drawn as a function of $R_0$ for two fixed wave-packet sizes of the Kaons of $\sqrt{\sigma_K} = 1\,\text{MeV}^{-1}$ (left panel) and $\sqrt{\sigma_K} = 0.1\,\text{MeV}^{-1}$ (right panel).
The experimental result is provided by the PDG~\cite{ParticleDataGroup:2022pth} [shown in Eq.~\eqref{PDG ratio result}].
}
\label{fig:R-WP-phi}
\end{figure}
\begin{figure}
\centering
\includegraphics[width=.43\textwidth]{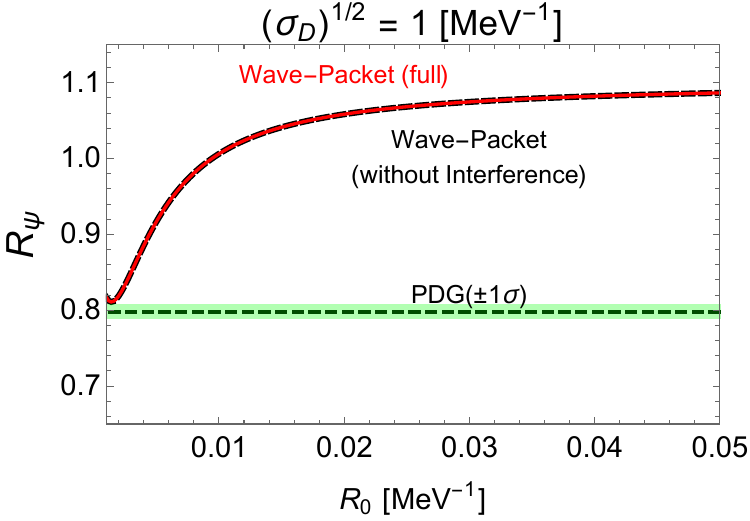} \ \
\includegraphics[width=.43\textwidth]{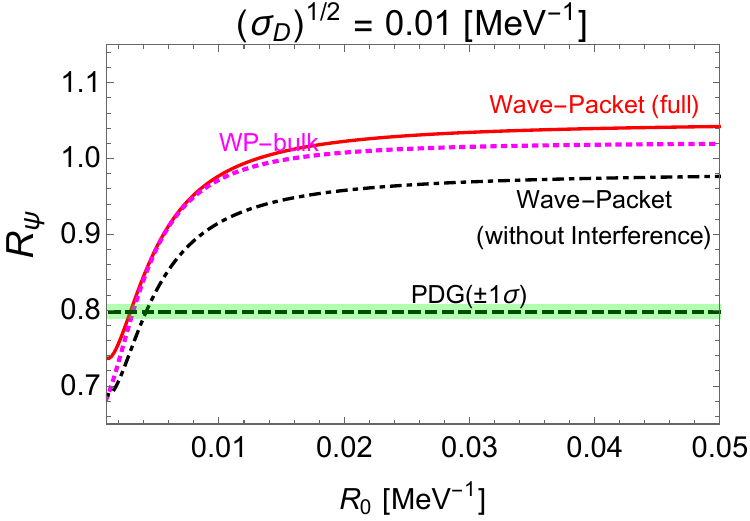} \\
\caption{
The ratio comparing the decay rates of $\psi \to D^+D^-$ to $\psi \to D^0\ol{D^0}$ is drawn as a function of $R_0$ for two fixed wave-packet sizes of the D-mesons of $\sqrt{\sigma_D} = 1\,\text{MeV}^{-1}$ (left panel) and $\sqrt{\sigma_D} = 0.01\,\text{MeV}^{-1}$ (right panel).
The experimental result is provided by the PDG~\cite{ParticleDataGroup:2022pth} [shown in Eq.~\eqref{PDG ratio result}].
}
\label{fig:R-WP-psi}
\end{figure}
\begin{figure}
\centering
\includegraphics[width=.43\textwidth]{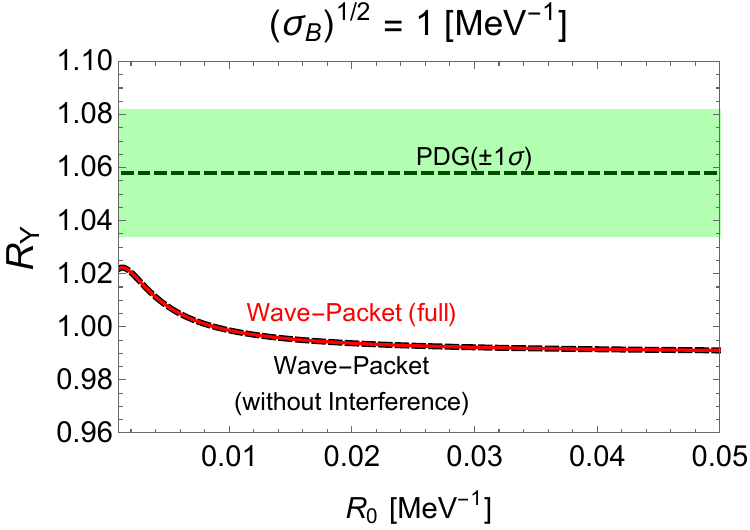} \ \
\includegraphics[width=.43\textwidth]{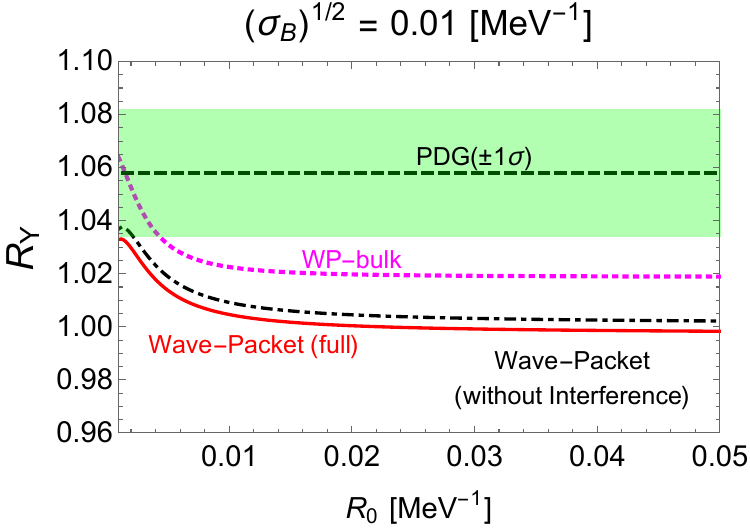} \\
\caption{
The ratio comparing the decay rates of $\Upsilon \to B^+B^-$ to $\Upsilon \to B^0\ol{B^0}$ is drawn as a function of $R_0$ for two fixed wave-packet sizes of the B-mesons of $\sqrt{\sigma_B} = 1\,\text{MeV}^{-1}$ (left panel) and $\sqrt{\sigma_B} = 0.01\,\text{MeV}^{-1}$ (right panel).
The experimental result is provided by the PDG~\cite{ParticleDataGroup:2022pth} [shown in Eq.~\eqref{PDG ratio result}].
}
\label{fig:R-WP-Upsilon}
\end{figure}

\subsection{Wavepacket results}
In Figs.~\ref{fig:R-WP-phi}, \ref{fig:R-WP-psi}, and \ref{fig:R-WP-Upsilon}, we show the wave-packet results for the decay probability ratios of $\phi$, $\psi$, and $\Upsilon$, respectively.
Several comments are in order:
\begin{itemize}
\item For the $\phi$ decay in Fig.~\ref{fig:R-WP-phi}, the full wave-packet result in the red solid line can fit the PDG result around the form-factor size $R_0\simeq 2\times 10^{-3}\,\tx{MeV}^{-1}$ and $6\times 10^{-3}\,\tx{MeV}^{-1}$ for the wave-packet size of the decay product ${\sqrt{\sigma_K}} = 1\,\tx{MeV}^{-1}$ and $ 0.1\,\tx{MeV}^{-1}$, respectively.
\item For the $\psi$ decay in Fig.~\ref{fig:R-WP-psi}, the full wave-packet result in the red solid line can fit the PDG result around the form-factor size $R_0\simeq 2 \times 10^{-3}\,\tx{MeV}^{-1}$ and $3 \times 10^{-3}\,\tx{MeV}^{-1}$ for the wave-packet size of the decay product ${\sqrt{\sigma_D}}= 1\,\tx{MeV}^{-1}$ and $0.01\,\tx{MeV}^{-1}$, respectively.
\item For the $\Upsilon$ decay in Fig.~\ref{fig:R-WP-Upsilon}, the full wave-packet result in the red solid line can fit the PDG result around the form-factor size $R_0\simeq 2 \times 10^{-3}\,\tx{MeV}^{-1}$ and $2 \times 10^{-3}\,\tx{MeV}^{-1}$ for the wave-packet size of the decay product ${\sqrt{\sigma_B}}=1\,\tx{MeV}^{-1}$ and $0.01\,\tx{MeV}^{-1}$, respectively.
\item
As discussed in Fig.~\ref{fig:P-factor}, if $\sqrt{\sigma_P}$ is sufficiently large,
the bulk contribution becomes exponentially suppressed compared to the boundary one.
In this regime, we may still formally evaluate the ratio between the (exponentially small) bulk contributions of $P^0$ and $P^-$:
\al{
R_V^\tx{bulk}
	\sim
		\frac{  e^{- F^0_\tx{bulk}|_{\tx{for }P^+} } }{  e^{- F^0_\tx{bulk}|_{\tx{for } P^0} }  }
	\sim
		\begin{cases}
			e^{12\,\tx{MeV}^2\, \sigma_K} & \text{for } V = \phi, \\
			e^{- 1.0 \times 10^3\,\tx{MeV}^2 \, \sigma_D} & \text{for } V = \psi, \\
			e^{3.5 \times 10^2\,\tx{MeV}^2\, \sigma_B} & \text{for } V = \Upsilon,
		\end{cases}
}
where the exponent is from Eq.~\eqref{F0bulk}.
This ratio becomes either exponentially large or small due to the mass difference between $P^+$ and $P^0$,
where the magnitude of the exponents is much greater than ${\cal O}\fn{1}$.
For example, we obtain $\sigma_P\gtrsim10\,\tx{MeV}^{-2}$ and $100\,\tx{MeV}^{-2}$ if we estimate $\sqrt{\sigma_P}$ to be larger than the smallest radius of an electron in atoms that interact with decay products of $P$, namely, the Bohr radius divided by a typical atomic number of the detector atoms, say, $a_\tx{B}/Z\simeq3\,\tx{MeV}^{-1}$ and $10\,\tx{MeV}^{-1}$ for lead and iron with $Z=82$ and 26, respectively.
As introduced, the experimental results of $R_\phi$, $R_\psi$, and $R_\Upsilon$ are around unity, and 
they disagree with $R_V^\tx{bulk}$, both for $\phi$ and $\psi$.
So, the $R_V^\tx{bulk}$ curves are completely out of the depicted ranges of the left panels of Figs.~\ref{fig:R-WP-phi},~\ref{fig:R-WP-psi}, and \ref{fig:R-WP-Upsilon}.
\end{itemize}

\subsection{Planewave analysis\label{sec:planewave-format}}

For a comparison with the wave-packet results, we also show results with the plane-wave decay rate $\Gamma^\tx{plane}$ (see Eq.~\eqref{plane-wave decay rate with form factor} in Appendix~\ref{decay rate section}), taking into account the relativistic form factor~\eqref{relativistic form factor}.
The resultant plane-wave ratio becomes
\al{
R_V^\tx{plane} 
	&:= \frac{\Gamma^\tx{plane}_{V \to P^+P^-}}{\Gamma^\tx{plane}_{V \to P^0 \ol{P^0}}}
	=
		R_V^\tx{parton}
		\left| 
			\frac{ \pn{\frac{R_0 \paren{ m_V^2 - 4 m_{P^0}^2 }^{1/2} }{2}}^2 + 1 }
			       { \pn{\frac{R_0 \paren{ m_V^2 - 4 m_{P^+}^2 }^{1/2} }{2}}^2 + 1 }
		\right|^2,
	\label{eq:R-PW-relativistic}
}
where the parton-level contribution to the ratio is
\al{
R_V^\tx{parton} :=
\paren{ \frac{m_V^2 - 4 m_{P^+}^2}{m_V^2 - 4 m_{P^0}^2} }^{3/2},
	\label{eq:R-parton}
}
and the other factor is from the relativistic form factor~\eqref{relativistic form factor} written in terms of the masses and $R_0$.

For another comparison, we will also show analyses using its non-relativistic approximated form:
\al{
R_V^\tx{plane} 
	\Rightarrow
R_V^\tx{plane, non-rel}
	:=	 \frac{  m_{P^+}^{1/2} \paren{ m_V - 2 m_{P^+} }^{3/2}  }{  m_{P^0}^{1/2} \paren{ m_V - 2 m_{P^0} }^{3/2}  }
		\left|   
			\frac{ \paren{\frac{R_0 m_{P^0} \ab{\bs V_1 - \bs V_2}_{P^0} }{2}}^2 + 1 }
			       { \paren{\frac{R_0 m_{P^+} \ab{\bs V_1 - \bs V_2}_{P^+} }{2}}^2 + 1 }  
		\right|^2,
	\label{eq:R-PW-non-relativistic}
}
with
\al{
\ab{\bs V_1 - \bs V_2}_{P^+}
	&\approx
		\frac{ 2\paren{m_V - 2m_{P^+}}^{1/2} }{ m_{P^+}^{1/2} },&
\ab{\bs V_1 - \bs V_2}_{P^0}
	&\approx
		\frac{ 2\paren{m_V - 2m_{P^0}}^{1/2} }{ m_{P^0}^{1/2} },&
}
where ``$\Rightarrow$'' represents the operation of taking the non-relativistic approximation and $\approx$ denotes equality under the non-relativistic approximation.
The contributions from the form factor are not canceled out in $R_V^\tx{plane}$.\footnote{
A similar factor is taken into account in Ref.~\cite{Fischbach} as a purely phenomenological cutoff factor of a divergent integral within the plane-wave formalism.
}
Note that the ratio~\eqref{eq:R-PW-non-relativistic} can be obtained from the wave-packet counterpart by taking the limits $\Gamma_V\to0$ and $\sigma_P\to\infty$ in $\Gamma_VP_{V \to P\ol{P}}$; see Appendix~\ref{sec:Gamma-zero-limit} for details.

\begin{figure}
\centering
\includegraphics[width=.43\textwidth]{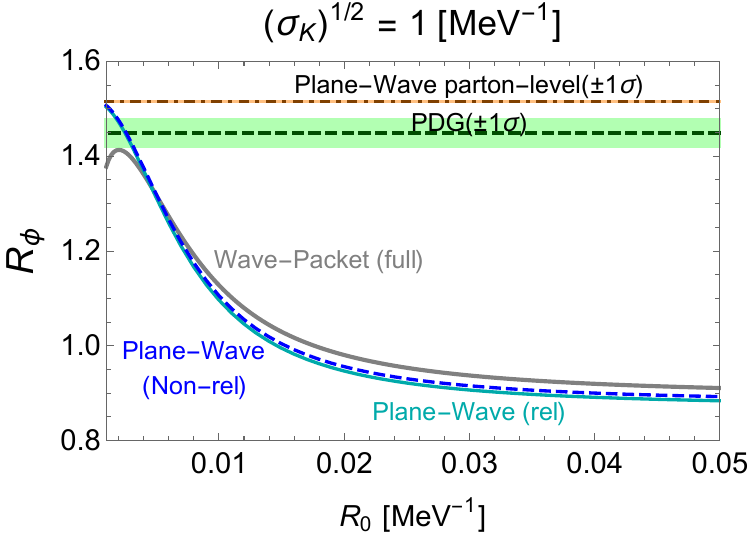} \ \
\includegraphics[width=.43\textwidth]{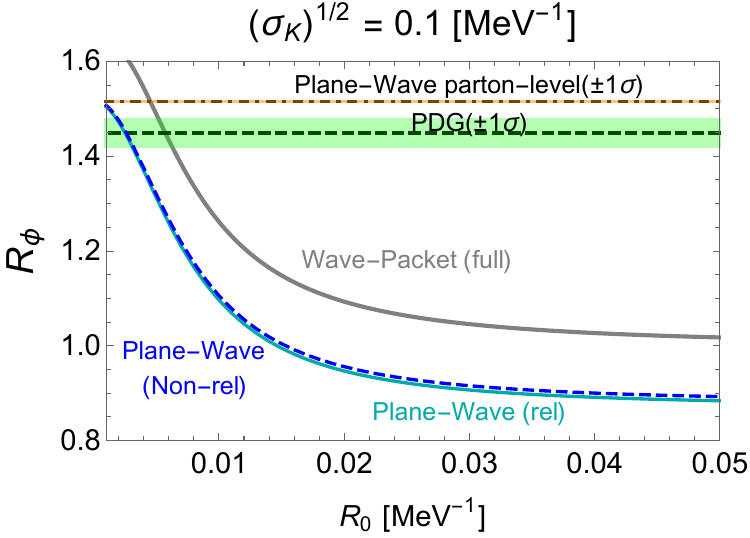} \\
\caption{
The plane-wave ratio comparing the decay rates of $\phi \to K^+K^-$ to $\phi \to K^0\ol{K^0}$ is drawn as a function of $R_0$, where the captions ``rel'' and ``Non-rel'' mean the relativistic and non-relativistic results shown in Eqs.~\eqref{eq:R-PW-relativistic} and \eqref{eq:R-PW-non-relativistic}, respectively.
For comparison, we also show the wave-packet results for two fixed wave-packet sizes of the Kaons of $\sqrt{\sigma_K} = 1\,\text{MeV}^{-1}$ (left panel) and $\sqrt{\sigma_K} = 0.1\,\text{MeV}^{-1}$ (right panel).
In both panels, the plane-wave results are the same since they are independent of $\sigma_K$.
The experimental result, shown in Eq.~\eqref{PDG ratio result}, is provided by the PDG~\cite{ParticleDataGroup:2022pth}.
}

\label{fig:R-PW-phi}
\end{figure}
\begin{figure}
\centering
\includegraphics[width=.43\textwidth]{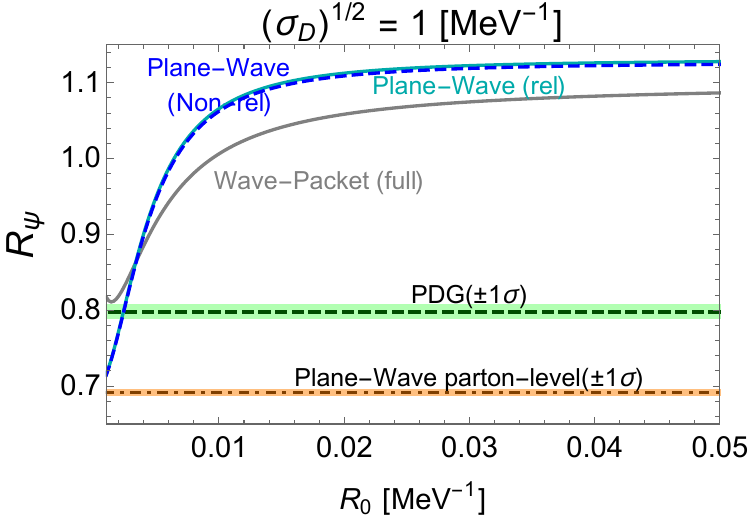} \ \
\includegraphics[width=.43\textwidth]{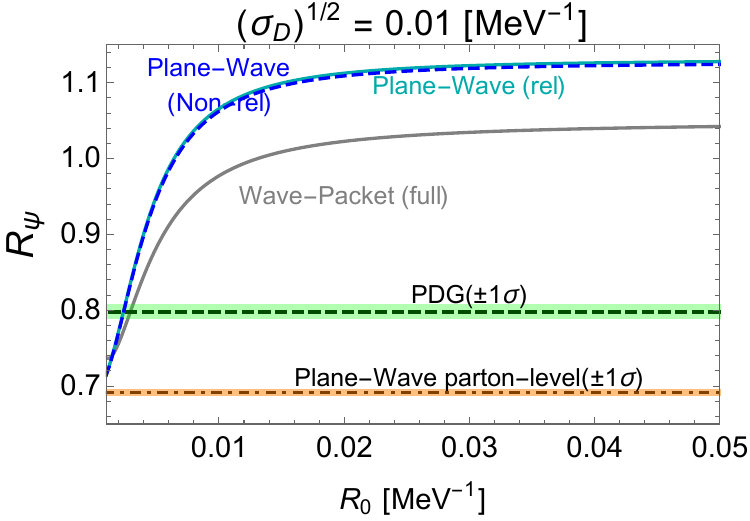} \\
\caption{
The plane-wave ratio comparing the decay rates of $\psi \to D^+D^-$ to $\psi \to D^0\ol{D^0}$ is drawn as a function of $R_0$.
For comparison, we also show the wave-packet results for two fixed wave-packet sizes of the D-mesons of $\sqrt{\sigma_D} = 1\,\text{MeV}^{-1}$ (left panel) and $\sqrt{\sigma_D} = 0.01\,\text{MeV}^{-1}$ (right panel).
The other conventions are the same as in Fig.~\ref{fig:R-PW-phi}.
}
\label{fig:R-PW-psi}
\end{figure}
\begin{figure}
\centering
\includegraphics[width=.43\textwidth]{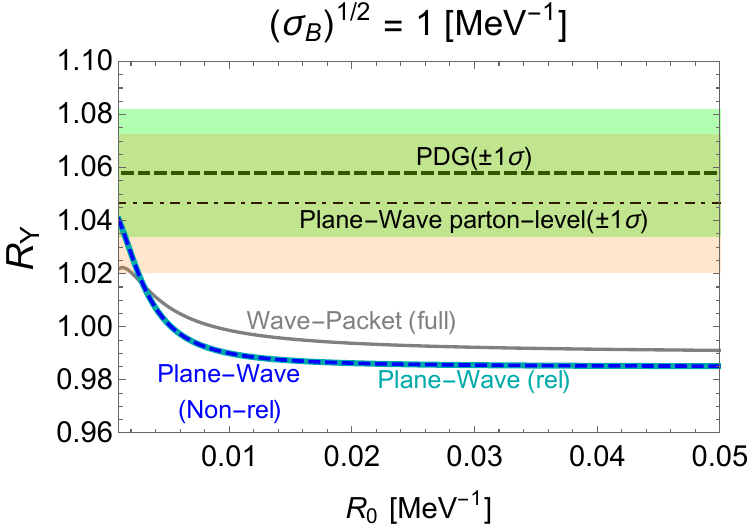} \ \
\includegraphics[width=.43\textwidth]{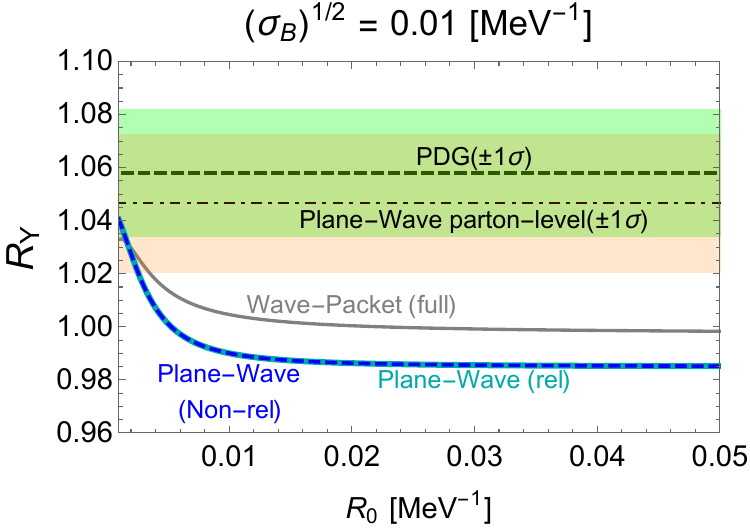} \\
\caption{
The plane-wave ratio comparing the decay rates of $\Upsilon \to B^+B^-$ to $\Upsilon \to B^0\ol{B^0}$ is drawn as a function of $R_0$.
For comparison, we also show the wave-packet results for two fixed wave-packet sizes of the B-mesons of $\sqrt{\sigma_B} = 1\,\text{MeV}^{-1}$ (left panel) and $\sqrt{\sigma_B} = 0.01\,\text{MeV}^{-1}$ (right panel).
The other conventions are the same as in Fig.~\ref{fig:R-PW-phi}.
}
\label{fig:R-PW-Upsilon}
\end{figure}

\subsection{Plane-wave results}
We provide comments on the plane-wave results shown in Figs.~\ref{fig:R-PW-phi}, \ref{fig:R-PW-psi}, and \ref{fig:R-PW-Upsilon} below:
\begin{itemize}
\item
For all of the vector mesons, $\phi$, $\psi$, and $\Upsilon$, the parton-level ratios~\eqref{tree plane} (under the isospin-symmetric limit for the couplings,\footnote{
Here, we are assuming the isospin-symmetric limit in the sense of Eq.~\eqref{isospin-symmetric limit} for both the wave-packet and plane-wave calculations.
}
without taking into account the form factor) are disfavored with the PDG's central values
at the level of $2.1\,\sigma$, $9.5\,\sigma$, and $0.32\,\sigma$, respectively.
\item
On the other hand, when the form-factor effect is included, which is compulsory since the vector mesons are composite particles,
we can see agreements with the PDG's results.
It suggests the importance of the form factor in addressing the ratio, where its effect is not fully canceled.
Also, we can confirm that the non-relativistic results approximate their relativistic counterparts well for the current system.
\item
We can find appropriate ranges of the form-factor parameter $R_0$, where theoretical predictions agree with the PDG's results for both the full wave-packet and plane-wave curves.
For each vector meson, the favored regions of $R_0$ for the wave packet and the plane wave are close to each other but different.
\end{itemize}
The calculation based on the plane wave works successfully,
even though the presumption of free plane waves characterizing initial and final states is, at most, a viable approximation.
The wave packet-based calculation provides a comprehensive approach, accounting for all aspects of the quantum nature inherent in the initial and final states, thereby enhancing its reliability.
It would be important to precisely discuss the theoretically valid region of $R_0$, which depends on many details on the strong interaction.
We leave this point for future research.

Note that we also briefly consider the isospin violation on the $\rho$ system, where the result is separately available in Appendix~\ref{sec:R-rho} since it might be out of our main interest.


\section{Constraint from the shape of vector-meson resonances\label{sec:resonance}}
In general, it is expected that the resonance shape of $V$ is modified by the inclusion of the wave packet effects.
That is, vector mesons, produced as resonances in electron-positron colliders, are subjected to a shape-fitting process. 
This section addresses the constraint on the wave-packet size of pseudoscalar mesons through the resonance shape of the process $e^-e^+ \to V \to P \overline{P}$. Sufficiently precise resonance data from experiments is available for $\phi$ and $\psi$, facilitating this purpose. However, detailed resonance data for $\Upsilon$ is currently unavailable.
Consequently, our focus is maintained on the instances of $\phi$ and $\psi$.
Our analysis in this section is meant to be a brief consistency check, assuming the factorization of the production and decay processes of $V$ in both the wave-packet and plane-wave formalisms, and hence is confined to data around the peak.

\subsection{Invariant mass distribution of decaying vector-meson wave packet}

First, we summarize the invariant mass distribution of the decaying vector-meson
when the Gaussian wave packet describes the decaying state.

We define a Lorentz-invariant mass squared $M^2$ for the pair of pseudo-scalars in the final state:
\al{
M^2
	&:= 
		\paren{E_1 + E_2}^2 - \paren{ {\bs P_1 + \bs P_2} }^2
	\Rightarrow
		m_P^2 \paren{4 + V_-^2},
	\label{eq:Meq-definition}
}
where
$V_-$ is the magnitude of $\bs V_-:=\bs V_1-\bs V_2$ with $\bs V_a=\bs P_a/E_a$ for $a=1,2$ (see Eq.~\eqref{plus minus notations defined} in Appendix~\ref{detailed decay probability section}).
We will use
\al{
V_-^2 \approx \frac{1}{m_P^2} \paren{ M^2 - 4 m_P^2 },
	\label{eq:Vm-to-Msq}
}
which results in
\al{
V_-\df V_-
	&\approx
		{M\df M\ov m_P^2}
	\simeq
		{m_V\df M\ov m_P^2},
	\label{eq:dVminus-to-dM}
}
where we have approximated that $V$ decays at rest in the last step.

It is straightforward to derive the following forms after integrating Eq.~\eqref{master dP} over the final state phase space, except for $V_-$, under the current non-relativistic approximation, which is easily rewritten as the invariant mass distribution by use of Eq.~\eqref{eq:dVminus-to-dM}:
\al{
\frac{\df P_{V \to P \ol{P}}}{\df M} 
	=
		 \frac{\df P_{V \to P \ol{P}}^\text{bulk}}{\df M} + \frac{\df P_{V \to P \ol{P}}^\text{bdry}}{\df M} + \frac{\df P_{V \to P \ol{P}}^\text{intf}}{\df M},
		 \label{eq:NR-WP-resonance}
}
where
\al{
\frac{\df P_{V \to P \ol{P}}^\text{bulk}}{\df M}
	&\simeq
		\frac{2m_V}{m_P^2}
		\paren{ \frac{1}{16\sqrt{2\pi}} {\cal C}_{V \to P \ol{P}} }
		{m_P^2\sqrt{\sigma_P}\ov\Gamma_V}
		V_-^2
		{  
			\frac
				{  e^{  -F^0_\text{bulk} - \frac{m_P^2\sigma_P}{2}\paren{V_- - V_-^\text{B}}^2  }  }
				{A_\text{bulk}^{3/2} }      
		}
		\ab{ \widetilde{F}\fn{V_-} }^2, \\
\frac{\df P_{V \to P \ol{P}}^\text{bdry}}{\df M}
	&\simeq
		\frac{2m_V}{m_P^2}
		\paren{  \frac{1}{4\pi} {\cal C}_{V \to P \ol{P}}  }
		\frac{\widetilde{f}_\text{bdry}\fn{V_-}}{V_-}
			\ab{ \widetilde{F}\fn{V_-} }^2, \\
\frac{\df P_{V \to P \ol{P}}^\text{intf}}{\df M}
	&\simeq
		-
		\frac{2m_V}{m_P^2}
		\paren{ \frac{1}{8\sqrt{2} \pi} {\cal C}_{V \to P \ol{P}} }
		\paren{ \sigma_P m_P^2 }
		V_-
		{
			\frac
				{e^{  -F^0_\text{intf} - \frac{1}{2} \paren{\frac{m_P^2\sigma_P}{2}} \paren{V_- - V_-^\text{I}}^2   }}
				{A_\text{intf}^{3/2}}
			} \notag \\
	&\quad \times
		\sqbr{
		\frac
		{ \paren{\Gamma_V^2 - \paren{\delta\omega}^2}
			\cos\fn{2 \Gamma_V\sigma_t\delta\omega}
		  	- 2 \Gamma_V\delta\omega \sin\fn{2 \Gamma_V\sigma_t\delta\omega} }
		{\paren{  \Gamma_V^2 + \paren{{\delta\omega}}^2     }^2\sigma_t}
		}_{V_+ \to 0}
		\ab{ \widetilde{F}\fn{V_-} }^2.
}
Here, we consider the distribution of $M$ instead of $M^2$ due to the convenience of comparing the wave-packet shape with the non-relativistic Breit-Wigner~(BW) shape, which is the well-known resonant shape for the decaying plane wave with the decay rate $\Gamma_V$; see the next subsection.\footnote{
Experimentally, one may perform a precision experiment by measuring the ratio per each bin $\Delta M$ near the resonance, in principle:
\als{
{{\df P_{V\to P^+P^-}\ov\df M}\Delta M\ov{\df P_{V\to P^0\ol{P^0}}\ov\df M}\Delta M}
	&=	{{\df P_{V\to P^+P^-}\ov\df M}\ov{\df P_{V\to P^0\ol{P^0}}\ov\df M}}.
}
}

We note that, under the current setup $T_\text{out} \to \infty$, the factor $N_V^2$ in ${\cal C}_{V \to P \ol{P}}$ (recall Eq.~\eqref{eq:common-C-factor}) can be determined by the normalization
\al{
P^\text{bulk}_{V \to P \ol{P}} + P^\text{bdry}_{V \to P \ol{P}} + P^\text{intf}_{V \to P \ol{P}} = \paren{\text{the corresponding branching ratio}}
}
that is obtained after integrating over $M$; see Eqs.~\eqref{eq:P-integrated_bulk}, \eqref{eq:P-integrated_bdry}, and \eqref{eq:P-integrated_intf}.

\subsection{Breit-Wigner shape}

For the plane-wave calculation, it is well-known that the non-relativistic Breit-Wigner distribution nicely describes the shape of a narrow resonance,\footnote{
The relativistic Breit-Wigner distribution takes
\als{
f_\text{R-BW}\fn{s}
	&=
		\frac{\paren{m_\tx{res} \, \Gamma}/\pi}{ \paren{s-m_\tx{res}^2}^2 + \paren{m_\tx{res}\,\Gamma}^2 } &
		\bigg( \int_{-\infty}^{+\infty} \df s \, f_\text{R-BW}\fn{s} &= 1 \bigg),
}
which is not used for the calculation.
Mandelstam's variable $s$ is equal to $E^2$.
}
\al{
f_\text{NR-BW}\fn{E}
	&=
		\frac{\Gamma/\paren{2\pi}}{ \paren{E-m_\tx{res}}^2 + {\Gamma}^2/4 } &
		\bigg( \int_{-\infty}^{+\infty} \df E \, f_\text{NR-BW}\fn{E} &= 1 \bigg),
	\label{eq:NR-BW-shape}
}
where $m_\tx{res}$, $E$, and $\Gamma$ are the resonant mass, the total energy in the center-of-the-mass frame, and the total width of
an intermediate resonant particle, respectively.
Note that
\al{
E = M.
	\label{eq:E-to-M}
}
Since we used the non-relativistic approximation, we are adopting the non-relativistic Breit-Wigner shape~\eqref{eq:NR-BW-shape} for comparison.

\subsection{Method of analyzing resonant shape}

We assume the following factorization for the resonant production, where
the cross-section of the resonant production of $V$ and its subsequent decay into $P$ and $\overline{P}$,
$\sigma_{e^-e^+ \to V \to P \ol{P}}$, can be factorized in the wave-packet (WP) and plane-wave (PW) formalisms, respectively as
\al{
\sigma_{e^-e^+ \to V \to P \ol{P}}^\tx{WP}\fn{M}
	&=
		{\cal N}^\tx{WP}_{e^-e^+ \to V}  \frac{\df P_{V \to P \ol{P}}}{\df M},
	\label{eq:sigma-V-WP} \\
\sigma_{e^-e^+ \to V \to P \ol{P}}^\tx{PW-Parton}\fn{M}
	&=
		{\cal N}^\tx{PW-Parton}_{e^-e^+ \to V}  f_\text{NR-BW}\fn{M},
	\label{eq:sigma-V-PW} \\
\sigma_{e^-e^+ \to V \to P \ol{P}}^\tx{PW-FF}\fn{M}
	&=
		{\cal N}^\tx{PW-FF}_{e^-e^+ \to V}  f_\text{NR-BW}\fn{M}
		\ab{ \frac{1}{1+ {\frac{R_0^2 \paren{M^2 - 4m_P^2}}{4}} } }^2,
	\label{eq:sigma-V-PWFF}
}
where we consider the two cases for PW with and without the form factor (FF); the two cases are discriminated by the short-hand notations ``PW-FF'' and ``PW-Parton''.
For the form-factor part of~\eqref{eq:sigma-V-PWFF}, we used the relation in Eq.~\eqref{eq:Vm-to-Msq} to convert $V_-$ to $M$.

${\cal N}^\tx{WP}_{e^-e^+ \to V}$, ${\cal N}^\tx{PW-Parton}_{e^-e^+ \to V}$, and ${\cal N}^\tx{PW-FF}_{e^-e^+ \to V}$
 possess the mass dimension of minus one and describe the factorized production part $e^-e^+ \to V$ via the $e^-e^+$ collision at the center-of-the-mass energy $M$.
Here, we take these three factors to be independent of $M$ since the primal structure of the resonance is in ${\df P_{V \to P \ol{P}}}/{\df M}$ or $f_\text{NR-BW}$,
and we use only the data points near the peak of a resonance.\footnote{
As is widely known, under the narrow-width approximation in the plane-wave calculation at the resonant peak $M = m_\tx{res}$,
we can derive the factorized form explicitly:
\als{
\sigma_{e^-e^+ \to V \to P \ol{P}}^\tx{PW}\fn{M}
	\simeq
		\sigma^\tx{PW}_{e^-e^+ \to V} \, \tx{Br}\fn{V \to P \overline{P}};
}
refer to, e.g., Chapter~16 of~\cite{Thomson:2013zua}.
And at this point, ${\cal N}^\tx{PW}_{e^-e^+ \to V}$ is determined as
\als{
{\cal N}^\tx{PW}_{e^-e^+ \to V}
	\simeq
		\sigma^\tx{PW}_{e^-e^+ \to V} \, \frac{\pi}{2} \Gamma_{V \to P \overline{P}}.
}
Also, we note that the width-to-mass ratios of the vector mesons take
$\Gamma_\phi/m_\phi \simeq 0.42\%$ and $\Gamma_\psi/m_\psi \simeq 0.72\%$, where the adaptation of the narrow-width approximation is justified.
}
We will take $m_\tx{res}$ and $\Gamma$ for $f_\text{NR-BW}\fn{M}$ in Eq.~\eqref{eq:NR-BW-shape} as $m_V$ and $\Gamma_V$, respectively; see also Eq.~\eqref{eq:E-to-M}.
The actual analysis for $e^-e^+ \to V \to P \ol{P}$ will be done in the following manner:

\begin{itemize}
\item
We focus on the values of the experimentally-given cross sections only around the resonant peak, namely in $[m_V- \Gamma_V/2, m_V+ \Gamma_V/2]$ since the factorized forms in Eqs.~\eqref{eq:sigma-V-WP}, \eqref{eq:sigma-V-PW}, and \eqref{eq:sigma-V-PWFF} may work only around the peak.
Here, we will adopt the PDG values for $m_V$ and $\Gamma_V$~\cite{ParticleDataGroup:2022pth}.
\item
In the analysis, we fix the values of $\Gamma_V$ and $m_P$ as confirmed by the PDG group~\cite{ParticleDataGroup:2022pth}, while we treat $m_V$ as an unfixed parameter and will determine it through our statistical fit.
The isospin-symmetric coupling $g_V$ and the wave-function renormalization factor for $V$, $N_V$ are taken as unity since it can be absorbed into the factor ${\cal N}_{e^-e^+ \to V}$.
Furthermore, for simplicity, we focus on $T_\tx{in} = T_0$, where the exponential decay factor in ${\cal C}_{V \to P \ol{P}}$ in Eq.~\eqref{eq:common-C-factor} does not work.
\item
Under the current scheme,
$\sigma_{e^-e^+ \to V \to P \ol{P}}^\tx{WP}$ has
six parameters $\{ {\cal N}^\tx{WP}_{e^-e^+ \to V},\, m_V, \, \Gamma_V, \, m_P, \, R_0, \, \sigma_P \}$,
$\sigma_{e^-e^+ \to V \to P \ol{P}}^\tx{PW-FF}$ has
five parameters $\{ {\cal N}^\tx{PW-FF}_{e^-e^+ \to V}, \, m_V, \, \Gamma_V, \, m_P,\, R_0 \}$,
and $\sigma_{e^-e^+ \to V \to P \ol{P}}^\tx{PW-Parton}$ has
three parameters $\{ {\cal N}^\tx{PW-Parton}_{e^-e^+ \to V}, \, m_V, \, \Gamma_V \}$, respectively.
We will determine them through statistical analysis.
We remind ourselves that the vector-meson wavepacket size $\sigma_V$ does not appear in $\sigma_{e^-e^+ \to V \to P \ol{P}}^\tx{WP}$ under the saddle-point approximation.
\end{itemize}

\subsection{Result of $\phi$}

In Ref.~\cite{Kozyrev:2017agm}, 
the latest result of the resonant shape of $\phi$ through $e^-e^+ \to \phi \to K^+ K^-$ measured with the CMD-3 detector in the center-of-mass energy range $1010$--$1060\,\tx{MeV}$ was reported, where the Born cross sections of $e^-e^+ \to \phi \to K^+ K^-$ around the resonance are available in Table~I of~\cite{Kozyrev:2017agm}.
According to our guideline, we adopt the seven data points from $1018.0\,\tx{MeV}$ to $1021.3\,\tx{MeV}$ and adopt the $\chi^2$ functions:
\al{
\chi^2_{\phi,\, \tx{WP}}
	&:=
		\sum_{i=1}^{7}
			\frac{\paren{\sigma_i^\tx{WP} - \sigma_i^\tx{exp}}^2}{\paren{\delta\sigma_i^\tx{exp}}^2},& \notag \\
\chi^2_{\phi,\, \tx{PW-Parton}}
	&:=
		\sum_{i=1}^{7}
			\frac{\paren{\sigma_i^\tx{PW-Parton} - \sigma_i^\tx{exp}}^2}{\paren{\delta\sigma_i^\tx{exp}}^2},&
\chi^2_{\phi,\, \tx{PW-FF}}
	&:=
		\sum_{i=1}^{7}
			\frac{\paren{\sigma_i^\tx{PW-FF} - \sigma_i^\tx{exp}}^2}{\paren{\delta\sigma_i^\tx{exp}}^2},&
	\label{eq:chisq-form-phi}
}
where $i$ discriminates the seven points of $M$ where experimental data is available;
$\sigma_i^\tx{exp}$ and $\delta\sigma_i^\tx{exp}$ are the central and error of the experimentally-determined cross section at the point $i$, respectively.
$\sigma_i^\tx{WP}$, $\sigma_i^\tx{PW-Parton}$ and $\sigma_i^\tx{PW-FF}$ represent the theoretical values of the corresponding cross sections at the energy point identified by $i$.

\begin{figure}[t]
\centering
\includegraphics[width=.49\textwidth]{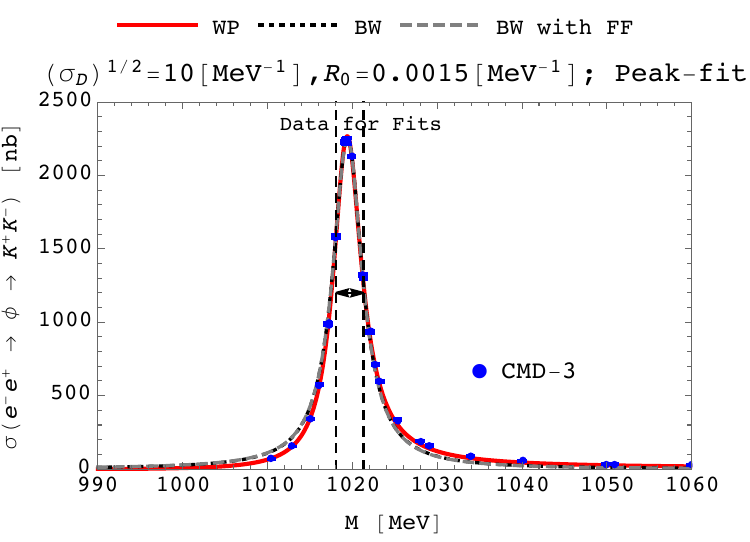} \ \ 
\includegraphics[width=.47\textwidth]{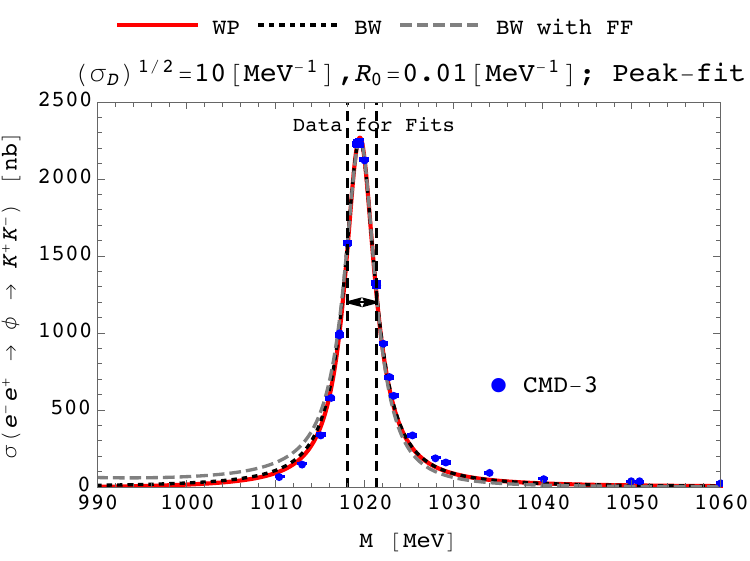}
\caption{
The fitted resonance distributions of $\phi$ in $e^-e^+ \to \phi \to K^+K^-$ are drawn
for the two sets of the fixed parameters
$\sqrt{\sigma_K} = 10\,\tx{MeV}^{-1}$ and $R_0 = 0.0015 \, \tx{MeV}^{-1}$ [Left panel] and
$\sqrt{\sigma_K} = 10\,\tx{MeV}^{-1}$ and $R_0 = 0.01 \, \tx{MeV}^{-1}$ [Right panel],
where the mass of $\phi$ and the normalization factor ${\cal N}_{e^-e^+ \to \phi}$ are determined through our statistical analysis based on the $\chi^2$ function defined in Eq.~\eqref{eq:chisq-form-phi}.
The best-fit parameters and the $\chi^2$ functions for the left/right panel are shown in Eqs.~\eqref{eq:phi-domain-Left} and~\eqref{eq:chi-eq-concrete-Left} / in Eqs.~\eqref{eq:phi-domain-Right} and~\eqref{eq:chi-eq-concrete-Right}, respectively, for the wave-packet (WP) and plane-wave without/with form factor (BW/BW with FF).
}
\label{fig:phi-peak}
\end{figure}

As examples, we show the fitted distributions for the two sets of the fixed parameters
$\sqrt{\sigma_K} = 10\,\tx{MeV}^{-1}$ and $R_0 = 0.0015 \, \tx{MeV}^{-1}$ for the left panel of Fig.~\ref{fig:phi-peak},\footnote{
Note that the value of $R_0 = 0.0015 \, \tx{MeV}^{-1}$ is a typical value in the current scheme of the form factor (see Appendix~\ref{form factor session}), and its magnitude is favored by the analysis on $R_V$ (see Section~\ref{sec:R_V}).
}
$\sqrt{\sigma_K} = {10}\,\tx{MeV}^{-1}$ and $R_0 = 0.01 \, \tx{MeV}^{-1}$ for the right panel of Fig.~\ref{fig:phi-peak},
where the two remaining parameters $\{ {\cal N}_{e^-e^+ \to \phi},\, m_\phi \}$ take the best-fit values,
and the values of the $\chi^2$ over the degrees of freedoms~(DOFs), which is currently five, at the best-fit points are calculated as
\begin{itemize}
\item
for $\sqrt{\sigma_K} = 10\,\tx{MeV}^{-1}$ and $R_0 = 0.0015 \, \tx{MeV}^{-1}$ [Left panel of Fig.~\ref{fig:phi-peak}]:
\al{
\left.m_\phi^\tx{WP}\right|_\tx{best-fit}
	&=
		 1019.8 \, \tx{MeV},&
\left.{\cal N}_{e^-e^+ \to \phi}^\tx{WP}\right|_\tx{best-fit}
	&=
		6.23 \times 10^{5}  \,\text{MeV}^{-1},&
	\notag \\
\left.m_\phi^\tx{PW-Parton}\right|_\tx{best-fit}
	&=
		 1019.4 \, \tx{MeV},&
\left.{\cal N}_{e^-e^+ \to \phi}^\tx{PW-Parton}\right|_\tx{best-fit}
	&=
		1.50 \times 10^{4} \,\text{MeV}^{-1},&
	\notag \\
\left.m_\phi^\tx{PW-FF}\right|_\tx{best-fit}
	&=
		 1019.4 \, \tx{MeV},&
\left.{\cal N}_{e^-e^+ \to \phi}^\tx{PW-FF}\right|_\tx{best-fit}
	&=
		1.62 \times 10^{4} \,\text{MeV}^{-1},&
	\label{eq:phi-domain-Left}
}
\vspace{-6mm}
\al{
\left.\frac{\chi^2_{\phi,\, \tx{WP}}}{(\tx{DOFs})}\right|_\tx{best-fit}
	&\simeq
		5.5,&
\left.\frac{\chi^2_{\phi,\, \tx{PW-Parton}}}{(\tx{DOFs})}\right|_\tx{best-fit}
	&\simeq
		6.3,&
\left.\frac{\chi^2_{\phi,\, \tx{PW-FF}}}{(\tx{DOFs})}\right|_\tx{best-fit}
	&\simeq
		6.7,&
	\label{eq:chi-eq-concrete-Left}
}

\item
for $\sqrt{\sigma_K} = 10\,\tx{MeV}^{-1}$ and $R_0 = 0.01 \, \tx{MeV}^{-1}$  [Right panel of Fig.~\ref{fig:phi-peak}]:
\al{
\left.m_\phi^\tx{WP}\right|_\tx{best-fit}
	&=
		 1019.9 \, \tx{MeV},&
\left.{\cal N}_{e^-e^+ \to \phi}^\tx{WP}\right|_\tx{best-fit}
	&=
		3.94 \times 10^{6}  \,\text{MeV}^{-1},&
	\notag \\
\left.m_\phi^\tx{PW-Parton}\right|_\tx{best-fit}
	&=
		 1019.4 \, \tx{MeV},&
\left.{\cal N}_{e^-e^+ \to \phi}^\tx{PW-Parton}\right|_\tx{best-fit}
	&=
		1.51 \times 10^{4} \,\text{MeV}^{-1},&
	\notag \\
\left.m_\phi^\tx{PW-FF}\right|_\tx{best-fit}
	&=
		 1019.6 \, \tx{MeV},&
\left.{\cal N}_{e^-e^+ \to \phi}^\tx{PW-FF}\right|_\tx{best-fit}
	&=
		1.03 \times 10^{5} \,\text{MeV}^{-1},&
	\label{eq:phi-domain-Right}
}
\vspace{-6mm}
\al{
\left.\frac{\chi^2_{\phi,\, \tx{WP}}}{(\tx{DOFs})}\right|_\tx{best-fit}
	&\simeq
		14,&
\left.\frac{\chi^2_{\phi,\, \tx{PW-Parton}}}{(\tx{DOFs})}\right|_\tx{best-fit}
	&\simeq
		6.3,&
\left.\frac{\chi^2_{\phi,\, \tx{PW-FF}}}{(\tx{DOFs})}\right|_\tx{best-fit}
	&\simeq
		14.&
	\label{eq:chi-eq-concrete-Right}
}
\end{itemize}
We comment on the difference between the plane-wave resonant shapes with and without the form factor.
Without the factor, the shape obeys the Breit--Wigner distribution~\eqref{eq:NR-BW-shape} and is symmetric under the reflection around the peak ($M = m_\phi$), while taking into account it, the resonant shape becomes asymmetric under the reflection around the peak.
The magnitude of the asymmetry is governed by the part $R_0^2 \paren{M^2 - 4 m_{K^+}^2}$ of the form factor.
So, for a greater $R_0$, a more significant asymmetry will be realized, as observed in Fig.~\ref{fig:phi-peak}.

\begin{figure}[t]
\centering
\includegraphics[width=.50\textwidth]{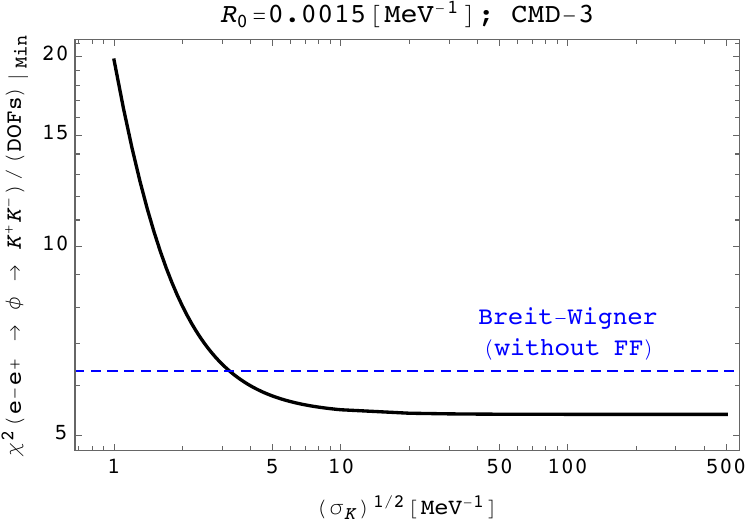} \ \ 
\caption{
We plot the variable $\left.{\chi^2_{\phi}}/{(\tx{DOFs})}\right|_\tx{min}$ defined in Eq.~\eqref{eq:chisq-ov-DOFs_phi} to compare the significance of the wave-packet calculation with the plane-wave one for various $\sqrt{\sigma_K}$.
Here, $R_0$ is fixed as $0.0015 \, \tx{MeV}^{-1}$ and for each $\sqrt{\sigma_K}$, $m_\phi$ and ${\cal N}_{e^-e^+ \to \phi}$ are determined to (locally) minimize the corresponding $\chi^2$ function in Eq.~\eqref{eq:chisq-form-phi}.
The black curve and blue dashed horizontal line describe the values in the wave-packet and plane-wave calculations without form factor, where the latter is manifestly independent of $\sqrt{\sigma_K}$.
}
\label{fig:chieqdist_phi}
\end{figure}

Here, we comment on the origin of the ``over-$5\sigma$'' values of $\chi^2/(\tx{DOFs})$: this is because
the resolution of the experimental results near the peak is very high, and the current simple scheme for $\sigma_{e^-e^+ \to V \to P\ol{P}}$ in Eqs.~\eqref{eq:sigma-V-WP}, \eqref{eq:sigma-V-PW} and \eqref{eq:sigma-V-PWFF} is not enough for discussing statistical significance precisely.
On the other hand, however, we are able to discuss the relative significance between the wave-packet and plane-wave results.
According to Eq.~\eqref{eq:chi-eq-concrete-Left}, the shape of the wave-packet resonant distribution is at least as good as that of the plane-wave resonant distribution at the focused parameter point, where we conclude that the wave-packet result at the first parameter point (for the left panel of Fig.~\ref{fig:phi-peak}) is consistent with the experiment.
Note that at the first parameter point, $R_0$ is taken as a typical value in the current scheme of the form factor (see Appendix~\ref{form factor session}), and a wave packet with a greater size looks similar to the plane wave.

We also see the significance of the wave-packet results over a broad range of $\sqrt{\sigma_K}$ under $R_0 = 0.0015 \, \tx{MeV}^{-1}$.
In Fig.~\ref{fig:chieqdist_phi}, we plot the ``minimized'' $\chi^2/(\tx{DOFs})$ defined by
\al{
\left.\frac{\chi^2_{\phi,\, \tx{WP/PW-Parton}}}{(\tx{DOFs})}\right|_\tx{min}
	:=
		\min_{m_\phi,\ {\cal N}_{e^-e^+ \to \phi}^\tx{WP/PW-Parton}}
		\sqbr{
		\frac{\chi^2_{\phi,\, \tx{WP/PW-Parton}}}{(\tx{DOFs})}
		},
	\label{eq:chisq-ov-DOFs_phi}
}
which measures the statistical significance for $\sqrt{\sigma_K}$.
We do not consider the PW-FF case since no sizable difference is generated when $R_0 = 0.0015 \, \tx{MeV}^{-1}$, as shown in the left panel of Fig.~\ref{fig:phi-peak}, and the form-factor part does not depend on $\sqrt{\sigma_K}$.
Under the simple guideline that a wave-packet result is at least as good as the ordinary plane-wave one,
from Fig.~\ref{fig:chieqdist_phi}, we can put the lower bound on $\sqrt{\sigma_K}$ as
\al{
\sqrt{\sigma_K} \gtrsim 3\,\text{MeV}^{-1}
}
for $R_0 = 0.0015 \, \tx{MeV}^{-1}$.

\subsection{Result of $\psi$}

The BaBar, Belle, BES, and CLEO experimental collaborations have provided recent experimental data of the $\psi$'s resonance produced by the $e^-e^+$ collision.

\begin{itemize}
\item
We will adopt the results on~\cite{Shamov:2016mxe} of the converting experimental results
to the exclusive initial-state-radiation scattering cross-section,
$e^+ e^- \to 2D$, where the following experimental papers are taken into account
by BaBar~\cite{BaBar:2006qlj}, by Belle~\cite{Belle:2007hht}, and by CLEO~\cite{CLEO:2005mpm,CLEO:2013rjc}.
``$2D$'' means the inclusive final states of $D^+ D^-$ and $D^0 \overline{D^0}$.\footnote{
The exclusive initial-state-radiation scattering cross-sections of $e^+ e^- \to D^+ D^-$ and $e^+ e^- \to D^0 \overline{D^0}$ are also
reported by Babar~\cite{BaBar:2009elc} and by Belle~\cite{Belle:2007qxm}.
Since few data points are available inside the focused range $[m_\psi-\Gamma_\psi/2,\, m_\psi+\Gamma_\psi/2]$, we do not adopt them for our analysis.
}
\item
We also take account of the experimental results by BES of the inclusive hadron-production cross-section in~\cite{Ablikim:2006mb,BES:2006piq}.
The original data is provided in the form,
\al{
R(s) = \frac{\sigma^0_\text{had}(s)}{\sigma^0_{\mu^+\mu^-}(s)},
}
where $\sigma^0_{\mu^+\mu^-}(s) = 4\pi \alpha^2\fn{0}/3s$ is the lowest-order QED cross section for muon pair production at the total center-of-mass energy $E = \sqrt{s}$. $\alpha\fn{0} \simeq 1/137$ is the QED fine structure constant at the Thomson limit.
$R_{uds(c)+\psi(3770)}$ and $R_{uds}$ are reported in~\cite{Ablikim:2006mb} and in~\cite{BES:2006piq}, respectively.
Through the approximation $R_{\psi(3770)} \simeq R_{uds(c)+\psi(3770)} - R_{uds}$, we can recast the cross section of $e^+ e^- \to 2D$,
as done in~\cite{Achasov:2021adv}.
\end{itemize}

Since the final state is inclusive as $2D$, we adopt the following factorized form for the production cross section:
\al{
\sigma_{e^-e^+ \to \psi \to 2D}^\tx{WP}\fn{M}
	&=
		{\cal N}^\tx{WP}_{e^-e^+ \to \psi} 
			\paren{
			\frac{\df P_{\psi \to D^+ D^-}}{\df M} +
			\frac{\df P_{\psi \to D^0 \ol{D^0}}}{\df M} 
			},
	\label{eq:sigma-psi-WP} \\
\sigma_{e^-e^+ \to \psi \to 2D}^\tx{PW-Parton}\fn{M}
	&=
		{\cal N}^\tx{PW-Parton}_{e^-e^+ \to \psi} \, f_\text{NR-BW}\fn{M},
	\label{eq:sigma-psi-PW} \\
\sigma_{e^-e^+ \to \psi \to 2D}^\tx{PW-FF}\fn{M}
	&=
		{\cal N}^\tx{PW-FF}_{e^-e^+ \to \psi} \, f_\text{NR-BW}\fn{M} \notag \\
	&\quad
		\times
		\sqbr{
			\tx{Br}_{\psi \to D^+D^-} \ab{ \frac{1}{1+ {\frac{R_0^2 \paren{M^2 - 4m_{D^+}^2}}{4}} } }^2 +
			\tx{Br}_{\psi \to D^0\ol{D^0}} \ab{ \frac{1}{1+ {\frac{R_0^2 \paren{M^2 - 4m_{D^0}^2}}{4}} } }^2
		},
	\label{eq:sigma-psi-PWFF}
}
where $\tx{Br}_{\psi \to D^+D^-} = 0.41$ and $\tx{Br}_{\psi \to D^0\ol{D^0}} = 0.52$ are the corresponding branching ratios~\cite{ParticleDataGroup:2022pth}.
The $\chi^2$ functions are defined as
\al{
\chi^2_{\psi,\, \tx{WP}}
	&:=
		\sum_{\tx{I: BaBar, Belle, BES, CLEO}} \sum_{i_\tx{I}}
			\frac{\paren{\sigma_{i_\tx{I}}^\tx{WP} - \sigma_{i_\tx{I}}^\tx{exp,I} }^2}{\paren{\delta\sigma_{i_\tx{I}}^\tx{exp,I} }^2},
			\notag \\
\chi^2_{\psi,\, \tx{PW-Parton}}
	&:=
		\sum_{\tx{I: BaBar, Belle, BES, CLEO}} \sum_{i_\tx{I}}
			\frac{\paren{\sigma_{i_\tx{I}}^\tx{PW-Parton} - \sigma_{i_\tx{I}}^\tx{exp,I} }^2}{\paren{\delta\sigma_{i_\tx{I}}^\tx{exp,I} }^2},
			\notag \\
\chi^2_{\psi,\, \tx{PW-FF}}
	&:=
		\sum_{\tx{I: BaBar, Belle, BES, CLEO}} \sum_{i_\tx{I}}
			\frac{\paren{\sigma_{i_\tx{I}}^\tx{PW-FF} - \sigma_{i_\tx{I}}^\tx{exp,I} }^2}{\paren{\delta\sigma_{i_\tx{I}}^\tx{exp,I} }^2},
	\label{eq:chisq-form-psi}
}
where $\sigma_{i_\tx{I}}^\tx{exp,I}$ and $\delta\sigma_{i_\tx{I}}^\tx{exp,I}$ are the central and error of the experimentally-determined cross section at the point $i_\tx{I}$ of the experiment $\tx{I}$, respectively.
$\sigma_{i_\tx{I}}^\tx{WP}$, $\sigma_{i_\tx{I}}^\tx{PW-Parton}$ and $\sigma_{i_\tx{I}}^\tx{PW-FF}$ represent the theoretical values of the corresponding cross sections at the energy point identified by ${i_\tx{I}}$.

\begin{figure}[t]
\centering
\includegraphics[width=.49\textwidth]{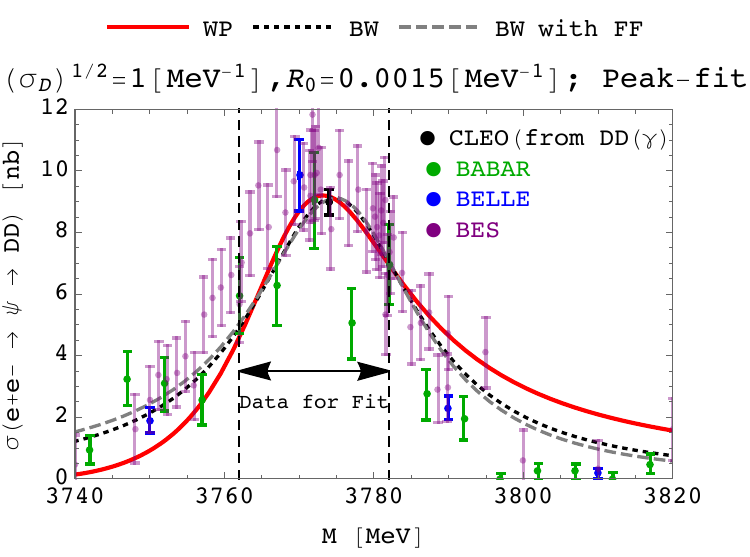} \ \ 
\includegraphics[width=.47\textwidth]{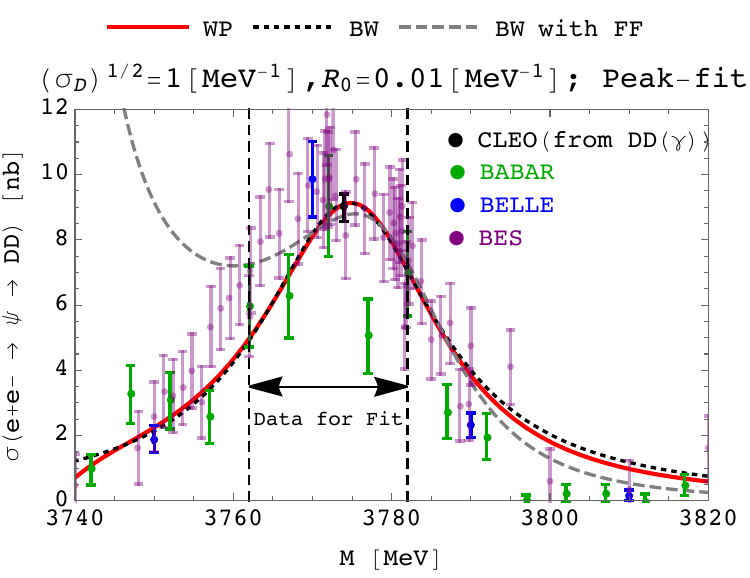}
\caption{
The fitted resonance distributions of $\phi$ in $e^-e^+ \to \psi \to 2D$ are drawn for the fixed parameters $\sqrt{\sigma_D} = 1\,\tx{MeV}^{-1}$ and $R_0 = 0.0015 \, \tx{MeV}^{-1}$ (for the left panel), and $\sqrt{\sigma_D} = 1\,\tx{MeV}^{-1}$ and $R_0 = 0.01 \, \tx{MeV}^{-1}$ (for the right panel), where the mass of $\psi$ and the normalization factor ${\cal N}_{e^-e^+ \to \psi}$ are determined through our statistical analysis based on the $\chi^2$ function defined in Eq.~\eqref{eq:chisq-form-psi}.
The best-fit parameters and the $\chi^2$ functions for the left/right panel are shown
in Eqs.~\eqref{eq:D-data-1} and~\eqref{eq:D-data-2} / in Eqs.~\eqref{eq:D-data-3} and~\eqref{eq:D-data-4}, respectively.
The other conventions are the same as those of Fig.~\ref{fig:phi-peak}.
}
\label{fig:psi-peak}
\end{figure}

Here, we will see the two examples for the same wave-packet size $\sqrt{\sigma_D} = 1\,\text{MeV}^{-1}$ but two different values for the form-factor parameter $R_0$.
In the left and right panels of Fig.~\ref{fig:psi-peak}, the fitted distributions about the parameters $\{ {\cal N}_{e^-e^+ \to \psi},\, m_\psi \}$ are shown for $R_0 = 0.0015\,\text{MeV}^{-1}$ and $R_0 = 0.01\,\text{MeV}^{-1}$, respectively, where the valid ranges of ${\cal N}_{e^-e^+ \to \psi}$ and $m_\psi$ are fixed as

\begin{itemize}
\item
for $\sqrt{\sigma_D} = 1\,\text{MeV}^{-1}$ and $R_0 = 0.0015\,\text{MeV}^{-1}$ [Left panel of Fig.~\ref{fig:psi-peak}]:
\al{
\left.m_\psi^\tx{WP}\right|_\tx{best-fit}
	&=
		 3770.4 \, \tx{MeV},&
\left.{\cal N}_{e^-e^+ \to \psi}^\tx{WP}\right|_\tx{best-fit}
	&=
		1.02 \times 10^5 \,\text{MeV}^{-1},&
		\notag \\
\left.m_\psi^\tx{PW-Parton}\right|_\tx{best-fit}
	&=
		 3774.6 \, \tx{MeV},&
\left.{\cal N}_{e^-e^+ \to \psi}^\tx{PW-Parton}\right|_\tx{best-fit}
	&=
		3.90 \times 10^2 \,\text{MeV}^{-1},&
		\notag \\
\left.m_\psi^\tx{PW-FF}\right|_\tx{best-fit}
	&=
		 3775.5 \, \tx{MeV},&
\left.{\cal N}_{e^-e^+ \to \psi}^\tx{PW-FF}\right|_\tx{best-fit}
	&=
		5.77 \times 10^2 \,\text{MeV}^{-1},&
	\label{eq:D-data-1}
}
\vspace{-6mm}
\al{
\left.\frac{\chi^2_{\phi,\, \tx{WP}}}{(\tx{DOFs})}\right|_\tx{best-fit}
	&\simeq
		0.92,&
\left.\frac{\chi^2_{\phi,\, \tx{PW-Parton}}}{(\tx{DOFs})}\right|_\tx{best-fit}
	&\simeq
		0.91,&
\left.\frac{\chi^2_{\phi,\, \tx{PW-FF}}}{(\tx{DOFs})}\right|_\tx{best-fit}
	&\simeq
		0.91.&
	\label{eq:D-data-2}
}
\item
for $\sqrt{\sigma_D} = 1\,\text{MeV}^{-1}$ and $R_0 = 0.01\,\text{MeV}^{-1}$ [Right panel of Fig.~\ref{fig:psi-peak}]:
\al{
\left.m_\psi^\tx{WP}\right|_\tx{best-fit}
	&=
		 3775.7 \, \tx{MeV},&
\left.{\cal N}_{e^-e^+ \to \psi}^\tx{WP}\right|_\tx{best-fit}
	&=
		4.92 \times 10^6 \,\text{MeV}^{-1},&
		\notag \\
\left.m_\psi^\tx{PW-Parton}\right|_\tx{best-fit}
	&=
		 3774.6 \, \tx{MeV},&
\left.{\cal N}_{e^-e^+ \to \psi}^\tx{PW-Parton}\right|_\tx{best-fit}
	&=
		3.90 \times 10^2 \,\text{MeV}^{-1},&
		\notag \\
\left.m_\psi^\tx{PW-FF}\right|_\tx{best-fit}
	&=
		 3780.0 \, \tx{MeV},&
\left.{\cal N}_{e^-e^+ \to \psi}^\tx{PW-FF}\right|_\tx{best-fit}
	&=
		3.36 \times 10^4 \,\text{MeV}^{-1},&
	\label{eq:D-data-3}
}
\vspace{-6mm}
\al{
\left.\frac{\chi^2_{\phi,\, \tx{WP}}}{(\tx{DOFs})}\right|_\tx{best-fit}
	&\simeq
		0.92,&
\left.\frac{\chi^2_{\phi,\, \tx{PW-Parton}}}{(\tx{DOFs})}\right|_\tx{best-fit}
	&\simeq
		0.91,&
\left.\frac{\chi^2_{\phi,\, \tx{PW-FF}}}{(\tx{DOFs})}\right|_\tx{best-fit}
	&\simeq
		0.92.&
	\label{eq:D-data-4}
}
\end{itemize}
From Fig.~\ref{fig:psi-peak}, when $R_0$ is large as $\sim 10^{-2}\,\tx{MeV}^{-1}$, the resonant distribution of the wave packet becomes identical with that of the plane-wave without taking into account the form factor. Meanwhile, when $R_0 \sim 10^{-3}\,\tx{MeV}^{-1}$, where this size is favored with the agreement in $R_\psi$, we observe the deviation from the BW shape in the wave-packet distribution.
Note that all three kinds of distributions agree with the experimental data for the larger and smaller $R_0$.

We comment on the large asymmetry observed in the right panel of Fig.~\ref{fig:psi-peak}, namely the large deviation of ``BW with FF'' in the low $M$ range.
As mentioned in the previous subsection, the asymmetry under the reflection around the peak originates from
the parts $R_0^2 \paren{M^2 - 4 m_{D^+}^2}$ and $R_0^2 \paren{M^2 - 4 m_{D^0}^2}$ of the form factors.
The realized asymmetry in $R_0 = 0.01 \, \tx{MeV}^{-1}$ becomes extensive when $M$ is less than the range used for the statistical fit,
so this case is considered to be disfavored even though the limited part near the resonant peak is fitted to the experimental results well.

\begin{figure}[t]
\centering
\includegraphics[width=.5\textwidth]{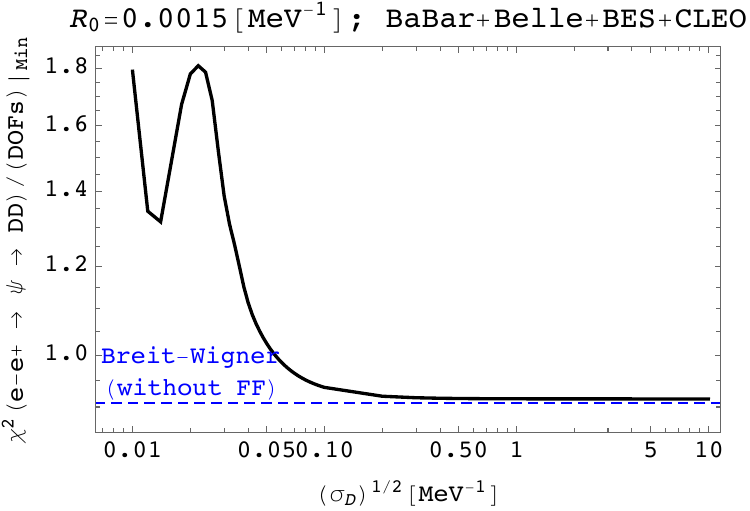}
\caption{
We plot the variable $\left.{\chi^2_{\psi}}/{(\tx{DOFs})}\right|_\tx{min}$ defined in Eq.~\eqref{eq:chisq-ov-DOFs_psi} to compare the significance of the wave-packet calculation with the plane-wave one for various $\sqrt{\sigma_D}$.
Here, $R_0$ is fixed as $0.0015 \, \tx{MeV}^{-1}$ and for each $\sqrt{\sigma_D}$, $m_\psi$ and ${\cal N}_{e^-e^+ \to \psi}$ are determined to (locally) minimize the corresponding $\chi^2$ function in Eq.~\eqref{eq:chisq-form-psi}.
The black curve and the blue dashed horizontal line describe the values in the wave-packet and plane-wave calculations, respectively, where the latter is manifestly independent of $\sqrt{\sigma_D}$.
}
\label{fig:chieqdist-psi}
\end{figure}

To clarify the experimentally-valid range for $\sqrt{\sigma_D}$, we see the curve of the ``minimized'' $\chi^2/(\tx{DOFs})$ defined by
\al{
\left.\frac{\chi^2_{\psi,\, \tx{WP/PW-Parton}}}{(\tx{DOFs})}\right|_\tx{min}
	:=
		\min_{m_\psi,\ {\cal N}_{e^-e^+ \to \psi}^\tx{WP/PW-Parton}}
		\sqbr{
		\frac{\chi^2_{\psi,\, \tx{WP/PW-Parton}}}{(\tx{DOFs})}
		},
	\label{eq:chisq-ov-DOFs_psi}
}
for $R_0 = 0.0015\,\text{MeV}^{-1}$.
Due to the same reason with the case of $\phi$ for $R_0 = 0.0015\,\text{MeV}^{-1}$, we skip to consider the PW-FF case.
From Fig.~\ref{fig:chieqdist-psi}, we recognize that the range for $\sqrt{\sigma_D}$ is consistent with the constraint,
\als{
\left.\frac{\chi^2_{\psi,\, \tx{WP/PW-Parton}}}{(\tx{DOFs})}\right|_\tx{min} < 2,
}
even though the wave-packet shape does not exceed the BW shape in the goodness of fit.
To summarize, within the current scheme for the production cross section, no significant bounds on $\sqrt{\sigma_D}$ are imposed.
This is because, as recognized from Fig.~\ref{fig:psi-peak}, the experimental results still have sizable errors for the $\psi$'s resonant shape.

\section{Summary and discussion\label{sec:summary}}

In this paper, we have discussed the long-standing anomaly in the ratio of the decay rates of the vector mesons $\phi$ and $\psi$, namely,
$R_\phi = \Gamma\fn{\phi \to K^+ K^-}/\Gamma\fn{\phi \to K^0_\tx{L} K^0_\tx{S}}$ and
$R_\psi = \Gamma\fn{\psi \to D^+ D^-}/\Gamma\Fn{\psi \to D^0 \ol{D^0}}$, where the strong interaction causes the decay channels, and they measure isospin breakings.
If we estimate their theoretical values in the plane-wave formalism without considering the effects originating from the composite nature of the initial-state vector mesons, they are disfavored with the PDG's central values
at the level of $2.1\,\sigma$ and $9.5\,\sigma$.
In particular, there has been no explanation for the latter 9.5\,$\sigma$ anomaly so far.

The decay channels that we focus on are near the mass thresholds, where the velocities in the final state are small, and hence the localization of the overlap of the wave packets is more significant.
Here, we fully take into account such effects in the Gaussian-wave-packet formalism.
We carefully clarified the properties of one-to-two-body non-relativistic quantum transitions between normalizable physical states described by Gaussian wave packets under the presence of the decaying nature of the initial state, which is a full-fledged calculation taking into account the essences that are missing in the plane-wave calculations.

The result shows agreement with the PDG's combined results within $\sim1\,\sigma$ confidence level.
We conclude that the long-standing anomalies in $R_\phi$ and $R_\psi$ are resolved.

In the calculation, the above-mentioned compositeness has been described by the form factor. The agreement is achieved when we appropriately take the form-factor parameter at around the physically reasonable value $R_0\sim\pn{500\,\tx{MeV}}^{-1}$.

We also analyzed and made a comment on the $b\ol{b}$-vector-meson counterpart $\Upsilon$, namely 
$R_\Upsilon = \Gamma\fn{\Upsilon \to B^+ B^-}/\Gamma\Fn{\Upsilon \to B^0 \ol{B^0}}$, where the plane-wave calculation without considering the above-mentioned composite nature already agrees at the $0.32\,\sigma$ level with the corresponding PDG result due to the smallness of the mass difference between $B^\pm$ and $B^0$.
The wave-packet result agrees well with the PDG result around the same value of $R_0$.

We mention that the same form factors can be formally multiplied on the ratio of the plane-wave decay rates in order to partially take into account the wave-packet effects, though the wave-packet approach is more comprehensive in describing quantum transitions.
By doing so, around the same value of $R_0$, the plane-wave results can also be made to agree with the PDG ones.

In general, the shape of a wave-packet resonance deviates from the Briet-Wigner shape, where the magnitude of the deviation depends on the size of wave packets. For $\phi$ and $\psi$, experimental data is available, and we put constraints on the size.
We found that when the size of the wave packets is small, the derivation from the Briet-Wigner shape tends to be sizable.
Both for $\phi$ and $\psi$, wide ranges of the wave-packet size are consistent with the experimental data.

In the decay channels of the vector mesons, the non-relativistic approximation works fine, which simplifies the integrations in the $S$-matrices and the final-state phase spaces in the wave-packet formalism.
Many other quantum transitions in high-energy physics are relativistic, and it is worthwhile to establish the general method to perform such integrations without relying on the non-relativistic approximation.
Also, analyzing resonant productions precisely requires the full transition probabilities, including production parts.
Doing more dedicated analyses on resonant shapes will be another important task.

\subsection*{Acknowledgments}
K.I.\ thanks for useful communications with Dr.~Elliott~D.~Bloom, Dr.~Walter~Toki, and Dr.~Tomasz~Skwarnicki on the crystal-ball functions.
K.N.\ thanks Dr.~Akimasa~Ishikawa for the useful information on experimental details on $R_\Upsilon$ and Dr.~Ryosuke~Sato for useful comments on form factors.
This work is partially supported by the JSPS KAKENHI Grant Nos.~JP21H01107 (K.I., O.J., K.N., K.O.) and JP19H01899 (K.O.).

\appendix 
\section*{Appendix}

\section{Vector-meson form factor}\label{form factor section}
\label{form factor session}

We focus on the following form for a non-relativistic bound state $V\fn{Q\ol Q}$ composed of a heavy constituent quark $Q$ and its anti-particle $\overline{Q}$ at a certain time:
\al{
\ket{V(Q\overline{Q}), \bs P} :=
	{\sum_{s_1,s_2}}
	\int \df^3\bs x_1\, \df^3\bs x_2\, e^{i \bs P \cdot \frac{\bs x_1 + \bs x_2}{2}}
	F_{s_1,s_2}\fn{\bs x_1 - \bs x_2}
	Q^\dagger\fn{\bs x_1, s_1} \overline{Q}^\dagger\fn{\bs x_2, s_2} \ket{0},
}
where $F_{s_1,s_2}\fn{\bs x_1 - \bs x_2}$ is a wave function for the bound state;
$Q^\dagger\fn{\bs x_1, s_1}$ and $\overline{Q}^\dagger\fn{\bs x_2, s_2}$ are the Fourier transforms of the momentum-space creation operators of $Q$ and $\ol Q$, respectively, with $\bs x_{1,2}$ and $s_{1,2}$ being their positions and spins;
and $\ket{0}$ is the vacuum; see e.g.\ Ref.~\cite{Eichten:1978tg}.

We consider the following matrix element
\al{
\Braket{Q\paren{\bs p_1, s_1} \overline{Q}\paren{\bs p_2, s_2} | V(Q\overline{Q}), \bs P}
	&=
		\int \df^3\bs x_1\, \df^3\bs x_2\, e^{i \bs P \cdot \frac{\bs x_1 + \bs x_2}{2}}
		F_{s_1,s_2}\fn{\bs x_1 - \bs x_2}
		\paren{\frac{1}{\sqrt{2\pi}}}^6 e^{- i \bs p_1 \cdot \bs x_1 -i \bs x_2 \cdot \bs p_2} \notag \\
	&= 
		\int \df^3\bs X\, \df^3\bs r\,e^{i \paren{\bs P - \bs p_1 -\bs p_2} \cdot \bs X -i \paren{\frac{\bs p_1 - \bs p_2}{2}} \cdot \bs r}
		\frac{1}{\paren{2\pi}^3}
		F_{s_1,s_2}\fn{\bs r} \notag \\
	&=
		\delta^{3}\fn{\bs P - \bs p_1 - \bs p_2} \underbrace{\int \df^3\bs r\,
		e^{ -i \paren{\frac{\bs p_1 - \bs p_2}{2}} \cdot \bs r}
		F_{s_1,s_2}\paren{\bs r}}_{=: \widetilde{F}_{s_1,s_2} \fn{\frac{\bs p_1 - \bs p_2}{2}} },
}
where $\bs X := (\bs x_1 + \bs x_2)/2$, $\bs r := \bs x_1 - \bs x_2$, and the normalization of the form factor $\widetilde{F}_{s_1,s_2} \fn{\frac{\bs p_1 - \bs p_2}{2}}$ is irrelevant for our purpose.\footnote{
What is relevant is only the product of the normalization of the effective coupling and that of the form factor. Such a normalization factor will be dropped out of the final ratio of the decay probabilities under the isospin-symmetric limit~\eqref{isospin-symmetric limit}.
}

Hereafter, we assume the separable form
\al{
\widetilde{F}_{s_1,s_2} \fn{\frac{\bs p_1 - \bs p_2}{2}}
	=
		S_{s_1,s_2} \,\wt{F} \fn{\frac{\bs p_1 - \bs p_2}{2}}
}
for the heavy and non-relativistic quarks $Q\ol Q$, which have negligible spin-orbital angular momentum interaction.
Further, we drop the spin structure $S_{s_1,s_2}$, which will be canceled out in the ratio of the neutral to charged rates, and focus on the momentum part.\footnote{
Concretely, the orbital and total angular momenta $\ell_J$ is $S_1$ ($\ell=1$), $D_1$ ($\ell=2$), and $S_1$ for $\phi\fn{1020}$, $\psi\fn{3770}$ and $\Upsilon\fn{4S}$, respectively; see e.g.\ ``Quark Model'' section in Ref.~\cite{ParticleDataGroup:2022pth}.
}

We adopt an approximate form of the wave function of the ($s$-wave) ground state under a Coulomb potential in the position space:
\al{
F\fn{\bs r}
	=
		\frac{N}{\sqrt{2\pi R_0}} \frac{e^{- \frac{r}{R_0}}}{r},
		\label{eq:aprrox-F(r)}
}
where $r:=\ab{\bs r}$, the parameter $R_0$ describes a typical length scale of the bound state discussed below, and $N$ is the irrelevant normalization factor mentioned above.
Its Fourier-transform is
\al{
\widetilde{F}\fn{\bs p} 
	&=
		N\int{\df^3\bs r\ov\pn{2\pi}^{3/2}}\,e^{-i \bs p \cdot \bs r} F\fn{\bs r}
	=	{N\ov\pi\sqrt{R_0}}{1\ov\bs p^2+{1\ov R_0^2}},
}
where we used $\int_0^\infty \df r \sin\fn{pr} e^{- \frac{r}{R_0}} = \frac{p}{p^2 + \frac{1}{R_0^2}}$.
In this paper, we choose $N= \pi/R_0^{3/2}$ such that $\wt F\fn{\bs 0}=1$:
\al{
\wt F\fn{\bs p}
	&=	{1\ov R_0^2}{1\ov\bs p^2+{1\ov R_0^2}}.
}
This form is also introduced in Ref.~\cite{Fischbach} to cut off a UV divergence in the plane-wave computation.
The treatment in Refs.~\cite{FloresBaez:2008jp,Kuksa:2009zz} is equivalent to taking this form factor to be unity.

Here, a vector meson $V\fn{Q\ol Q}$ decays into two light pseudo-scalar mesons $P\fn{Q\ol q}$ and $\ol P\fn{q\ol Q}$.
We approximate the mass and momentum for each psuedoscalar $P$ ($\ol P$) by those of the constituent quark $Q$ ($\ol Q$): $m_P\simeq m_Q$ and $\bs p_P\simeq \bs p_1$ ($\bs p_{\ol P}\simeq\bs p_2$), respectively.
In this paper, we focus on the situation where the masses of the two pseudo-scalar mesons are almost the same (due to the approximated flavor-isospin $SU(2)$ symmetry), and the mass relation is near the decay threshold,
\al{
m_V \approx 2 \, m_P.
	\label{eq:NRLimit-1}
}
Therefore, we can treat the process as a non-relativistic one, and thus, we conclude that
\al{
\bs p_1 - \bs p_2 \approx
m_P \paren{\bs V_1 - \bs V_2},
	\label{eq:NRLimit-2}
}
thereby,
\al{
\widetilde{F} \fn{\frac{\bs p_1 - \bs p_2}{2}}
	&=
		\frac{1}{\paren{\frac{R_0 \paren{\bs p_1 - \bs p_2}}{2}}^2 + 1 }
		\label{relativistic form factor}\\
	&\Rightarrow
		\frac{1}{\paren{\frac{R_0 m_P \paren{\bs V_1 - \bs V_2}}{2}}^2 + 1 },
}
where $\bs V_1$ and $\bs V_2$ are the (non-relativistic) velocities of $P$ and $\ol P$, respectively.

Finally, we reach the spin-independent dimensionless function suitable for our purpose,
\al{
\widetilde{F}\fn{\ab{\bs V_1 - \bs V_2}} 
	:=
		\frac{1}{\paren{\frac{R_0 m_P \paren{\bs V_1 - \bs V_2}}{2}}^2 + 1 }.
		\label{eq:form-factor-final-form}
}
This is the form factor shown in Eq.~\eqref{eq:form-factor-for-calc} for the matrix elements of the meson decays (with $m_V \approx 2 \, m_P$) in the rest system.

Now we estimate a typical value of the parameter $R_0$ in Eq.~\eqref{eq:aprrox-F(r)}.
The quarkonium potential can be approximated by a sum of the confining linear potential and the QCD Coulomb potential
\al{
V(r) = \frac{r}{a_s^2} - \frac{\alpha_s}{r},
	\label{Cornell potential}
}
where $a_s$ is known as $a_s = 1.95\,\text{GeV}^{-1}$~\cite{Eichten:1978tg} and $\alpha_s$ is the QCD fine structure constant.
For the domain where $r\lesssim r_c := a_s \sqrt{\alpha_s}\simeq1.5\,\tx{GeV}^{-1}$,\footnote{
Here, we have put $\alpha_s\simeq0.6$ at the scale $0.5$\,GeV~\cite{Deur:2016tte}.
}
the wave function can be approximated by the Coulomb form~\eqref{eq:aprrox-F(r)}.
One can estimate $r$ by equating the potential and kinetic energies, $V\fn{r}\sim K_Q$, where $K_Q\sim m_V-2m_Q={\cal O}(10)\,\text{MeV}$.
Since $K_Q$ is much smaller than the typical energy scale $a_s^{-1}=0.5$\,GeV of the potential, the typical $Q\ol Q$ distance can be estimated by equating two terms in the right-hand side of Eq.~\eqref{Cornell potential}: $r\sim r_c$.
The use of Coulomb wave function~\eqref{eq:aprrox-F(r)} is marginally justified, which suffices for our current consideration. 
See e.g.\ Ref.~\cite{Brambilla:2010cs} for further refinement.

Finally, a typical value of the parameter in Eq.~\eqref{eq:aprrox-F(r)} is
\al{
R_0\sim r_c\simeq 1.5\,\tx{GeV}^{-1}=0.0015\,\tx{MeV}^{-1}={1\ov660\,\tx{MeV}}.
}

\section{Details on calculations of $P_{V \to P \ol{P}}$}
\label{detailed decay probability section}

We present detailed computation to obtain the decay probability integrated over the final-state positions and momenta under the rest-frame assumption $\bs P_0=0$.

\subsection{Variables under non-relativistic limit}\label{variables in appendix}

As preparation, we show concrete forms of the kinetic variables and parameters, whose physical meaning is given in Section~\ref{sec:$S$-matrix}, under the non-relativistic limit $\ab{\bs V_{1,2}} \ll 1$.
At first, we define the `light-cone' variables for later convenience:
\al{
\bs P_\pm &:= \bs P_1 \pm \bs P_2, & \bs V_\pm &:= \bs V_1 \pm \bs V_2.
	\label{plus minus notations defined}
}
The kinetic variables are
\al{
\bs P_0 = 0 &\Rightarrow \bs V_0 = 0, &
\bs P_1 &\Rightarrow m_P \bs V_1, & 
\bs P_2 &\Rightarrow m_P \bs V_2, \notag \\
E_0 = m_V&\Rightarrow m_V,&
E_1 &\Rightarrow m_P + \frac{m_P}{2} \bs V_1^2, &
E_2 &\Rightarrow m_P + \frac{m_P}{2} \bs V_2^2, \notag
}
\al{
\overline{|g_\tx{eff}|^2}
	=
		{\frac{g^2}{3}} (\bs P_1 - \bs P_2)^2 &\Rightarrow
		{\frac{g^2}{3}} \paren{m_P^2 \bs V_-^2},
}
where the symbol $\Rightarrow$ represents the non-relativistic approximation, which we will take later.
We also have
\al{
\overline{\bs V} &= \sigma_s \paren{ \frac{\bs V_1}{\sigma_P} + \frac{\bs V_2}{\sigma_P}
}
= \frac{\sigma_s}{\sigma_P} \bs V_+, \\
\overline{\bs V^2} &= \sigma_s \paren{ \frac{\bs V_1^2}{\sigma_P} + \frac{\bs V_2^2}{\sigma_P}
}
= \frac{\sigma_s}{2\sigma_P} \paren{ \bs V_+^2 + \bs V_-^2 }, \\
\delta \bs P &=
\bs P_1 + \bs P_2 \Rightarrow  m_P \bs V_+ , \\
\delta E &= - m_V + E_1 + E_2 \Rightarrow - \paren{ m_V - 2m_P} + \frac{1}{2} m_P \paren{ \bs V_1^2 + \bs V_2^2 } \notag \\
&\phantom{= - E_0 + E_1 + E_2}\
= - \paren{m_V - 2m_P} + \frac{1}{4} m_P \paren{ \bs V_+^2 + \bs V_-^2 }, \\
\delta \omega &= \delta E - \overline{\bs V} \cdot \delta \bs P \Rightarrow \delta E - m_P \frac{\sigma_s}{\sigma_P} \bs V_+^2, \\
\sigma_s &= \frac{\sigma_V \sigma_P}{2\sigma_V + \sigma_P},   \\
\sigma_t &=
\frac{\sigma_s}{\overline{\bs V^2} - \paren{\overline{\bs V}}^2} =
\frac{\sigma_P}{ \frac{1}{2} \paren{ \bs V_+^2 + \bs V_-^2} - \frac{\sigma_s}{\sigma_P} \bs V_+^2 }.
}
Also, we define the variables $V_+$ and $V_-$
\al{
V_+ &:= \ab{\bs V_+}, &
V_- &:= \ab{\bs V_-}.
}

\subsection{Bulk contribution}

We compute the bulk contribution in Eq.~\eqref{dP given}.
First, we perform the position integrals $\int_{-\infty}^{+\infty} \df^3 \bs X_1 \df^3 \bs X_2$. 
As in Ref.~\cite{Ishikawa-Oda}, we obtain
\al{
\df^3 \bs X_1 \df^3 \bs X_2
	&=
		\df^5 y\, \df y_0, \\
\int \df^5 y\, e^{ - \mc R}
	&=
		\sqrt{ \frac{\pi^5}{\sigma_t} \paren{ \frac{\sigma_0\sigma_1\sigma_2}{\sigma_s} }^3 }
		\frac{1}{ \sqrt{  \paren{\delta \bs V_1}^2 + \paren{\delta \bs V_2}^2  } },  \\
{\mf T - {\Gamma_V\sigma_t\ov2}}
	&=
		- \frac{y_0}{ \sqrt{\paren{\delta \bs V_1}^2 + \paren{\delta \bs V_2}^2} } + \cdots, 
		\label{eq:Tmathflak_definition} \\
\df \mf T
	&=
		- \frac{\df y_0}{ \sqrt{\paren{\delta \bs V_1}^2 + \paren{\delta \bs V_2}^2} },
}
where $y_0$ becomes a flat direction under the (unphysical) no-decay limit $\Gamma_V \to 0$ (as considered in~\cite{Ishikawa-Oda}),
while the other five directions are not flat directions irrespective of $\Gamma_V$:
\al{
\int \df^3 \bs X_1 \df^3 \bs X_2 e^{ - \mc R -\Gamma_V\pn{\mf T-T_0} + \frac{\Gamma_V^2\sigma_t}{4} } \Pn{W(\mf T)}^2
	&=
		\sqrt{ \frac{\pi^5}{\sigma_t} \paren{ \frac{\sigma_0\sigma_1\sigma_2}{\sigma_s} }^3 } \,
		\int_{ {\Tin + {\Gamma_V\sigma_t\ov2}} }^{ {\Tout + {\Gamma_V\sigma_t\ov2}  }  } \df \mf T\,
		e^{ -\Gamma_V\pn{\mf T-T_0} + \frac{\Gamma_V^2\sigma_t}{4} }   \notag \\
	&=
		\sqrt{ \frac{\pi^5}{\sigma_t} \paren{ \frac{\sigma_0\sigma_1\sigma_2}{\sigma_s} }^3 } \,
		\Gamma_V^{-1} e^{  -\Gamma_V\pn{\Tin-{T_0}} {- \frac{\Gamma_V^2\sigma_t}{4}}  },
\label{eq:Integration_bulk}
}
where we also used Eq.~\eqref{eq:bulk-window-function} and took the limit $\Tout \to \infty$.
The range of the integration is given by the bulk window function $W(\mf T)$ in Eq.~(\ref{eq:bulk-window-function}).
After the position integrations, we obtain
\al{
\df P^\text{bulk}_{V \to P \ol{P}}	
	&=
		\ab{g_\tx{eff}}^2 N_V^2 \Gamma_V^{-1}  e^{  -\Gamma_V\pn{\Tin-{T_0}} {- \frac{\Gamma_V^2\sigma_t}{4}}  } \frac{1}{2 E_0}
		\frac{\df^3 \bs P_1}{(2\pi)^3 2E_1} \frac{\df^3 \bs P_2}{(2\pi)^3 2E_2} \notag \\
	&\quad
		\times
		\pn{2 \pi}^4
		\paren{ \sqrt{\frac{\sigma_t}{\pi}} \paren{\frac{\sigma_s}{\pi}}^{3/2} 
		e^{-\sigma_t\paren{\delta \omega}^2  -\sigma_s\paren{\delta \bs P}^2  }
		 } 
		 \ab{ \widetilde{F}\fn{V_-} }^2,
		 \label{dP in appendix}
}
where the damping factors $\sqrt{\sigma_t\ov\pi} \, e^{-\sigma_t\paren{\delta \omega}^2}$ and 
$\paren{\sigma_s\ov\pi}^{3/2} e^{-\sigma_s\paren{\delta \bs P}^2  }$ provide the approximate conservation for the mean energy and momentum, respectively.\footnote{
Of course the energy-momentum conservation is fulfilled in itself as a fundamental physical law of nature, in particular, for each partial plane-wave component in the Fourier transforms.
}
So far the expression~\eqref{dP in appendix} does not assume $\bs P_0=0$ nor the non-relativistic approximation given in Sec.~\ref{variables in appendix}.

Next, we perform the momentum integrals under the saddle-point approximation with $\bs P_0=0$ in the non-relativistic approximation given in Sec.~\ref{variables in appendix}:
\al{
P^\text{bulk}_{V \to P \ol{P}}	
	&\Rightarrow
		\pn{ N_V^2 \Gamma_V^{-1}  e^{  -\Gamma_V\pn{\Tin-{T_0}} } }
		\frac{g^2}{3} \frac{1}{2 m_V} \frac{1}{(2\pi)^2 4 E_1 E_2} 
		\frac{m_P^6}{8} \paren{ \int_{0}^{\infty} 4 \pi V_+^2 \df V_+ } 
		\paren{ \int_0^\infty 4\pi V_-^2  \df V_- } \notag \\
	&\quad
		\times \frac{1}{\pi^2} \sigma_t^{1/2}
		\sigma_s^{3/2} \paren{m_P^2 V_-^2}
		e^{ - F_\text{bulk}(V_+, V_-) }
		 \ab{ \widetilde{F}\fn{V_-} }^2,
}
where the factor $1/8 = 1/2^3$ is from the Jacobian and the exponent is
\al{
F_\text{bulk}(V_+, V_-) 
	&:=
		\sigma_s\paren{\delta \bs P}^2 + \sigma_t\paren{\delta \omega}^2 { + \frac{\Gamma_V^2\sigma_t}{4}}\nn
	&=	\sigma_s m_P^2 V_+^2 \notag \\
	&\quad
		+
		\frac{2 \sigma_P}{ \paren{ 1 - \frac{2\sigma_s}{\sigma_P} } V_+^2 +  V_-^2 }
		\br{
		\sqbr{
		m_P \paren{ \frac{1}{4} - \frac{\sigma_s}{\sigma_P} } V_+^2 + \frac{1}{4} m_P V_-^2 - \paren{m_V - 2m_P}
		}^2 + { {\Gamma_V^2\ov4} } }.
}

We see that
\al{
1 - \frac{2\sigma_s}{\sigma_P}
	=
		\frac{1}{ 2 \sigma_V/\sigma_P + 1}
	>	0,
}
which implies
\al{
\sigma_P - {2\sigma_s} > 0.
	\label{eq:sigma-inequality}
}
Also, we see that
\al{
\sigma_P - 4\sigma_s
=
\frac{ \sigma_P \paren{ \sigma_P - 2\sigma_s } }{  \sigma_P + 2\sigma_s } > 0,
}
which leads to
\al{
\frac{1}{4} - \frac{\sigma_s}{\sigma_P}
=
\frac{1}{4\sigma_P} \paren{ \sigma_P - 4 \sigma_s } > 0.
}
Thereby,
\al{
e^{ - F_\text{bulk} } &\underbrace{\to}_{V_+ \to \infty} 0, &
e^{ - F_\text{bulk} } &\underbrace{\to}_{V_- \to \infty} 0.
}

The stationary point $(V_+^s, V_-^s)$ that satisfies
\al{
\frac{\pal F_\text{bulk} (V_+^s, V_-^s)}{\pal V_+} = 0 \quad \text{and} \quad
\frac{\pal F_\text{bulk} (V_+^s, V_-^s)}{\pal V_-} = 0
}
is found to be
\al{
\paren{V_+^s, V_-^s}
	&\in
		\left\{\paren{0, \, 
			\frac{ 2\pn{\pn{m_V-2m_P}^2+{\Gamma_V^2\ov4}}^{1/4}  }{\sqrt{m_P}}
			}\right. , 
	&
		&\paren{0, \, 
			\frac{ -2\pn{\pn{m_V-2m_P}^2+{\Gamma_V^2\ov4}}^{1/4}  }{\sqrt{m_P}}
			}, \notag \\
	&\qquad
		\paren{0, \, 
			\frac{ 2i\pn{\pn{m_V-2m_P}^2+{\Gamma_V^2\ov4}}^{1/4}  }{\sqrt{m_P}}
			}, 
	&
		&\paren{0, \, 
			\frac{ -2i\pn{\pn{m_V-2m_P}^2+{\Gamma_V^2\ov4}}^{1/4}  }{\sqrt{m_P}}
			}, \notag \\
	&\qquad
		\paren{
			\frac{ 2\pn{\pn{m_V-2m_P}^2+{\Gamma_V^2\ov4}}^{1/4}  }{\sqrt{m_P}}
			,\, 0},
	&
		&\paren{
			\frac{ -2\pn{\pn{m_V-2m_P}^2+{\Gamma_V^2\ov4}}^{1/4}  }{\sqrt{m_P}}
			,\, 0}, \notag \\
	&\qquad
		\paren{
			\frac{ 2i\pn{\pn{m_V-2m_P}^2+{\Gamma_V^2\ov4}}^{1/4}  }{\sqrt{m_P}}
			,\, 0},
	&\qquad
		&\left.\paren{
			\frac{ -2i\pn{\pn{m_V-2m_P}^2+{\Gamma_V^2\ov4}}^{1/4}  }{\sqrt{m_P}}
			,\, 0}\right\}.
}
Now, we focus on the kinetic region ($V_+ \simeq 0$ and $V_- > 0$) and thus only the first one is relevant in our calculation,
\al{
\paren{V_+^\text{B}, V_-^\text{B}}
	&:=
		\paren{0, \, 
			\frac{ 2\pn{\pn{m_V-2m_P}^2+{\Gamma_V^2\ov4}}^{1/4}  }{\sqrt{m_P}}
			=
			2 \sqbr{ \frac{\paren{m_V-2m_P}^2}{m_P^2} + {\Gamma_V^2\ov4m_P^2} }^{1/4}
			}.
}
Note that
\al{
{F_\text{bulk}^0} :=
F_\text{bulk}\paren{V_+^\text{B}, V_-^\text{B}}
	& = 
		m_P \sigma_P
		\paren{ -\paren{m_V - 2m_P}  + \sqrt{ \paren{m_V - 2m_P}^2+{\Gamma_V^2\ov4} }  },
}
which goes to zero in the (unphysical) limit $\Gamma_V\to0$.
Also, we can see
\al{
\left.\frac{\pal^2 F_\text{bulk} }{\pal V_+^2}\right|_{\paren{V_+^\text{B}, V_-^\text{B}}}
	&= 
		2 m_P^2 \sigma_s 
		\sqbr{
		\half + \frac{m_V - 2m_P}{2\sqrt{\paren{m_V-2m_P}^2+{\Gamma_V^2\ov4}}}
		}=: 2 m_P^2 \sigma_s {A_\text{bulk}}\quad (>0) ,
		\notag \\
\left.\frac{\pal^2 F_\text{bulk} }{\pal V_-^2}\right|_{\paren{V_+^\text{B}, V_-^\text{B}}}
	   &= 
	   	m_P^2 \sigma_P \quad(>0),  \notag \\
\left.\frac{\pal^2 F_\text{bulk} }{\pal V_+ \pal V_-}\right|_{\paren{V_+^\text{B}, V_-^\text{B}}}
	   &=
	   	0,
}
where $A_\text{bulk} \to 1$ under the (unphysical) limit $\Gamma_V \to 0$.

We perform the integrals under the saddle-point approximation (putting $E_1=E_2=m_P$ in the overall factor) as follows:
\al{
P^\text{bulk}_{V \to P\ol{P}}
	&\simeq
		\frac{g_V^2}{3} \frac{1}{2 m_V} \frac{1}{(2\pi)^2 4m_P^2} \,
		\frac{m_P^6}{8} \paren{ \int_{0}^{\infty} 4 \pi V_+^2 \df V_+ } 
		\paren{ \int_0^\infty 4\pi \paren{V_-^\text{B}}^2  \df V_- } \notag \\
	&\quad
		\times \frac{1}{\pi^2} \sqbr{\sigma_t^{1/2}}_{V_+ \to V_+^\text{B}, \, V_- \to {V_-^\text{B}}}\,
		\sigma_s^{3/2} \sqbr{m_P^2 \paren{V_-^\text{B}}^2}  \notag \\
	&\quad
		\times
		{e^{-F^0_\text{bulk}}}
		e^{  - \frac{1}{2} \paren{2 m_P^2 \sigma_s} {A_\text{bulk}} 
			\paren{ V_+ - V_+^\text{B} }^2  }
		e^{  - \frac{1}{2} \paren{m_P^2 \sigma_P} \paren{ V_-  -  V_-^\text{B} }^2  }
		\pn{ {N_V^2\ov\Gamma_V}  e^{  -\Gamma_V\pn{\Tin-{T_0}} } }
		 \ab{ \widetilde{F}\fn{V_-^\text{B}} }^2   \notag \\
	&=
		\frac{g^2 m_P^3  N_V^2  e^{  -\Gamma_V\pn{\Tin-{T_0}} }   }{12 \pi m_V m_P^2}\,
		{m_P\ov\Gamma_V}
		\sqbr{ \frac{\paren{m_V-2m_P}^2}{m_P^2} + {\Gamma_V^2\ov4m_P^2} }^{3/4}
		\notag \\
	&\qquad
		\times\frac{1}{2}
		\sqbr{  1 + \text{erf}\paren{  \frac{  m_P\sqrt{\sigma_P}  {V_-^\text{B}}  }{\sqrt{2}}  }  } \
		{\frac{e^{-F^0_\text{bulk}}}{A_\text{bulk}^{3/2}}}     
		\ab{ \widetilde{F}\fn{V_-^\text{B}} }^2,
		\label{eq:Gamma-phi-to-KKbar-V1_Nonrelativistic}
}
where we have used the formulas for $a > 0$,
\al{
\int_0^\infty 4 \pi \, r^2 e^{ - \frac{a}{2} r^2 } \df r &= \frac{ 2\sqrt{2} \, \pi^{3/2} }{ a^{3/2} },  
	\label{eq:integral_formula_1} \\
\int_0^\infty e^{ - \frac{a}{2} \paren{r - r_0 }^2 } \df r &= \sqrt{\frac{\pi}{2 a}} 
	\sqbr{  1 + \text{erf}\paren{  \frac{\sqrt{a} r_0}{\sqrt{2}}  }  },
	\label{eq:integral_formula_2} 
}
and the Taylor expansion around one of the stationary points $(V_+, V_-) = (V_+^s, V_-^s)$
\al{
F(V_+, V_-)
\simeq
F(V_+^s, V_-^s) 
+ \frac{1}{2} \sum_{i,j = +, -}  \frac{\pal^2 F(V_+^s, V_-^s)}{\pal V_i\,\pal V_j} \paren{V_i - V_i^s} \paren{V_j - V_j^s}.
}

\subsection{Boundary contribution}
We compute the boundary contribution in Eq.~\eqref{dP given}:
\al{
& \int \df^3\bs X_1 \df^3\bs X_2 
	e^{
	-{\cal R} - \Gamma_V\pn{\mf T-\Tin} + \frac{\Gamma_V^2\sigma_t}{4}
	-\frac{ \paren{\mf T - \Tin - {\Gamma_V\sigma_t\ov2}}^2 }{\sigma_t}
	}
		\frac{2\sigma_t}{\pi} 
		\frac{1}{ \paren{ \mf T -\Tin - {\Gamma_V\sigma_t\ov2} }^2 + \paren{ \sigma_t \delta\omega }^2 } \notag \\
&=
	\sqrt{ \frac{\pi^5}{\sigma_t} \paren{ \frac{\sigma_0\sigma_1\sigma_2}{\sigma_s} }^3 }
	e^{- \Gamma_V\pn{\Tin-T_0} }
		\frac{2\sigma_t}{\pi}
		\int_{-\infty}^{\infty} \df \mf T' e^{ - \frac{1}{\sigma_t} \paren{\mf T'}^2}
		\frac{1}{ \paren{ \mf T' - {\Gamma_V\sigma_t\ov2} }^2 + \paren{ \sigma_t \delta\omega }^2 },
	\label{eq:boundary-Tprime-integral}
}
where $\mf T':=\mf T-\Tin$ and the range of the integration is $(-\infty,+\infty)$ since there exists no window function $W(\mf T)$ for this boundary term other than the exponential factor.

Now we perform the Taylor expansion of $\mf T'$ around the saddle point $\mf T'=0$ up to the second order
\al{
\frac{ e^{ - \frac{1}{\sigma_t} \paren{\mf T'}^2} }{ \paren{ \mf T' - {\Gamma_V\sigma_t\ov2} }^2 + \paren{ \sigma_t \delta\omega }^2 }
	&\simeq
		\left(
			\frac{1}
				{ \paren{{\Gamma_V\sigma_t\ov2}}^2 + \paren{\sigma_t \delta\omega}^2 } +
			\frac{ 2 \paren{{\Gamma_V\sigma_t\ov2}} \mf T' }
				{ \pn{\paren{{\Gamma_V\sigma_t\ov2}}^2 + \paren{\sigma_t \delta\omega}^2}^2 }\right. \notag \\
	&\quad+
		\left.
			\frac{ \pn{ 3\paren{{\Gamma_V\sigma_t\ov2}}^2 - \paren{\sigma_t \delta\omega}^2 } \paren{\mf T'}^2 }
				{ \pn{\paren{{\Gamma_V\sigma_t\ov2}}^2 + \paren{\sigma_t \delta\omega}^2}^3 }	
		\right)
		e^{ - \frac{1}{\sigma_t} \paren{\mf T'}^2},
}
to obtain
\al{
\int_{-\infty}^{\infty} \df\mf T'
\frac{ e^{ - \frac{1}{\sigma_t} \paren{\mf T'}^2} }{ \paren{ \mf T' - {\Gamma_V\sigma_t\ov2} }^2 + \paren{ \sigma_t \delta\omega }^2 }
	&\simeq
		\frac{\sqrt{ \pi \sigma_t }}{ \sigma_t^2 }
		\left(
			\frac{1}
				{ \paren{{\Gamma_V\ov2}}^2 + \paren{\delta\omega}^2 } +
			\frac{ \pn{ 3\paren{{\Gamma_V\ov2}}^2 - \paren{\delta\omega}^2 } }
				{ \pn{\paren{{\Gamma_V\ov2}}^2 + \paren{\delta\omega}^2}^3 2\sigma_t }
		\right),
}
which yields
\al{
P^\text{bdry}_{V \to P\ol{P}}
	&\simeq
		\pn{ N_V^2  e^{  -\Gamma_V\pn{\Tin-{T_0}} } }
		\frac{g_V^2}{3} \frac{1}{2 m_V} \frac{1}{(2\pi)^2 4 E_1 E_2}
		\frac{m_P^6}{8} \paren{ \int_{0}^{\infty} 4 \pi V_+^2 \df V_+ } 
		\paren{ \int_0^\infty 4\pi \df V_- } \notag \\
	&\quad
		\times \frac{1}{2\pi^{5/2}}
		\sigma_s^{3/2}\,{m_P^2}\,e^{- \sigma_s m_P^2 V_+^2}
		\notag \\
	&\quad
		\times
		\left\{
			\frac{V_-^4}
				{ \paren{{\Gamma_V\ov2}}^2 + \paren{\delta\omega}^2 } +
			\frac{ V_-^4 \sqbr{ 3\paren{{\Gamma_V\ov2}}^2 - \paren{\delta\omega}^2 } }
				{ \sqbr{\paren{{\Gamma_V\ov2}}^2 + \paren{\delta\omega}^2}^3 2\sigma_t }
		\right\}
		 \ab{ \widetilde{F}\fn{V_-} }^2.
}
There is no $V_-$ in the exponent, and we will perform the numerical computation for the $V_-$ integral.
On the other hand, the saddle point of $V_+$ is located at $V_+ = 0$.
With it in mind, we approximate the polynomial part of the integrand by setting $V_+ = 0$ other than $V_+^2$:
\al{
P^\text{bdry}_{V \to P\ol{P}}
	&\simeq
		\pn{ N_V^2  e^{  -\Gamma_V\pn{\Tin-{T_0}} } }
		\frac{g_V^2}{3} \frac{1}{2 m_V} \frac{1}{(2\pi)^2 4 m_P^2}
		\frac{m_P^6}{8} 
		\int_{0}^{\infty} 4 \pi V_+^2 \df V_+
		\int_0^\infty 4\pi  \df V_-	\notag \\
	&\quad
		\times \frac{1}{2\pi^{5/2}}
		\sigma_s^{3/2} m_P^2 e^{- \sigma_s m_P^2 V_+^2}
		\paren{\frac{m_P}{4}}^{-2} 
		\widetilde{f}_\text{bdry}\fn{V_-},
}
where we define a dimensionless function
\al{
\widetilde{f}_\text{bdry}\fn{V_-}
	&:=
		\paren{\frac{m_P}{4}}^{2}
		\left\{
			\frac{V_-^4}
				{ \paren{{\Gamma_V\ov2}}^2 + \paren{\delta\omega}^2 } +
			\frac{ V_-^4 \sqbr{ 3\paren{{\Gamma_V\ov2}}^2 - \paren{\delta\omega}^2 } }
				{ \sqbr{\paren{{\Gamma_V\ov2}}^2 + \paren{\delta\omega}^2}^3 2\sigma_t }
		\right\}_{V_+ \to 0} \ab{ \widetilde{F}\fn{V_-} }^2 \notag \\
	&=
		\left\{
			\frac
			{V_-^4}
			{\paren{V_-^2 - 4\frac{m_V-2m_P}{m_P}}^2 
				+ \frac{4^2}{m_P^2} 
					\paren{ 
							{\Gamma_V^2\ov4} 
						}
			}\right. \notag \\
	&\qquad
			-
			\left.
			\frac
			{  V_-^4 \paren{\sqbr{V_-^2 - 4\frac{m_V-2m_P}{m_P}}^2 -3 \frac{4\Gamma_V^2}{m_P^2}} }
			{
				2\paren{\sigma_t|_{V_+\to 0}}
				\paren{ \frac{m_P}{4} }^2
				\sqbr{ \paren{V_-^2 - 4\frac{m_V-2m_P}{m_P}}^2 + \frac{4\Gamma_V^2}{m_P^2} }^3
			}
		\right\} 
			\ab{ \widetilde{F}\fn{V_-} }^2,
}
with
\al{
\sigma_t|_{V_+\to 0}
	:=
		\frac{2 \sigma_P}{V_-^2}.	
}
Now we execute the $V_+$ integral using the formula~\eqref{eq:integral_formula_1}:
\al{
P^\text{bdry}_{V \to P\ol{P}}
	&=
		{\frac{g^2 m_P^3  N_V^2  e^{  -\Gamma_V\pn{\Tin-{T_0}} }   }{12 \pi m_V m_P^2}}
		\frac{I_\text{bdry}}{2\pi},
	\label{eq:P_boundary}
}
where the integral
\al{
I_\text{bdry}
	&:=
		\int_0^\infty \df V_- \widetilde{f}_\text{bdry}\fn{V_-}
}
is convergent.

\subsection{Interference contribution}
We compute the interference contribution in Eq.~\eqref{dP given}.
We focus on the part including the factor 
\als{
\frac{ e^{ +i \delta \omega \paren{ \mf T -\Tin - {\Gamma_V\sigma_t\ov2} } } }
{ \mf T -\Tin - {\Gamma_V\sigma_t\ov2} - i \sigma_t \delta\omega },
}
since the other part can be obtained by taking complex conjugation.
At first, we perform the square completion of the $\mf T$ part:
\al{
I_\tx{intf} := \
&
	\int \df^3\bs X_1 \df^3\bs X_2 
	e^{
	-{\cal R}
	- \Gamma_V\pn{ \mf T - {T_0}  } + \frac{\Gamma_V^2\sigma_t}{4}
	-\frac{ \paren{\mf T - \Tin - {\Gamma_V\sigma_t\ov2}}^2 }{\cblue{2}\sigma_t}
	+i \delta \omega \paren{ \mf T -\Tin - {\Gamma_V\sigma_t\ov2} } 
	}
		\frac{1}{ \mf T -\Tin - {\Gamma_V\sigma_t\ov2}  -i  \sigma_t \delta\omega} W\fn{\mf T} \notag \\
= \ &
	\sqrt{ \frac{\pi^5}{\sigma_t} \paren{ \frac{\sigma_0\sigma_1\sigma_2}{\sigma_s} }^3 }
	\int_{\Tin+{\Gamma_V\sigma_t\ov2}}^{\Tout+{\Gamma_V\sigma_t\ov2}} \df \mf T
	e^{
	- \frac{1}{2\sigma_t} \paren{\mf T - \paren{\Tin {-}  {\Gamma_V\sigma_t\ov2} + i\sigma_t\delta\omega }}^2
	- \Gamma_V\pn{\Tin-T_0}
	- \frac{1}{2} \sigma_t\pn{\delta\omega}^2
	+ \frac{\Gamma_V^2\sigma_t}{4}
	- i \Gamma_V\sigma_t\delta\omega
	} \notag \\
&\quad
	\times \frac{1}{ \mf T -\Tin - {\Gamma_V\sigma_t\ov2}  -i  \sigma_t \delta\omega} \notag \\
\to \ &
	\sqrt{ \frac{\pi^5}{\sigma_t} \paren{ \frac{\sigma_0\sigma_1\sigma_2}{\sigma_s} }^3 }
	\int_{\Tin}^{\infty} \df \mf T'
	e^{
	- \frac{1}{2\sigma_t} \paren{\mf T' - \paren{ \Tin   {-{\Gamma_V\sigma_t\ov2}}   + i\sigma_t\delta\omega }}^2
	- \Gamma_V\pn{\Tin-T_0}
	- \frac{1}{2} \sigma_t\pn{\delta\omega}^2
	+ \frac{\Gamma_V^2\sigma_t}{4}
	- i \Gamma_V\sigma_t\delta\omega
	} \notag \\
	&\quad
		\times \frac{1}{ \mf T' - \Tin  -i  \sigma_t \delta\omega},
}
where in the last fine, we changed the variable to $\mf T' := \mf T - \Gamma_V\sigma_t/2$ and took the limit $\Tout \to \infty$.

In order to use the analytic formula for $\sigma_t >0$ and $\alpha \in \mathbb{C}$,\footnote{
One of the necessary conditions for this relation is
``$\paren{\Tin+\alpha \not\in \mathbb{R}}\&\paren{\Im\fn{\Tin} \not= \Im\fn{\alpha}} \& \paren{\Re\fn{\Tin} \geq \Re\fn{\alpha}}$''.
In our case, the set of these conditions is manifestly fulfilled since $\alpha$ corresponds to ``$\Tin - \Gamma_V\sigma_t/2 + i \sigma_t\delta\omega$''.
}
\al{
\int_{\Tin}^\infty \df t \frac{1}{t - \alpha} e^{- \frac{1}{2\sigma_t} \paren{t-\alpha}^2}
	=
-\half \, \Ei\fn{ - \frac{1}{2\sigma_t} \paren{\Tin-\alpha}^2 },
}
where $\Ei\fn{z}$ is the exponential integral function defined by the principal value of
\al{
\Ei\fn{z} := - \int_{-z}^\infty \df t \frac{e^{-t}}{t},
}
we add an extra term in the denominator of the integrand of $I_\tx{intf}$ such that
\al{
I_\tx{intf}
	&\sim
		\sqrt{ \frac{\pi^5}{\sigma_t} \paren{ \frac{\sigma_0\sigma_1\sigma_2}{\sigma_s} }^3 }
		\int_{\Tin}^{\infty} \df \mf T'
		e^{
		- \frac{1}{2\sigma_t} \paren{\mf T' - \paren{ \Tin  {-{\Gamma_V\sigma_t\ov2}}   + i\sigma_t\delta\omega }}^2
		- \Gamma_V\pn{\Tin-T_0}
		- \frac{1}{2} \sigma_t\pn{\delta\omega}^2
		+ \frac{\Gamma_V^2\sigma_t}{4}
		- i \Gamma_V\sigma_t\delta\omega
		} \notag \\
	&\quad
		\times
		\frac{1}{ \mf T' - {\paren{ \Tin - {\Gamma_V\sigma_t\ov2}  + i\sigma_t\delta\omega }}   },
}
which would underestimate the integral to some extent.
Now, we reach the following analytic form
\al{
I_\tx{intf} \sim
	\sqrt{ \frac{\pi^5}{\sigma_t} \paren{ \frac{\sigma_0\sigma_1\sigma_2}{\sigma_s} }^3 }
	e^{
	- \Gamma_V\pn{\Tin-T_0}
	- \frac{1}{2} \sigma_t\pn{\delta\omega}^2
	+ \frac{\Gamma_V^2\sigma_t}{4}
	- i \Gamma_V\sigma_t\delta\omega
	}
	\times 
	\paren{ -\half \Ei\fn{ -X' } },
}
with
\al{
X' := \frac{\sigma_t}{2} \pn{  \Gamma_V + i \delta\omega    }^2.
}

Here, we will take the leading term of the expansion around infinity for $X'$,
\al{
- \Ei\fn{ -X' }
	=
		e^{ - X' + {\cal O}\paren{ \frac{1}{{X'}^2} } }
			\sqbr{ \frac{1}{X'} + {\cal O}\fn{ \frac{1}{{X'}^2} } },
}
which leads to
\al{
I_\tx{intf} + \paren{I_\tx{intf}}^\ast
	&\sim
		\frac{1}{2}
		\sqrt{ \frac{\pi^5}{\sigma_t} \paren{ \frac{\sigma_0\sigma_1\sigma_2}{\sigma_s} }^3 }
		e^{
		- \Gamma_V\pn{\Tin-T_0}
		- \frac{\Gamma_V^2\sigma_t}{4}
		}
		\sqbr{
			\frac{ e^{ -2i \Gamma_V\sigma_t \delta\omega } }
			       {{\sigma_t\ov2}\pn{\Gamma_V+i\delta\omega}^2}
			+
			\frac{ e^{ 2i \Gamma_V\sigma_t \delta\omega } }
			       {{\sigma_t\ov2}\pn{\Gamma_V-i\delta\omega}^2}
		},
}
and
\al{
dP^\tx{intf}_{V \to P\ol{P}}
	&\sim
		-
		|g_\tx{eff}|^2 N_V^2 \frac{1}{2 E_0}
		\frac{\df^3 \bs P_1}{(2\pi)^3 2E_1} \frac{\df^3 \bs P_2}{(2\pi)^3 2E_2}
		(2 \pi)^4
		\paren{ \sqrt{\frac{\sigma_t}{\pi}} e^{-\sigma_t\paren{\delta \omega}^2} }
		\sqbr{ \paren{\frac{\sigma_s}{\pi}}^{3/2} e^{-\sigma_s\paren{\delta \bs P}^2} } \notag \\
	&\quad
		\times e^{ - \Gamma_V\pn{\Tin-T_0} - \frac{\Gamma_V^2\sigma_t}{4} + \frac{1}{2} \sigma_t\pn{\delta\omega}^2  }
		\notag \\
	&\quad
		\times
		\sqrt{\frac{2\sigma_t}{\pi}}
		\frac
		{ \paren{\Gamma_V^2 - \pn{\delta\omega}^2} \cos\fn{2 \Gamma_V\sigma_t\delta\omega}
		  - 2 \Gamma_V\delta\omega \sin\fn{2 \Gamma_V\sigma_t\delta\omega} }
		{\sigma_t\paren{  \Gamma_V^2 + \pn{\delta\omega}^2    }^2}
		 \ab{ \widetilde{F}\fn{V_-} }^2.
}

We perform the momentum integrals in the non-relativistic limit:
\al{
P^\tx{intf}_{V \to P\ol{P}}
	&\Rightarrow
		-
		N_V^2   e^{  -\Gamma_V\pn{\Tin-{T_0}} } 
		\frac{g_V^2}{3} \frac{1}{2 m_V} \frac{1}{(2\pi)^2 4 E_1 E_2} \times
		\frac{m_P^6}{8} \paren{ \int_{0}^{\infty} 4 \pi V_+^2 \df V_+ } 
		\paren{ \int_0^\infty 4\pi V_-^2  \df V_- } \notag \\
	&\quad
		\times
		\frac{1}{\pi^2} \sigma_t^{1/2}
		\sigma_s^{3/2} \paren{m_P^2 V_-^2}
		e^{ - F_\tx{intf}(V_+, V_-) }
		\sqrt{\frac{2\sigma_t}{\pi}} \notag \\
	&\quad
		\times
		\frac
		{ \paren{\Gamma_V^2 - \pn{\delta\omega}^2} \cos\fn{2 \Gamma_V\sigma_t\delta\omega}
		  - 2 \Gamma_V\delta\omega \sin\fn{2 \Gamma_V\sigma_t\delta\omega} }
		{\sigma_t\paren{  \Gamma_V^2 + \pn{\delta\omega}^2    }^2}
		\ab{ \widetilde{F}\fn{V_-} }^2,
		\label{eq:mixed-nonrelativistic}
}
where
\al{
F_\tx{intf}(V_+, V_-)
	:=&\,
		\sigma_s\paren{\delta \bs P}^2 + \frac{1}{2} \sigma_t\paren{\delta \omega}^2  + \frac{\Gamma_V^2\sigma_t}{4}    \notag \\
	=&\,
		\sigma_s m_P^2 V_+^2 \notag \\
	&
		+
		\underbrace{\frac{2 \sigma_P}{ \paren{ 1 - \frac{2\sigma_s}{\sigma_P} } V_+^2 +  V_-^2 }}_{= \sigma_t}
		\br{
		\frac{1}{2}
		\sqbr{
		m_P \paren{ \frac{1}{4} - \frac{\sigma_s}{\sigma_P} } V_+^2 + \frac{1}{4} m_P V_-^2 - \paren{m_V - 2m_P}
		}^2 + {\Gamma_V^2\ov4}  }.
}
This function is similar to $F_\text{bulk}$, but a factor of the half comes in front of $\sigma_t\paren{\delta \omega}^2$.
As in the case of the bulk part, we can see
\al{
e^{ - F_\tx{intf} } \underbrace{\to}_{V_+ \to \infty} 0, \quad
e^{ - F_\tx{intf} } \underbrace{\to}_{V_- \to \infty} 0.
}
Stationary points $(V_+^s, V_-^s)$ are defined by
\al{
\frac{\pal F_\tx{intf} (V_+^s, V_-^s)}{\pal V_+} = 0 \quad \text{and} \quad
\frac{\pal F_\tx{intf} (V_+^s, V_-^s)}{\pal V_-} = 0,
}
and we find one stationary point in the kinetic region ($V_+ \simeq 0$ and $V_- > 0$),
\al{
\paren{V_+^\text{I}, V_-^\text{I}}
	&:=
		\paren{0, \, 
			\frac{ 2 \sqbr{\paren{m_V-2m_P}^2+{\Gamma_V^2\ov2}}^{1/4}  }{\sqrt{m_P}}
			}.
}
Note that 
\al{
{F_\tx{intf}^0} =
F_\tx{intf}\paren{V_+^\text{I}, V_-^\text{I}}
	& := 
		\frac
		{ m_P \sigma_P  }
		{2}
		\sqbr{ -\paren{m_V - 2m_P}  + \sqrt{ \paren{m_V - 2m_P}^2+{\Gamma_V^2 \ov2} }  },
}
which takes a positive value when $\Gamma_V$ is finite, and goes to zero in the (unphysical) limit $\Gamma_V \to 0$.
Also, we can see
\al{
\left.\frac{\pal^2 F_\tx{intf} }{\pal V_+^2}\right|_{\paren{V_+^\text{I}, V_-^\text{I}}}
	&= 
		\frac{1}{2} m_P^2 \sigma_s 
		\sqbr{
		3 + \frac{ m_V - 2m_P}{\sqrt{\paren{m_V-2m_P}^2+{\Gamma_V^2\ov2}}}
		}  =: \pn{\frac{1}{2} m_P^2 \sigma_s}4{A_\tx{intf}},
		\notag \\
\left.\frac{\pal^2 F_\tx{intf} }{\pal V_-^2}\right|_{\paren{V_+^\text{I}, V_-^\text{I}}}
	   &= 
	   	\frac{1}{2} m_P^2 \sigma_P,  \notag \\
\left.\frac{\pal^2 F_\tx{intf} }{\pal V_+ \pal V_-}\right|_{\paren{V_+^\text{I}, V_-^\text{I}}}
	   &=
	   	0,
}
where $A_\tx{intf} \to 1$ under the (unphysical) limit $\Gamma_V \to 0$.

Now, we evaluate the non-relativistic integral in Eq.~\eqref{eq:mixed-nonrelativistic}.
By use of Eqs.~\eqref{eq:integral_formula_1} and \eqref{eq:integral_formula_2}, we reach
\al{
P^\tx{intf}_{V \to P\ol{P}}
	&\sim
		-
		\paren{\frac{g_V^2 m_P^3  N_V^2  e^{  -\Gamma_V\pn{\Tin-{T_0}} }   }{12 \pi m_V E_1 E_2}}
		\frac{ m_P \sqrt{\sigma_P}}{ 2\sqrt{2} \sqrt{\pi}}
		\paren{ V_-^\text{I} }^2
		\notag \\
	&\quad
		\times
		\frac{1}{2}
		\sqbr{  1 + \text{erf}\paren{  \frac{  m_P\sqrt{\sigma_P}  {V_-^\text{I}}  }{2}  }  }
		{\frac{e^{-F^0_\tx{intf}}}{A_\tx{intf}^{3/2}}}\,
		 \ab{ \widetilde{F}\fn{V^\text{I}_-} }^2 \notag \\
	&\quad
		\times
		\sqbr{
		\frac
		{ \paren{\Gamma_V^2 - \paren{\wt{\delta\omega}}^2 }
			\cos\fn{2 \Gamma_V\wt{\sigma_t} \wt{\delta\omega}}
		  	- 2 \Gamma_V \wt{\delta\omega} \sin\fn{2 \Gamma_V\wt{\sigma_t} \wt{\delta\omega}} }
		{\wt{\sigma_t}\paren{  \Gamma_V^2 + \paren{\wt{\delta\omega}}^2     }^2}
		},
	\label{eq:P-intf}
}
where the factor $\sqrt{\sigma_P}$ comes in the overall factor (instead of $\Gamma_V^{-1}$ 
compared with the bulk result in Eq.~\eqref{eq:Gamma-phi-to-KKbar-V1_Nonrelativistic})
and we defined the parameters
\al{
\widetilde{\sigma_t}
	&:=
		\sigma_t \Big|_{V_+ \to V_+^\text{I},\, V_- \to V_-^\text{I}}
	=
		\left.
		\frac{2 \sigma_P}{ \paren{ 1 - \frac{2\sigma_s}{\sigma_P} } V_+^2 +  V_-^2 }
		\right|_{V_+ \to V_+^\text{I},\, V_- \to V_-^\text{I}}
	=
		\frac{2 \sigma_P}{ \paren{V_-^\text{I}}^2 }, \\
\widetilde{\delta\omega}
	&:=
		\delta\omega \Big|_{V_+ \to V_+^\text{I},\, V_- \to V_-^\text{I}}
	=
		\left.
		\sqbr{
		m_P \paren{ \frac{1}{4} - \frac{\sigma_s}{\sigma_P} } V_+^2 + \frac{1}{4} m_P V_-^2 - \paren{m_V - 2m_P}
		}
		\right|_{V_+ \to V_+^\text{I},\, V_- \to V_-^\text{I}} \notag \\
	&=
		\frac{1}{4} m_P \paren{V_-^\text{I}}^2 - \paren{m_V - 2m_P}.
}

\section{Plane-wave decay rate}
\label{decay rate section}

From the effective Hamiltonian~\eqref{eq:effective-Hamiltonian}, under the use of the form factor~\eqref{eq:form-factor-final-form},
it is immediate to get the following form in the plane-wave formalism for the resting $V$,
\al{
\Gamma^\tx{plane}_{V \to P \overline{P}}
	&=
		\frac{2}{3} \paren{ \frac{{g_{VP}}^2}{4\pi} } \frac{ |\bs k_{P}|^3 }{m_V^2}
		\left|   \widetilde{F} \fn{\frac{\bs k_1 - \bs k_2}{2}}   \right|^2 \notag \\
	&=
		\frac{{g_{VP}}^2}{48 \pi m_V^2} \paren{ m_V^2 - 4 m_{P}^2 }^{3/2}
		\left|   \frac{1}{\paren{\frac{R_0 \paren{\bs k_1 - \bs k_2}}{2}}^2 + 1 }  \right|^2,
		\label{plane-wave decay rate with form factor}
}
where $g_{VP}$ represents $g_{V+}$ (for $P = P^+$) or $g_{V0}$ (for $P = P^0$) and
the magnitude of the final-state momenta in the center-of-mass frame is given as
\al{
\ab{\bs k_P} = \ab{\bs k_1} = \ab{\bs k_2}
	&=
		\frac{1}{2} \paren{ m_V^2 - 4 m_{P}^2 }^{1/2}.
}
The form factor part does not take the non-relativistic limit as in Eq.~\eqref{eq:form-factor-final-form},
and currently, $\bs k_1 = - \bs k_2$ and thus
\al{
\ab{ \bs k_1 - \bs k_2 } = 2 \ab{\bs k_1} = \paren{ m_V^2 - 4 m_{P}^2 }^{1/2}.
}

Note that under the non-relativistic approximation taken in Eqs~\eqref{eq:NRLimit-1} and \eqref{eq:NRLimit-2}, it is easy to obtain the approximated form:
\al{
\Gamma^\tx{plane}_{V \to P \overline{P}}
	&\Rightarrow
	\Gamma^\tx{plane,non-rel}_{V \to P \overline{P}}
	:=
		\frac{{g_{VP}}^2 m_P^2}{12 \pi m_V} \paren{ \frac{m_V - 2 m_P}{m_P} }^{3/2}
		\left|   \frac{1}{\paren{\frac{R_0 m_P \paren{\bs V_1 - \bs V_2}}{2}}^2 + 1 }  \right|^2,
	\label{eq:PWDR-NRlimit}
	\\
\ab{\bs V_1 - \bs V_2}
	&\approx
		\frac{ 2\paren{m_V - 2m_P}^{1/2} }{ \sqrt{m_P} }.
}

\section{Comparison with plane-wave decay rate}
\label{sec:Gamma-zero-limit}

As shown in Sec.~\ref{magnitude section}, the boundary contribution dominates over the bulk and interference ones in all regions of the parameter space $R_0$ and $\sigma_P$.
This is because of the fast decay of the vector meson $V$ due to strong interactions.
The fast decay suppresses the contribution from the bulk so that the contribution from the initial time boundary becomes more significant compared to the bulk one.

As mentioned above, the plane-wave decay width $\Gamma_V$ is dependent on other theory parameters such as $g_V$, $m_P$, and $m_V$. Therefore, it is meaningless to take an ``on-shell'' limit $\Gamma_V\to0$ with other parameters being fixed. Nevertheless, one might pretend that one can take this unphysical limit, and then one extracts the plane-wave decay width~\eqref{eq:PWDR-NRlimit}:
\al{
\Gamma_{V\to P\ol P}^\tx{plane,non-rel}
	&=	\lim_{\sigma_P\to\infty}\pn{\lim_{\Gamma_V\to0}\Gamma_VP_{V\to P\ol P}}
	=	\lim_{\sigma_P\to\infty}\pn{\lim_{\Gamma_V\to0}\Gamma_VP_{V\to P\ol P}^\tx{bulk}},
	\label{plane, non-rel}
}
where the limit of large wave-packet size $\sigma_P\to\infty$ is taken after the limit $\Gamma_V\to0$.\footnote{
As said above, $\sigma_V$ anyway drops out of the result at this order of the saddle-point approximation.
}
The second equality in Eq.~\eqref{plane, non-rel} is derived as follows:
Under the limit $\Gamma_V\to0$, the two ratios $\Gamma_VP^\text{bdry}_{V \to P\ol{P}}$ and $\Gamma_VP^\text{intf}_{V \to P\ol{P}}$
approach zero since $P^\text{bdry}_{V \to P\ol{P}}$ and $P^\text{intf}_{V \to P\ol{P}}$ are not proportional to $\Gamma_V^{-1}$ as shown in
Eqs.~\eqref{eq:P-integrated_bdry} and \eqref{eq:P-integrated_intf}.
When taking the above limits $\Gamma_V\to0$ and $\sigma_P\to\infty$, the following is satisfied:
\al{
V_-^\text{B}
	&\to
		2 \frac{\sqrt{m_V-2m_P}}{\sqrt{m_P}},&
\text{erf}\paren{  \frac{  m_P\sqrt{\sigma_P}  {V_-^\text{B}}  }{\sqrt{2}}  }
	&\to
		1, \notag \\
F_\text{bulk}^0
	&\to
		0,&
A_\text{bulk}
	&\to
		1,
}
as well as the physical requirement $N_V \to 1$ from $\Gamma_V\to0$.

In the actual setup, the (unphysical) $\Gamma_V\to0$ limit is not good, and we see that the boundary contribution dominates over the bulk one for the parameters corresponding to the real experiments mainly because of the exponential suppression factor $e^{-F^0_\tx{bulk}}$.

\section{A brief comment on the isospin breaking of the $\rho$ system}
\label{sec:R-rho}

\begin{figure}
\centering
\includegraphics[width=.47\textwidth]{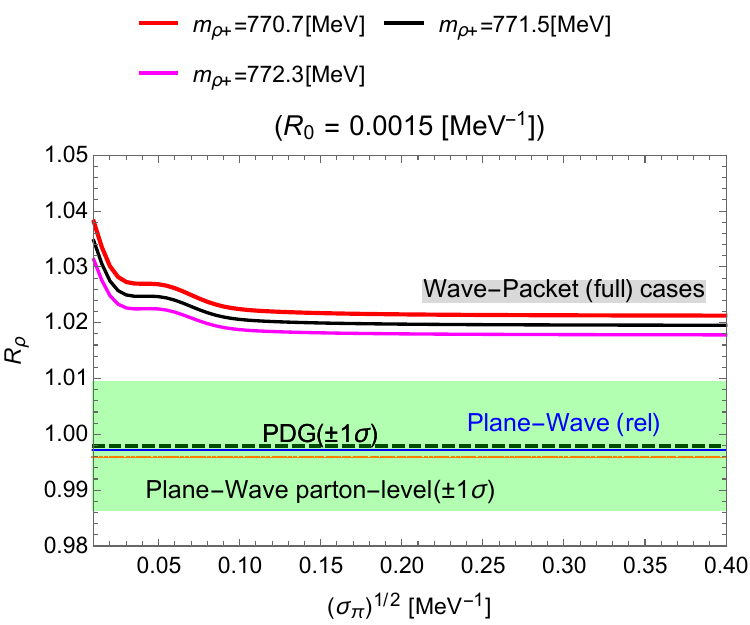} \ \
\includegraphics[width=.47\textwidth]{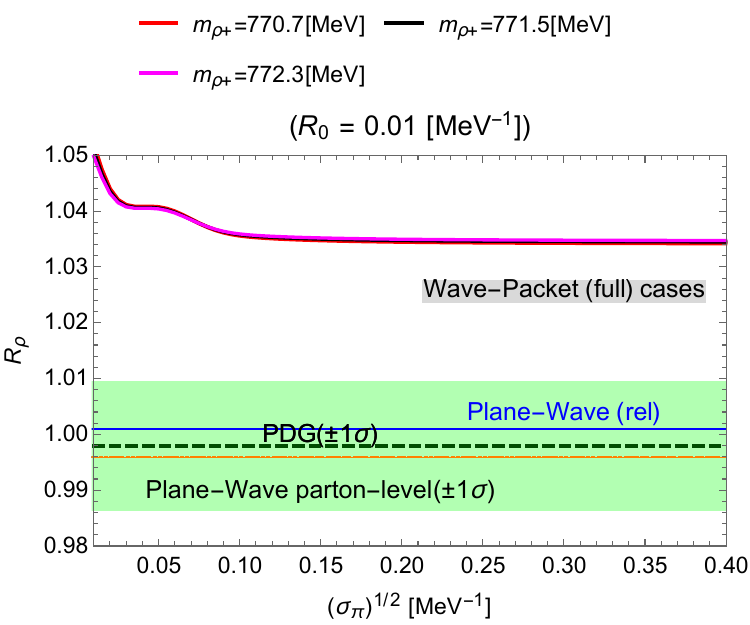} \\
\caption{
Values of $R_\rho$ defined in Eq.~\eqref{eq:R-rho} are depicted for $R_0 = 0.0015\,\tx{MeV}^{-1}$ (Left panel) and $R_0 = 0.01\,\tx{MeV}^{-1}$ (Right panel).
The captions ``Wave-Packet (full)'', ``Plane-Wave (rel)'', and ``Plane-Wave parton-level'' correspond to
$R_\rho^\tx{WP}$, $R_\rho^\tx{plane} $, and $R_\rho^\tx{parton}$, respectively.
See the main text of this section for other details.
}

\label{fig:R-rho}
\end{figure}

We will make a brief comment on the isospin breaking of the $\rho$ system.
Here, we define the following ratio,
\al{
R_\rho^\tx{WP}
	&:=
		{P_{\rho^+ \to \pi^+ \pi^0}\ov P_{\rho^0 \to \pi^+ \pi^-}},&
R_\rho^\tx{plane} 
	&:= \frac{\Gamma^\tx{plane}_{\rho^+ \to \pi^+ \pi^0}}{\Gamma^\tx{plane}_{\rho^0 \to \pi^+ \pi^-}},&
R_\rho^\tx{parton} 
	&:= \frac
		{\left.\Gamma^\tx{plane}_{\rho^+ \to \pi^+ \pi^0}\right|_\tx{without form factor}}
		{\left.\Gamma^\tx{plane}_{\rho^0 \to \pi^+ \pi^-}\right|_\tx{without form factor}},&
	\label{eq:R-rho}
}
where we replace $m_P$ to $\paren{m_{\pi^+} + m_{\pi^0}}/2$ for the calculations of $\rho^+ \to \pi^+ \pi^0$.
In Fig.~\ref{fig:R-rho}, the $R_\rho$ in terms of the wave packet and the relativistic plane wave is depicted
for $R_0 = 0.0015\,\tx{MeV}^{-1}$ (Left panel) and $R_0 = 0.01\,\tx{MeV}^{-1}$ (Right panel); see Sections~\ref{integrated decay probability section} and \ref{sec:planewave-format} for the details of
the wave-packet decay probabilities and the plane-wave decay rates, respectively.

First, we mention the experimental inputs that we adopt.
As official results reported by the PDG~\cite{ParticleDataGroup:2022pth},
\begin{itemize}
\item
Six digits are reported as typical mass scales of the broad-resonant $\rho$ system, where their central values are located from $769.0\,\text{MeV}$ to $775.26\,\text{MeV}$.
\item
Also, six digits are shown as typical width scales of the $\rho$ system,
where their central values are located from $147.4\,\text{MeV}$ to $151.5\,\text{MeV}$.
\item
The difference between $m_{\rho^0}$ and $m_{\rho^+}$ takes $m_{\rho^0} -  m_{\rho^+} = -0.7 \pm 0.8\,\tx{MeV}$.
\item
The difference between $\Gamma_{\rho^0}$ and $\Gamma_{\rho^+}$ takes $\Gamma_{\rho^0} -  \Gamma_{\rho^+} = 0.3 \pm 1.3\,\tx{MeV}$.
\end{itemize}
Since $R_\rho$ measures isospin-violating effects, and thus it is insensitive to what is a typical mass scale of the system.
So, we simply take $m_{\rho^0} = 770 \pm 1\,\text{MeV}$ and $\Gamma_{\rho^0} = 147.4 \pm 0.8\,\text{MeV}$. Based on the mass scale,
$m_{\rho^+}$ is estimated by use of the above data on $m_{\rho^0}$ and $m_{\rho^0} -  m_{\rho^+}$ as
\als{
m_{\rho^+} &= 770.7\,\tx{MeV}\,\tx{(central)},&
m_{\rho^+} &= 771.5\,\tx{MeV}\,(+1\sigma),&
m_{\rho^+} &= 772.3\,\tx{MeV}\,(+2\sigma),&
}
while the central value of $\Gamma_{\rho^+}$ is similarly estimated as $147.7\,\tx{MeV}$.
As expected and as shown in Fig.~\ref{fig:R-rho}, $R_\rho$ can depend on the mass difference sizably.

Next, we discuss the numerical result shown in Fig.~\ref{fig:R-rho}.
As expected, the experimental result is located very near the unity since the $\rho$ vector mesons do not contain heavy quarks.
The plane-wave theoretical predictions with and without the form-factor effect show similar results.
The wave-packet predictions deviate from the unity several percent upward, showing fewer agreements.
Nevertheless, this does not necessarily mean that wave-packet formalism works less effectively for the $\rho$ system than plane-wave formalism because the current wave-packet result is precise only for non-relativistic systems (due to the usage of non-relativistic approximations). Note that the decays $\rho^+ \to \pi^+ \pi^0$ and $\rho^0 \to \pi^+ \pi^-$ are fully relativistic due to the large difference between the total masses of the initial state and final state.
Several percent of theoretical errors are expected from the fully-relativistic wave-packet prediction, which is beyond the scope of this paper. The typical scale of the form factor under the current scheme $R_0 = 0.0015\,\tx{MeV}$ works well compared with $R_0 = 0.01\,\tx{MeV}$.

Also, we comment on the dependence on $R_0$ in $R_\rho$.
As typically observed in the curves of ``Plane-Wave (rel)'' of Fig.~\ref{fig:R-rho}, the form-factor part of $R_\rho$ is not sensitive to $R_0$ since the total mass differences between the initial and the final states take almost the same values in $\rho^+ \to \pi^+ \pi^0$ and $\rho^0 \to \pi^+ \pi^-$ and their final-state phase spaces are wide.

\bibliographystyle{JHEP}
\bibliography{manuscript}

\end{document}